\documentclass[twocolumn,superscriptaddress,nofootinbib,amsmath,amssymb,preprintnumbers,longbibliography]{revtex4-2}
\usepackage{amsmath,amsfonts,bm,graphicx,hyperref,color,physics}
\allowdisplaybreaks[1]

\usepackage{mathtools} 

\renewcommand\d{\partial}
\newcommand\+{\dagger}
\newcommand\<{\langle}
\renewcommand\>{\rangle}

\newcommand\rmR{\mathrm{R}}
\newcommand\rmL{\mathrm{L}}
\newcommand\rmT{\mathrm{T}}
\newcommand\rmC{\mathrm{C}}
\newcommand\ret{\mathrm{ret}}
\renewcommand\adv{\mathrm{adv}}

\newcommand\A{\mathcal{A}}
\newcommand\G{\mathcal{G}}
\renewcommand\S{\mathcal{S}}
\newcommand\T{\mathcal{T}}
\newcommand\N{\mathcal{N}}
\renewcommand\H{\mathcal{H}}
\newcommand\I{\mathcal{I}}

\renewcommand\r{{\bm r}}

\renewcommand\k{{\bm k}}

\newcommand\0{\mathbf{0}}

\usepackage{multirow}

\begin{document}
\title{Thermomagnetic anomalies in quantum magnon transport caused by tunable junction geometries in cold atomic systems}

\author{Yuta Sekino}
\affiliation{Interdisciplinary Theoretical and Mathematical Sciences Program (iTHEMS), RIKEN, Wako, Saitama 351-0198, Japan}
\affiliation{Nonequilibrium Quantum Statistical Mechanics RIKEN Hakubi Research Team, RIKEN Cluster for Pioneering Research (CPR), Wako, Saitama 351-0198, Japan}

\author{Yuya Ominato}
\affiliation{Waseda Institute for Advanced Study, Waseda University, Shinjuku, Tokyo 169-8050, Japan.}
\affiliation{Kavli Institute for Theoretical Sciences, University of Chinese Academy of Sciences, Beijing, 100190, China.}

\author{Hiroyuki Tajima}
\affiliation{Department of Physics, School of Science, The University of Tokyo, Tokyo 113-0033, Japan}

\author{Shun Uchino}
\affiliation{Faculty of Science and Engineering, Waseda University, Tokyo 169-8555, Japan}

\author{Mamoru Matsuo}
\affiliation{Kavli Institute for Theoretical Sciences, University of Chinese Academy of Sciences, Beijing, 100190, China.}
\affiliation{CAS Center for Excellence in Topological Quantum Computation, University of Chinese Academy of Sciences, Beijing 100190, China}
\affiliation{Advanced Science Research Center, Japan Atomic Energy Agency, Tokai, 319-1195, Japan}
\affiliation{RIKEN Center for Emergent Matter Science (CEMS), Wako, Saitama 351-0198, Japan}

\begin{abstract}
We study magnon-driven spin and heat transport in a magnetic linear junction (MLJ) formed by two ferromagnets in optical lattices linked via linearly aligned bonds. 
Using the Schwinger-Keldysh formalism, we uncover that under weak effective Zeeman fields, where Bose-Einstein statistics of magnons dominate, magnonic criticality dramatically enhances spin and thermal conductances.
These singular transport properties depend on the junction geometry, and the transport properties qualitatively differ between the linear junction in this study and the point contact in our previous work.
The quantum-enhanced conductances result in the breakdown of the magnonic Wiedemann-Franz (WF) law.
In the classical regime at temperatures much lower than magnon energy gaps, we find that a magnonic Lorenz number becomes independent of temperature yet dependent on junction geometry, sharply contrasting with the universal WF law for Fermi liquids.
We also find that the interface geometry of MLJ decouples spin and heat relaxations between ferromagnets with decay times insensitive to temperature and effective Zeeman fields.
These dynamics reveal junction-geometry-sensitive magnon transport distinct from Fermi liquids, paving the way for new avenues in thermomagnetic research leveraging the tunability of cold atomic systems.
\end{abstract}

\date{\today}
\maketitle

\section{Introduction}
There has been a surge of interest in investigating particle and heat currents in transport systems with the use of ultracold atoms~\cite{sidorenkov2013second,brantut2013thermoelectric,krinner2017,husmann2018breakdown,PhysRevX.10.011042,christodoulou2021observation,hoffmann2021second,yan2022thermography,Krinner2016-gd}.
Such atomic systems enable the creation of the ultraclean quantum many-body states, devoid of impurity scattering or phonon effects characteristic of solid-state systems, thus offering an ideal platform for exploring energy conversion mechanism in the case where particle and heat currents intersect.
The realization of thermoelectric transport in a two-terminal system of Fermi gases without lattice provides a controllable model for examining energy conversion mechanisms~\cite{brantut2013thermoelectric,husmann2018breakdown}.

Recent advances have heightened interest in spin-transport studies using ultracold atoms~\cite{enss2019universal,Schafer2020tools}, with works uncovering fascinating phenomena, including super-exchange dynamics in Mott insulators achieved with optical lattices~\cite{Nichols2019,Fukuhara2013-jv,Fukuhara2013-tj,Hild2014-hy,Jepsen2020-fg,Jepsen2021-ew,Jepsen2022-sf,Wei2022-qx} and spin diffusion in strongly-interacting Fermi gases~\cite{sommer2011universal,sommer2011spin}.
Inspired by the development of experimental techniques for spin transport and the progress in heat engine studies mentioned above, there is a growing impetus for quantum simulations of thermomagnetism, a research field where spin and heat currents interplay in two-terminal systems.
In solid-state physics, such cross-correlated spin-heat dynamics have been actively investigated from both theoretical and experimental viewpoints to pursue the mechanism of efficient spin-heat conversion~\cite{uchida2008observation,jaworski2010observation,uchida2010spin,xiao2010theory,adachi2011linear,adachi2013theory,ohnuma2017theory,matsuo2018spin,kato2019microscopic,Flipse2014-ne,Daimon2016-vy}, leading to a research field called `spin caloritronics'~\cite{bauer2012spin}.
However, the complexities inevitable in solid materials including interface disorder and impurity scatterings make it challenging to clarify factors in efficient spin-heat conversion mechanisms.
By contrast, the ultracold atoms without such complex factors offer a prime platform to study a coupling between spin and heat currents systematically.

Motivated by the controllability of cold atoms, our previous work has proposed thermomagnetic transport between two ferromagnetic insulators (FIs) linked via a magnonic quantum point contact (MQPC)~\cite{Sekino2024-vp}, which is a spin-system analog of a quantum point contact (QPC).
In this setup, we found anomalous thermomagnetic properties such as a divergence in spin conductance and a slowing down of spin relaxation that manifest in the weak effective-Zeeman-field limit.
These transport properties results from the magnonic critical point, at which magnons are regarded as a gapless Nambu-Goldstone mode.
We also discussed that the access to the vicinity of the magnonic critical point is feasible in cold-atomic systems yet elusive in conventional magnon transport setups of sold FIs.

The controllability of spatial degrees of freedom in ultracold atomic systems allows the systematic exploration of how interface geometry affects tunneling transport~\cite{krinner2017,doi:10.1116/5.0026178,amico2022atomtronic}.
The geometry of the interface can significantly affect the transport properties of junction systems, as exemplified by the quantization of conductance in QPCs~\cite{van-Houten1996-sx,krinner2017}. 
By utilizing the interaction between atoms and light, a single-atom potential can be controlled with high spatial resolution, leading to the realization of various junction systems for atomic gases such as QPCs and planar Josephson junctions~\cite{krinner2017}.
In solids materials, although a variety of junction geometries have been realized~\cite{Nazarov2009-ev}, the precise spatial control of the interfaces is usually challenging due to complicated factors such as lattice mismatch.
By contrast, the hopping rates for ultracold atoms in an optical lattice can be controlled individually at each site, which allows the formation of a lattice-mismatch-free interface with tunable spatial structure and tunneling strength, making it highly suitable for systematically studying the effects of interface geometry on tunneling transport.

\begin{figure}
    \centering
    \includegraphics[width=0.95\columnwidth,clip]{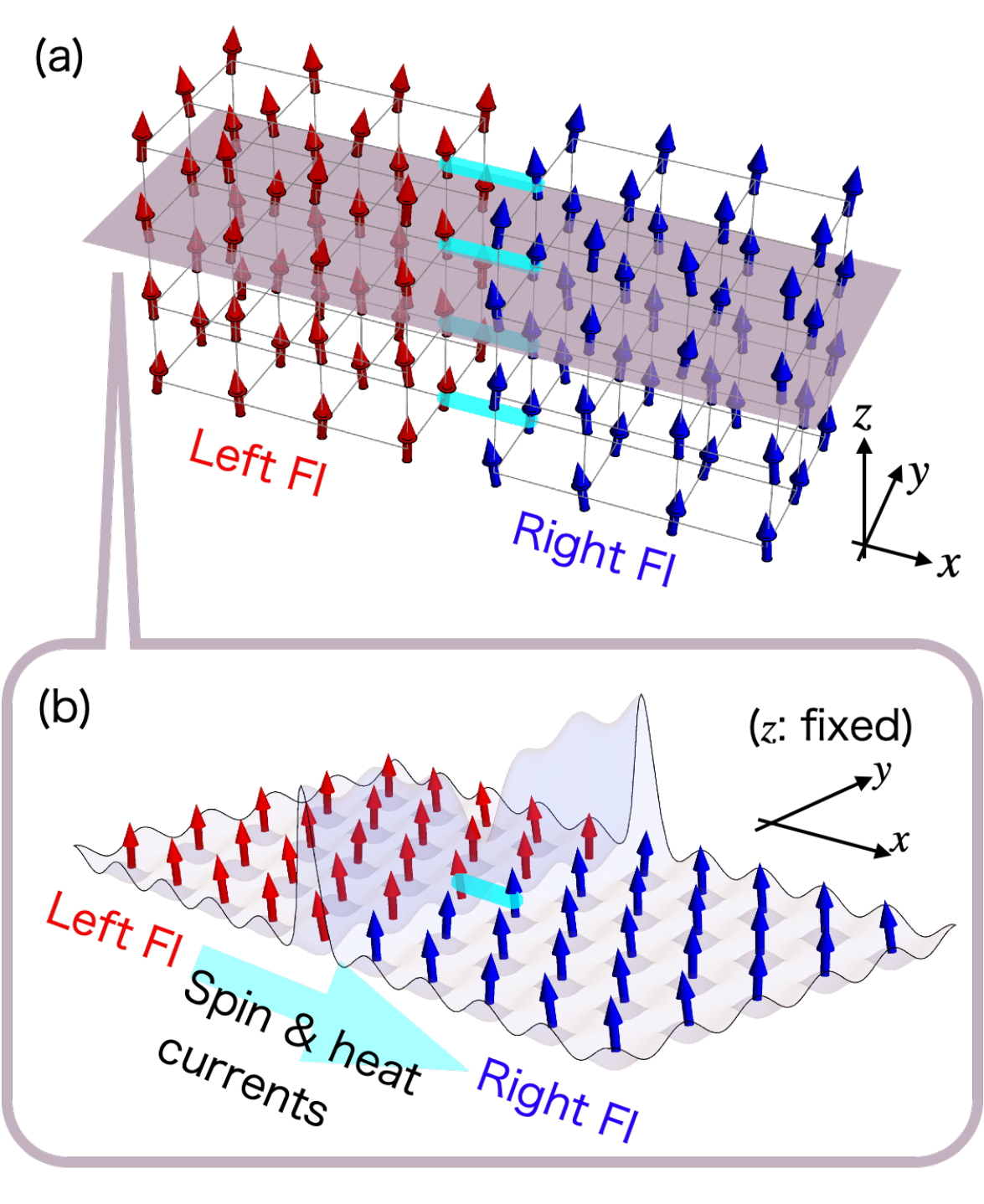}
    \caption{\label{fig:system}
    (a) Schematic for a magnetic linear junction (MLJ).
    Two ferromagnetic insulators (FIs) are coupled to each other via weak exchange interactions along the light blue bonds, which are aligned in the $z$-direction.
    (b) Potential profile on the purple plane in (a) with $z$ fixed to construct an MLJ in cold-atomic systems.
    The left and right FIs are separated by a barrier potential whose saddle point provides an exchange coupling between FIs.
    }
\end{figure}
In this paper, we microscopically explore cross-correlated spin and heat tunneling transport by magnons in a magnetic linear junction (MLJ) consisting of two FIs in optical lattices weakly linked via bonds linearly aligned in the $z$ direction as depicted in Fig.~\ref{fig:system}.
Using the nonequilibrium Green's function approach, we investigate the impact of the interface geometry of MLJ on magnon transport.
We uncover that, when effective Zeeman fields are weak, the magnonic criticality dramatically enhances spin and thermal conductances, and that these singular behaviors are influenced by the linear interface geometry, deviating from the MQPC case.
We then discuss the magnonic Wiedemann-Franz (WF) law~\cite{Nakata2015-ns,Nakata2017-hu,Nakata2017-ot,Nakata2017-iy,Sano2023-ff,Mook2018-oy,nakata2018magnonic,Nakata2022-ep}.
We point out its nonuniversality for classical magnon gases due to sensitivity to interface geometry, which is in a sharp contrast to the universal WF law for Fermi liquids~\cite{Franz1853-do}. 
On the other hand, the magnonic WF law is violated 
in regimes of weak effective Zeeman fields in which 
Bose-Einstein statistics of magnons plays a crucial role.
We also study relaxation dynamics of spin and heat, whose observations allow one to experimentally extract the conductances, and we find that these dynamics are decoupled, with decay times that are insensitive to effective Zeeman fields and temperature.
The above anomalous thermomagnetic phenomena result from the Bose-Einstein statistics of magnons which makes their transmittance sensitive to the linear interface geometry. 
This situation is in contrast to that for Fermi liquids such as electrons in metal and weakly interacting ultracold Fermi gases, for which the presence of a Fermi surface makes tunneling currents insensitive to the junction geometry when a tunneling coupling is weak.
Our results broaden the research horizon of thermomagnetic transport, traditionally confined to solid-state physics, spintronics, and mesoscopic systems, opening new avenues in quantum simulations with cold atomic systems. 

This paper is organized as follows:
Section~\ref{sec:model} is devoted to introducing our model and the quantum regime of magnons, where their Bose-Einstein statistics dominate.
Section~\ref{sec:magnon_transport} focuses on stationary spin and heat currents, discussing critical enhancement of conductances and the magnonic WF laws.
In Sec.~\ref{sec:relaxation}, we discuss relaxation dynamics of spin and heat between two FIs through MLJ.
We summarize the paper in Sec.~\ref{sec:summary}.
Throughout this paper, we set $\hbar=k_B=1$.

\section{\label{sec:model}Model}
We consider a system consisting of two ferromagnetic insulators (FIs) with spin $\S=1/2$ connected by the MLJ in Fig.~\ref{fig:system}.
Such a situation can be realized with two-component bosonic atoms in optical lattices under a sharp barrier potential~\cite{PhysRevLett.90.100401,Duan2003Controlling,Garcia-Ripoll:2003aa,Altman:2003aa,De_Hond2022-xi}.
To analyze the system, we use the tunneling Hamiltonian formalism~~\cite{mahan2000many}.
The corresponding Hamiltonian is given by
\begin{align}\label{eq:total_H}
\hat{\H}&=\hat{H}_\mathrm{L}+\hat{H}_\mathrm{R}+\hat{H}_\mathrm{T}.
\end{align}
The first two terms denote bulk of FIs:
\begin{align}\label{eq:H_alpha}
\hat{H}_{\alpha=\rmL,\rmR}&=-J\sum_{\<\r_\alpha,\r_\alpha'\>}{{\hat{\bm{s}}}}_{\r_\alpha}\cdot{{\hat{\bm{s}}}}_{\r_\alpha'},
\end{align}
where, $J>0$ is the nearest-neighbor coupling constant, ${{\hat{\bm{s}}}}_{\r_\alpha}=(\hat{s}_{\r_\alpha}^x,\hat{s}_{\r_\alpha}^y,\hat{s}_{\r_\alpha}^z)$ are spin-1/2 operators at site $\r_\alpha$ for the reservoir $\alpha=\mathrm{L},\mathrm{R}$.
The last term denotes the tunneling term
\begin{align}\label{eq:H_T}
\hat{H}_{\rmT}&=-J_\rmT\sum_{\<\r_\rmL,\r_\rmR\>'}{{\hat{\bm{s}}}}_{\r_\rmL}\cdot{{\hat{\bm{s}}}}_{\r_\rmR},
\end{align}
where $\<\r_\rmL,\r_\rmR\>'$ is a nearest-neighbor pair of spins on the interface denoted by light blue bonds in Fig.~\ref{fig:system}.
The coupling constant $J_\rmT$ between FIs is much smaller than that within the bulk
($0<J_\rmT\ll J$).
It is assumed that the left and right FIs have the same lattice structure.
We mention that recently a similar junction system, i.e., $XY$ spin ladders connected with MQPC has been realized with a superconducting quantum processor~\cite{Zhang2024-kz}.

We note that the Zeeman terms are not explicitly included in $\hat{H}_\alpha$ and that such external fields are not applied to the system.
On the other hand, we consider effective Zeeman fields $h_\alpha>0$ controlled by the population imbalance between two internal states of atoms as in our previous work~\cite{Sekino2024-vp}.
These fields works as Lagrange multipliers to fixed the $z$ component of the magnetization and are introduced through the thermal density operator $\hat{\rho}_{\hat{K}}=\hat{\rho}_{\hat{K}_\rmL}\otimes\hat{\rho}_{\hat{K}_\rmR}$, where $\hat{\rho}_{\hat{K}_\alpha}=e^{-\hat{K}_\alpha/T_\alpha}/\Tr[e^{-\hat{K}_\alpha/T_\alpha}]$, $T_\alpha$ is the temperature of FI $\alpha$,
\begin{align}\label{eq:K_alpha}
\hat{K}_\alpha=\hat{H}_\alpha
-h_\alpha \hat{M}_{\alpha}
\end{align}
is the grand canonical Hamiltonian, $\hat{M}_{\alpha}=\sum_{\r_\alpha}\hat{s}_{\r_\alpha}^z$ is the magnetization operator.

\subsection{\label{sec:quantum regime} Quantum regime of a magnon gas}
In this paper, we focus on the transport governed by quantum statistics of magnons, which are quasiparticles appearing in highly spin polarized FIs at sufficiently low temperature $T_\alpha\ll J$.
To this end, we investigate the case where both FIs are polarized along the $+z$ direction as in Fig.~\ref{fig:system}.
In such a situation, we can assume $h_\alpha\geq0$ and employ the spin-wave approximation (SWA), where the spin operators are expressed as
$
\hat{s}_{\r_i\alpha}^-
\simeq\hat{b}_{\r_i\alpha}^\+,
$
$
\hat{s}_{\r_i\alpha}^+
\simeq\hat{b}_{\r_i\alpha},
$
and
$
\hat{s}_{\r_i\alpha}^z
=1/2-\hat{b}_{\r_i\alpha}^\+\hat{b}_{\r_i\alpha},
$
in terms of bosonic operators $\hat{b}_{\r_i\alpha}^\+$ and $\hat{b}_{\r_i\alpha}$ describing magnons.
As a result, Eq.~\eqref{eq:K_alpha} is diagonalized as
\begin{align}\label{eq:K_SWA}
\hat{K}_\alpha
&\simeq\sum_{\k_\alpha} E_{\k_\alpha}\hat{b}_{\k_\alpha}^\+\hat{b}_{\k_\alpha},
\end{align}
where $\hat{b}_{\r_i\alpha}=\frac{1}{\sqrt{\N}}\sum_\k e^{i\k\cdot\r_i}\hat{b}_{\k_\alpha}$ with $\N$ being the number of lattice sites in each FI, 
\begin{align}\label{eq:E_k}
    E_{\k_\alpha}=J\S\k_\alpha^2+ h_\alpha
\end{align}
is the magnon energy, and the ground-state energy in Eq.~\eqref{eq:K_SWA} irrelevant to transport is neglected.
The details of SWA is reviewed in Appendix~\ref{appendix:SWA}.
Equation~\eqref{eq:E_k} shows that the effective Zeeman field $h_\alpha$ provides both the energy gap $E_{\k_\alpha=\0}=h_\alpha$ and chemical potential of magnons $\mu_\alpha=-h_\alpha$.

\begin{figure}
    \centering
    \includegraphics[width=0.95\columnwidth,clip]{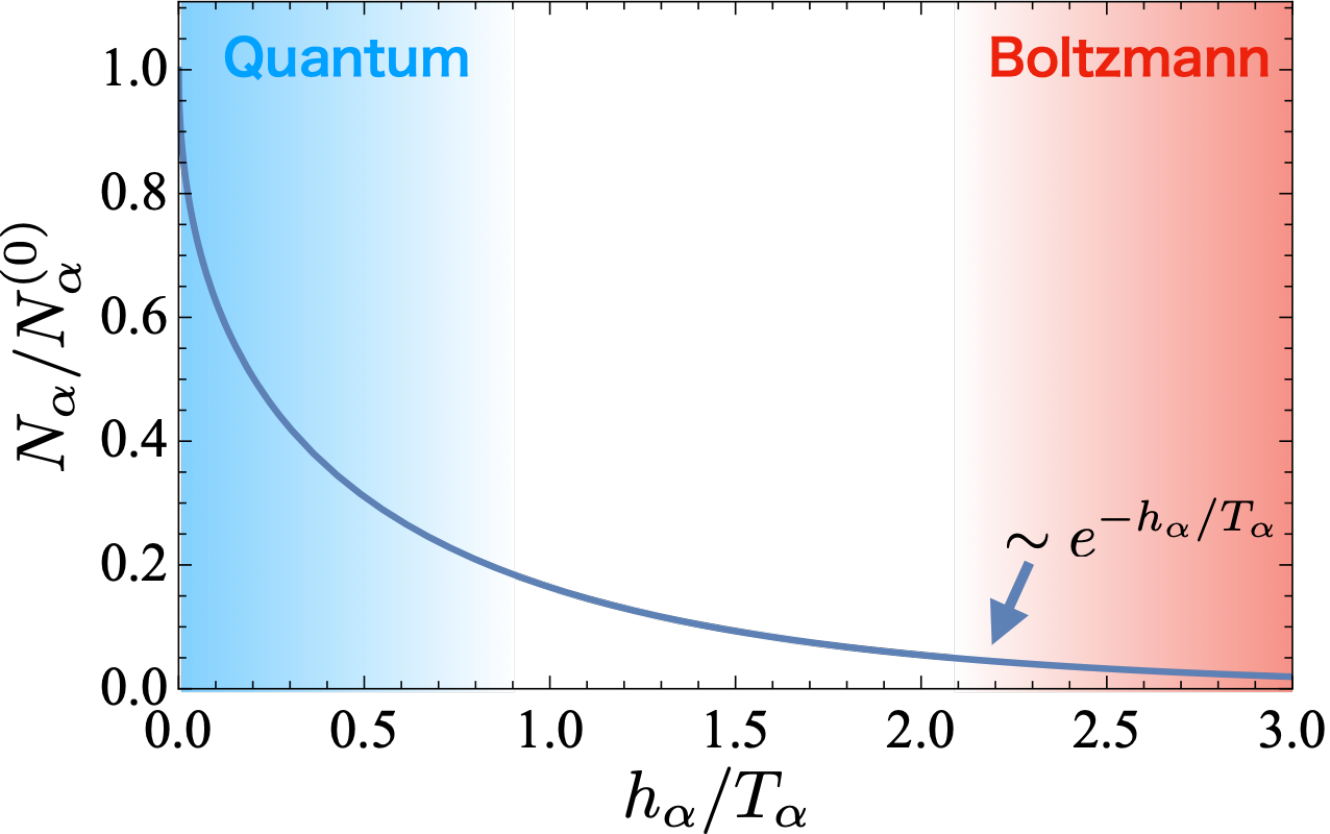}
    \caption{\label{fig:N_alpha}
    Equilibrium magnon number $N_\alpha$ [Eq.~\eqref{eq_app:N_SWA}] as a function of effective Zeeman field $h_\alpha$ scaled by $N_\alpha^{(0)}=N_\alpha|_{h_\alpha=0}$ with fixed temperature $T_\alpha>0$.
    The weak field region $h_\alpha\ll T_\alpha$ (blue region) corresponds to a quantum regime of magnons, where their Bose-Einstein statistics is relevant due to enhanced magnon density.
    In the strong field case $h_\alpha\gg T_\alpha$ (red region), on the other hand,  a large magnon gap $E_{\k_\alpha=\0}=h_\alpha$ exponentially suppresses the magnon density, leading to a classical Boltzmann gas with negligible overlap of magnon wave packets.}
\end{figure}

The quantum statistics of thermally excited magnons becomes significant when the effective Zeeman field becomes much smaller than temperature $(h_\alpha\ll T_\alpha)$~\cite{Sekino2024-vp}.
This is because the enhanced magnon density requires temperature larger than the magnon energy gap $E_{\k_\alpha=\0}=h_\alpha$ as shown in Fig.~\ref{fig:N_alpha}.
The gapless point $h_\alpha=0$ corresponds to the critical point of magnons, where the gap closing results from the spontaneous symmetry breaking of O(3) spin rotation.
In the vicinity of the critical point, it is known that the differential spin susceptibility in bulk FI and the spin conductance and spin relaxation time in the MQPC setup exhibit divergent behaviors~\cite{Sekino2024-vp}.
In this paper, we discuss how the junction geometry of MLJ affects these kinds of critical behavior in the quantum regime.
On the other hand, in the low temperature regime $T_\alpha\ll h_\alpha$ (in other words, dilute regime), the thermal excitations of magnons form a classical Boltzmann gas due to a large energy gap $E_{\k_\alpha=\0}=h_\alpha\gg T_\alpha$.

\section{\label{sec:magnon_transport}Tunneling magnon transport}
In this section, we evaluate the stationary spin and heat transport through MLJ.
Spin and heat currents flowing from the left to right FIs are described by the operators
\begin{align}\label{eq:hat_I_S}
    \hat{I}_\mathrm{S}&\equiv\d_t \hat{M}_{\rmL}=-i[\hat{M}_{\rmL},\hat{\H}]=-i[\hat{M}_{\rmL},\hat{H}_\rmT],\\
    \label{eq:hat_I_H}
    \hat{I}_\mathrm{H}&\equiv-\d_t \hat{K}_{\rmL}
    =-i[\hat{H}_\rmT,\hat{H}_{\rmL}]+h_\rmL\hat{I}_\mathrm{S},
\end{align}
respectively.
The corresponding currents are evaluated by taking the average of  Eqs.~\eqref{eq:hat_I_S} and \eqref{eq:hat_I_H} with respect to $\hat{\rho}_{\hat{K}}$.

By using the leading order perturbation theory with respect to the tunneling coupling $J_\mathrm{T}$ as well as SWA, we evaluate the currents transported by magnons.
After a straightforward analysis (see Appendix~\ref{appendix:currents}), we obtain the following expressions:
\begin{subequations}\label{eqs:I_SWA}
    \begin{align}
    I_\mathrm{S}&=\int_{-\infty}^{\infty}\dd{\omega}\T(\omega)\Delta n_\mathrm{B}(\omega),\\
    I_\mathrm{H}
    &=\int_{-\infty}^{\infty}\dd{\omega}(\omega+h_\rmL)\T(\omega)\Delta n_\mathrm{B}(\omega).
    \end{align}
\end{subequations}
Here, 
\begin{align}\label{eq:T_+-}
    \T(\omega)
    &=\frac{(J_\rmT)^2\N_z}{2\pi}\int\frac{\dd \k_\rmL\dd \k_\rmR}{(2\pi)^6}\delta(k_\rmR^z-k_\rmL^z)\nonumber\\
    &\qquad\qquad\qquad\times\A(\k_\rmL,\omega_\rmL)
    \A(\k_\rmR,\omega_\rmR)
\end{align}
is the frequency-dependent transmittance with $\omega_\alpha=\omega + h_\alpha$, $\N_z$ is the number of lattice sites in the $z$ direction corresponding to the number of blue bonds in Fig.~\ref{fig:system}a, 
\begin{align}\label{eq:A(k,omega)_SWA}
    \A(\k_\alpha,\omega_\alpha)&
    =\pi\,\delta(\omega_\alpha-E_{\k_\alpha})
\end{align}
is the magnon spectral function $\A(\k_\alpha,\omega_\alpha)$ in FI $\alpha$.
The delta function in Eq.~\eqref{eq:T_+-} results from the momentum conservation in the $z$ direction for magnons the tunneling through MLJ, while such a constraint on momentum never appears in MQPC~\cite{Sekino2024-vp}.
The function 
$\Delta n_\mathrm{B}(\omega)=n_{\mathrm{B},\rmL}(\omega_\rmL)-n_{\mathrm{B},\rmR}(\omega_\rmR)$
in Eqs.~\eqref{eqs:I_SWA} denotes the difference of the Bose-Einstein distribution function $n_{\mathrm{B},\alpha}(\omega_\alpha)=1/(e^{\omega_\alpha/T_\alpha}-1)$ between both FIs.
Performing the integration, we obtain the power law behavior of $\T(\omega)$,
\begin{align}\label{eq:T_+-_power}
\T(\omega)
&=\frac{A\, \theta(\omega)}{\Gamma(3/2)}\sqrt{\omega},
\end{align}
where 
$A=J_\mathrm{T}^2\N_z/(8\sqrt{2\pi J^5})$
and $\Gamma(d)$ is the Gamma function.

Substituting Eq.~\eqref{eq:T_+-_power} into $I_\mathrm{S}$ and $I_\mathrm{H}$ in Eqs.~\eqref{eqs:I_SWA}, we obtain
\begin{subequations}\label{eq:I_S/E/H_SWA2}
\begin{align}\label{eq:I_S_SWA}
I_\mathrm{S}&=A\Delta F_{3/2},\\
\label{eq:I_H_SWA}
I_\mathrm{H}&=A[(3/2)\Delta F_{5/2}+h_\mathrm{L}\Delta F_{3/2}],\\
\label{eq:F_d_SWA}
\Delta F_d&=(T_\rmL)^d F_{d}(x_\rmL)-(T_\rmR)^d F_{d}(x_\rmR),
\end{align}
\end{subequations}
where $x_\alpha=h_\alpha/T_\alpha$ is defined and $\Gamma(d)=(d-1)\Gamma(d-1)$ was used.
The term $\Delta F_d=(T_\rmL)^d F_{d}(x_\rmL)-(T_\rmR)^d F_{d}(x_\rmR)$ encoding the quantum statistics of magnons can be written in term of the Bose-Einstein integral
\begin{align}\label{eq:F_d(x)}
F_d(x_\alpha)=\int_0^\infty\!\!\dd{\omega}\,\frac{\omega^{d-1}}{\Gamma(d)T_\alpha^{d}}n_{\mathrm{B},\alpha}(\omega_\alpha)
&=\mathrm{Li}_d(e^{-x_\alpha}),
\end{align}
where $\mathrm{Li}_d(z)=\sum_{k=1}^\infty z^k/k^d$ is the polylogarithm.

\subsection{Critical enhancement of conductances in quantum regime}
We now show that the magnonic criticality enhances the transport efficiency of spin and heat.
To this end, we first focus on the conductances characterizing spin and heat currents under small biases.
Expanding Eqs.~\eqref{eq:I_S/E/H_SWA2} to the biases $\Delta h=-(h_\mathrm{L}-h_\mathrm{R})$ and $\Delta T = T_\mathrm{L}-T_\mathrm{R}$, we obtain
\begin{align}\label{eq:linear currents}
\begin{pmatrix}
I_\mathrm{S}\\
I_\mathrm{H}
\end{pmatrix}
=
\begin{pmatrix}
L_{11}&L_{12}\\
L_{21}&L_{22}
\end{pmatrix}
\begin{pmatrix}
\Delta h\\
\Delta T
\end{pmatrix},
\end{align}
where the Onsager coefficients are
\begin{subequations}\label{eqs:L_ij}
\begin{align}\label{eq:L_11}
    \frac{L_{11}}{A\sqrt{T}}&=F_{1/2}(x),\\
    \frac{L_{12}}{A\sqrt{T}}&=\frac{L_{21}}{AT^{3/2}}=\frac{3}{2}\,F_{3/2}(x)
    +x\,F_{1/2}(x),\\
    \frac{L_{22}}{AT^{3/2}}&=\frac{15}{4}\,F_{5/2}(x)+3x\,F_{3/2}(x)
    +x^2\,F_{1/2}(x),
\end{align}
$x=h/T$, and $h=(h_\mathrm{L} + h_\mathrm{R})/2$ and $T=(T_\mathrm{L} + T_\mathrm{R})/2$ are mean field strength and temperature, respectively.
The thermal conductance $K$ is defined by
\begin{align}\label{eq:K}
K=L_{22}-\frac{L_{12}L_{21}}{L_{11}}=L_{22}-\frac{T(L_{12})^2}{L_{11}}.
\end{align}
\end{subequations}

\begin{figure}
    \centering
    \includegraphics[width=0.9\columnwidth,clip]{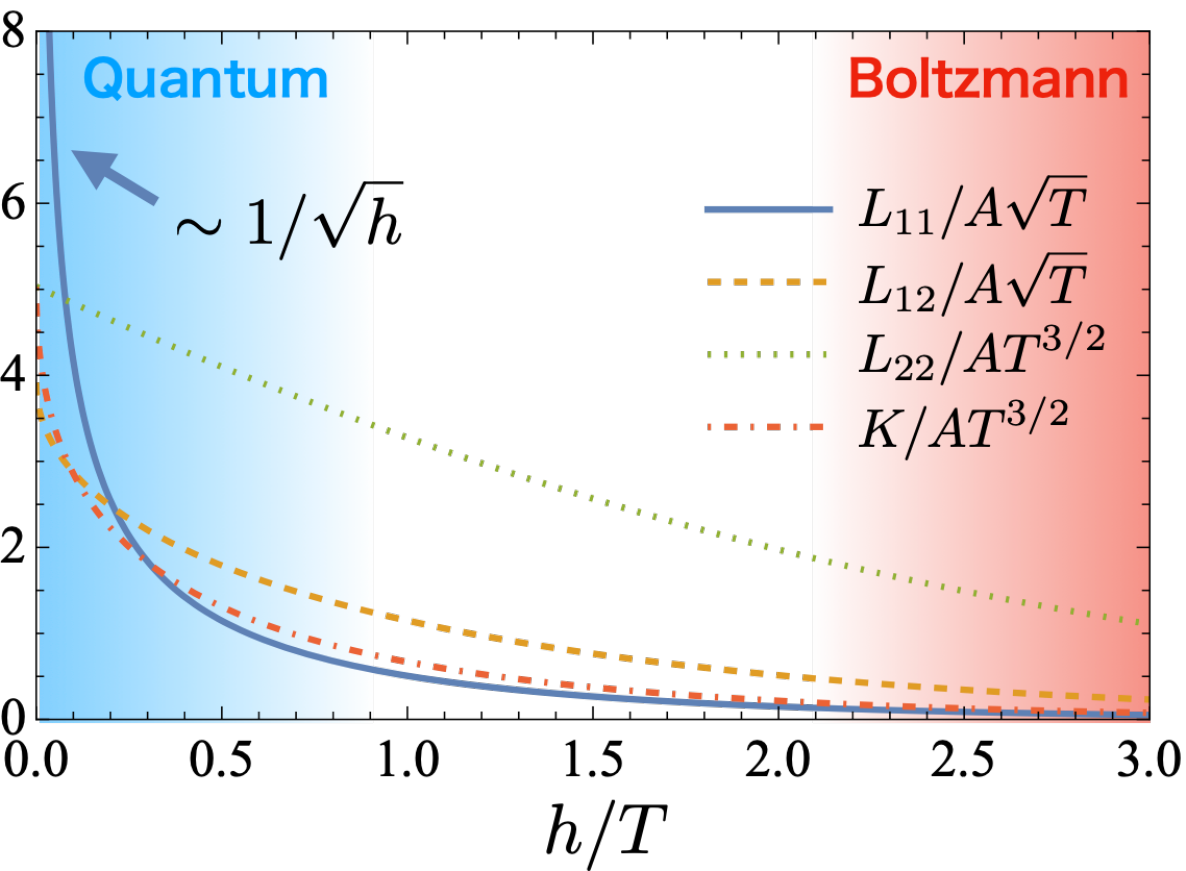}    
    \caption{Quantum enhancement of the transport coefficients $L_{ij}$ and $K$ in MLJ by the magnonic criticality.
    As the effective Zeeman field $h$ decreases, Bose-Einstein statistics of magnons becomes more significant, leading to the increase of the transport coefficients.
    In particular, the spin conductance $L_{11}\sim1/\sqrt{h}$ shows a divergent behavior in the vicinity of the magnonic critical point at $h=0$.
    }
    \label{fig:conductance}
\end{figure}
Figure~\ref{fig:conductance} exhibits the enhancement of conductances $L_{ij}$ and $K$ in the quantum regime due to degeneracy of magnons.
As $h$ decreases with $T$ fixed, the density of magnons increases, enhancing the transport efficiency of spin and heat.
In particular, the spin conductance in the vicinity of the magnonic critical point $h/T\to+0$ becomes divergently large
\begin{align}\label{eq:L_11_critical}
\frac{L_{11}}{AT\sqrt{T}}\simeq
\sqrt{\frac{\pi T}{h}}
\end{align}
because of the criticality.
We note that the divergence in MLJ is much significant than that in MQPC with $L_{ij}\sim 1/|\ln h|$~\cite{Sekino2024-vp}.
This is because the conservation of momentum $k_z$ for magnons tunneling through MLJ results in rapid increase of transmittance $\T(\omega)\sim \sqrt{\omega}$ at low $\omega$ [see Eq.~\eqref{eq:T_+-_power}] compared to the MQPC case $\T(\omega)\sim \omega$ (see Table~\ref{tab:results}).
In contract to $L_{11}$, the other conductances at $h\to+0$ approach constant values $L_{12}/A\sqrt{T}=L_{21}/AT^{3/2}=3\zeta(3/2)/2$ and $L_{22}/AT^{3/2}=K/AT^{3/2}=6\zeta(3)$ with the zeta function $\zeta(s)$, similarly to MQPC case.
We mention that similar enhancement is also discussed in tunneling transport between Bose gases~\cite{Papoular2016quantized}.

\subsection{\label{sec:Lorenz number}Nonuniversality and breakdown of the magnonic WF law}
Here, we discuss the nonuniversality of the magnonic WF law~\cite{Nakata2015-ns,Nakata2017-hu,Nakata2017-ot,Nakata2017-iy,nakata2018magnonic,Nakata2022-ep,Sano2023-ff,Mook2018-oy} indicating its sensitivity to junction systems.
To this end, we focus on the dimensionless Lorenz number of magnons defined by 
\begin{align}\label{eq:Lorenz number}
    \mathcal{L}\equiv\frac{K}{L_{11}T},
\end{align}
and discuss how magnonic criticality and junction geometry affects $\mathcal{L}$.
When $\mathcal{L}$ is constant, the ratio of thermal to spin conductances becomes proportional to the temperature
\begin{align}\label{eq:WF_law}
    \frac{K}{L_{11}}\propto T,
\end{align}
where the constant of proportionality is given by $\mathcal{L}$.
Following terminology in previous works, we call this linear-in-$T$ behavior magnonic WF law~\cite{Nakata2015-ns,Nakata2017-hu,Nakata2017-ot,Nakata2017-iy,nakata2018magnonic,Nakata2022-ep,Sano2023-ff,Mook2018-oy}.
This relation is similar to the celebrated WF law in Fermi liquids~\cite{Franz1853-do} in the sense that the ratios of two transport coefficients are proportional to temperature.
However, we below show that the magnonic WF law is not universal unlike the conventional WF law:
The values of $\mathcal{L}$ for magnons depend on details such as junction geometry and imbalance of parameters between reservoirs, which sharply contrasts to Fermi liquids with the universal constant $\mathcal{L}=\pi^2/3$  (in the unit of $e=k_B=1$).

In the tunneling magnon transport, the WF law~\eqref{eq:WF_law} holds when temperature scale is lower than that of the magnetic field strength ($T\ll h$), as pointed out in earlier works~\cite{Nakata2015-ns,Sekino2024-vp}.
With the notation of this paper, the condition $h\gg T$ corresponds to the Boltzmann regime where thermally excited magnons obey the Maxwell-Boltzmann distribution as discussed in Sec.~\ref{sec:quantum regime}.
This is similar to the situation, where Drude evaluated the Lorenz number $\mathcal{L}_\mathrm{bulk}$ for bulk transport in metal based on classical argument~\cite{Drude1900-bi,Drude1902-qx,Ashcroft1976-hn} before Sommerfeld computed more accurate value by considering Fermi-Dirac statistics of electrons~\cite{Sommerfeld1928-re}.

We will now derive the magnonic WF law in MLJ.
By taking the classical limit of $x=h/T\to\infty$ in Eqs.~\eqref{eq:L_11} and \eqref{eq:K}, we obtain
\begin{subequations}
\begin{align}
    \frac{L_{11}}{A\sqrt{T}}&\simeq e^{-x},\\
    \frac{K}{AT^{3/2}}&\simeq \frac{3}{2}e^{-x}.
\end{align}
\end{subequations}
These exponential suppression of conductances are seen in Fig.~\ref{fig:conductance}, arising from the diluteness of the magnon Boltzmann gas (see also Fig.~\ref{fig:N_alpha}).
Thus, we obtain the constant Lorenz number leading to the WF law resulting from the Maxwell-Bolzmann statistics of magnons in this regime:
\begin{align}\label{eq:Lorenz_classical}
\mathcal{L}=\frac{3}{2}.
\end{align}
As shown in Table~\ref{tab:WF_laws}, this value deviates from $\mathcal{L}=2$ in MQPC~\cite{Sekino2024-vp} and $\mathcal{L}=1$ in the magnetic planar junction (MPJ)~\cite{Nakata2015-ns}, indicating the sensitivity to the junction geometry.
Therefore, the magnonic WF law is not universal but depends on details of interface differently from the universal WF law for Fermi liquids discussed below.
This nonuniversality results from the fact that $\mathcal{L}$ for magnons is sensitive to the frequency dependence of $\T(\omega)$.
Indeed, $\mathcal{L}=d+1$ is given by the power of frequency in $\T(\omega)\propto \omega^{d}$ with $d=1$ for MQPC, $d=1/2$ for MLJ, and $d=0$ for MPJ [see also Fig.~\ref{fig:L for fermions} in Appendix~\ref{appendix:fermionic Lorenz number}].
In addition, the WF law in tunneling magnon transport could also depend on differences of bulk parameters between FIs such as external Zeeman fields and magnetic anisotropy, which change the magnon transmittance $\T(\omega)$.

We provide two remarks on the magnonic WF law.
First, the dependence of magnonic $\mathcal{L}$ on junction geometry contrasts to universality of the WF law for Fermi liquids.
In the tunneling transport for degenerate fermions, $\mathcal{L}=\pi^2/3$ is independent from microscopic details, i.e., both interface geometry and band structures in bulk as long as the Fermi liquid picture holds.
This universality results from the fact that fermions near the Fermi surface only contribute to transport properties, leading to insensitivity to frequency dependence of transmittance.
In Appendix~\ref{appendix:fermionic Lorenz number}, we demonstrate the robustness of $\mathcal{L}=\pi^2/3$ for free fermions against junction geometry [see Eq.~\eqref{eq_app:Lorenz_degenerate_fermions} and Fig.~\ref{fig:L for fermions}].
It is also worthy of noting that quantum degeneracy of fermions makes $\mathcal{L}$ identical with $\mathcal{L}_\mathrm{bulk}=\pi^2/3$, which is found by Wiedemann and Franz~\cite{Franz1853-do} and then derived by Sommerfeld based on quantum statistical mechanics~\cite{Sommerfeld1928-re}.
Therefore, the presence of the Fermi surface is crucial to the universal WF law for Fermi liquids. 

Second, unlike for fermions, the Bose-Einstein statistics generally breaks the WF law in tunneling magnon transport~\cite{Nakata2015-ns,Sekino2024-vp}.
Indeed, by substituting Eq.~\eqref{eq:L_11_critical} and $K/AT^{3/2}=6\zeta(3)$ at $h\to+0$ into Eq.~\eqref{eq:Lorenz number}, we see that the critical behavior of $L_{11}$ makes 
the Lorenz number in the quantum regime vanishingly small:
\begin{align}
    \mathcal{L}\sim\sqrt{h/T}.
\end{align}
This critical breakdown of the WF law sharply contrasts to the bulk transport, where the magnonic WF law with $\mathcal{L}_\mathrm{bulk}=35\zeta(7/2)/[4\zeta(3/2)]-25\zeta^2(5/2)/[4\zeta^2(3/2)]=2.126$ is predicted~\cite{Nakata2017-ot,Nakata2017-iy,Sano2023-ff,Mook2018-oy}.
It is noteworthy that the WF law in bulk magnon transport is broken by nonlinear bias effect~\cite{Nakata2022-ep} and magnon hydrodynamics~\cite{Sano2023-ff}.
The violations of the WF laws are also discussed for strongly interacting fermions~\cite{husmann2018breakdown,Li2002-hg, Garg2009-hl, Wakeham2011-ob, Mahajan2013-kw, Filippone2016-an} and Bose-Einstein condensates~\cite{PhysRevResearch.2.023284,PhysRevResearch.2.023340}.

\begin{table}
    \centering
    \begin{tabular}{ccccc}
    \hline\hline
        \multirow{3}{*}{}   & \multicolumn{4}{c}{Lorenz number $\mathcal{L}$} \\
                                    & MLJ & MQPC & MPJ & Bulk \\
                                    & Present & \cite{Sekino2024-vp} & \cite{Nakata2015-ns} & \cite{Nakata2017-ot,Nakata2017-iy,Sano2023-ff,Mook2018-oy}\\\hline\hline
        Classical magnons           & $3/2$ & $2$ & $1$ & $5/2$\\
        Quantum magnons             & $\sim\sqrt{h}$ & $\sim1/|\ln h|$ & $\sim h$ & $2.126$\\\hline
        Fermi liquids               & $\pi^2/3$ & $\pi^2/3$ & $\pi^2/3$ & $\pi^2/3$\\\hline\hline
    \end{tabular}
    \caption{\label{tab:WF_laws}
    Dimensionless Lorenz numbers $\mathcal{L}$ on several transport setups.
    The classical magnons in the Boltzmann regime follow the so-called magnonic WF laws, in which constant $\mathcal{L}$ are not universal but sensitive to the setups.
    For magnons in the quantum regime, the magnonic criticality makes $\mathcal{L}$ for tunneling transport (MLJ, MQPC, and MPJ) depend on $h$, violating the WF law.
    As references, we show the results of Fermi liquids, which obey the universal WF law with $\mathcal{L}=\pi^2/3$ independent from transport setups.
    }
\end{table}

\subsection{Nonlinear spin current by criticality}
\begin{figure}[tb]
    \centering
    \includegraphics[width=0.9\columnwidth,clip]{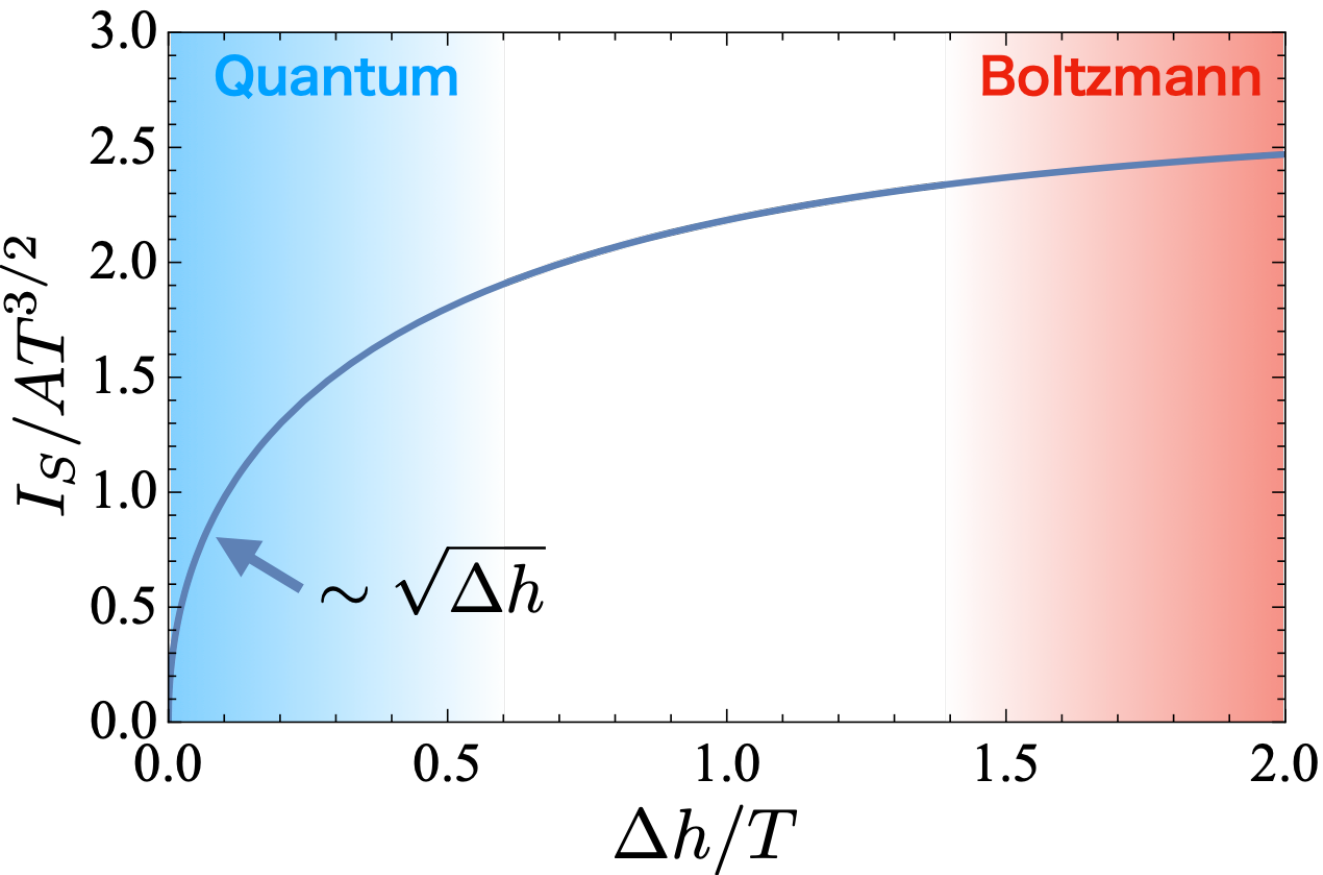}    
    \caption{
    Breaking down of Ohm's law of spin current in the vicinity of the magnonic critical points.
    At $h_\rmL=\Delta T=0$, the spin current in the quantum regime nonlinearly depends on the bias $\Delta h$ due to the criticality.
    }
    \label{fig:I_s}
\end{figure}
The divergent behavior of the spin conductance in Eq.~\eqref{eq:L_11_critical} means the breakdown of Ohm's law for spin in the quantum regime.
To this end, we focus on the bias dependence of the spin current in the case where the left FI is at the critical point $h_\mathrm{L}=0$.
As depicted in Fig.~\ref{fig:I_s}, the spin current driven by spin bias in the isothermal condition $\Delta T=0$ shows the nonlinear dependence on $\Delta h>0$
\begin{align}\label{eq:nonlinear_I_S}
    \frac{I_\mathrm{S}}{AT^{3/2}}\simeq 2\sqrt{\frac{\pi \Delta h}{T}},
\end{align}
which has the different form of the critical behavior of a spin current $I_\mathrm{S}\sim \Delta h|\ln \Delta h|$ in MQPC~\cite{Sekino2024-vp}.
On the other hand, $I_\mathrm{S}$ varies linearly with the temperature bias, and $I_\mathrm{H}$ has a similar linear dependence on $\Delta h$ and $\Delta T$.

\subsection{\label{sec:interaction}Magnon-magnon interactions}
Here, we briefly comment on the effects of magnon-magnon interaction beyond SWA, which would be relevant to transport properties for dense magnon gases.

We consider that the magnon-magnon interaction is not so significant even in the quantum regime for our highly spin polarized FIs at sufficiently low temperature $T_\alpha\ll J$.
As shown in Fig.~\ref{fig:N_alpha}, the magnon density $N_\alpha/\mathcal{N}$ at fixed temperature increases as $h_\alpha\geq0$ decreases and takes the maximum value at $h_\alpha=0$.
Note that $N_\alpha$ in Fig. 2 is scaled by not the  number of lattice sites $\mathcal{N}$ but the maximum magnon number $N_\alpha^{(0)}=N_\alpha|_{h_\alpha=0}$.
However, the maximum magnon density $N_\alpha^{(0)}/\mathcal{N}\propto(T_\alpha/J)^{3/2}$ [see Eq.~\eqref{eq_app:N_SWA}] remains small at sufficiently low temperature $T_\alpha\ll J$, which would allow us to neglect the magnon-magnon interaction.
In addition, the O(3) spin rotation symmetry in the Hamiltonian would suppress the interaction effect on dilute magnon gases in the quantum regime.
Indeed, it is shown that, in the continuum limit, the isotropy of the Heisenberg interaction in Eq.~\eqref{eq:H_alpha} results in a vanishing two-magnon contact interaction, which is naively expected to be leading-order corrections beyond SWA~\cite{Nakata2015-ts}.

On the other hand, the magnon-magnon interaction could modify the transport behaviors from our results when magnons form dense gases at relatively high temperatures $T_\alpha\sim J$ yet below the Curie temperature under weak effective fields $h_\alpha\ll T_\alpha$.
In the context of transport, this temperature regime seems interesting because there exist many magnons working as spin and heat carriers.
A detailed examination of the transport between dense magnon gases is left for future work.

\section{\label{sec:relaxation}Relaxation dynamics relevant to cold-atom experiments}
In this section, we investigate relaxation dynamics of spin and heat between FIs, which enables one to experimentally determine conductances $L_{ij}$ in cold-atomic systems.
We start by reviewing this experimental scheme in Sec.~\ref{sec:quantum regime}.
In Sec.~\ref{sec:anomalous_relaxation}, we then discuss the junction geometry of MLJ results in anomalous relaxation dynamics with three strong constraints: (i) decoupling of spin and heat relaxations, (ii) correspondence of relaxation time between spin and heat, and (iii) insensitivity of relaxation time to the effective Zeeman field and temperature.
Such dynamics is quite different from that in MQPC case~\cite{Sekino2024-vp}.

\subsection{\label{sec:quasistationary}Review of quasistationary relaxation}
Cold atoms are isolated systems, so that initially prepared spin and temperature biases gradually change through the exchange of spin and heat between FIs.
Unlike in solid-state systems, where the biases can be externally sustained and Eq.~\eqref{eq:linear currents} is directly applicable, it is difficult to directly measure the stationary currents studied in Sec.~\ref{sec:magnon_transport}.
Instead, the scheme to determine $L_{ij}$ in cold-atom experiments has been proposed originally for Fermi gases without lattice~\cite{brantut2013thermoelectric,grenier2012probing} and then extended to FIs~\cite{Sekino2024-vp}.
We review this formalism below.

In this scheme, we first consider the relaxation dynamics of magnetization and temperature differences between FIs.
At time $t<0$, the channel between FIs (the light blue bonds in Fig~\ref{fig:system}) is closed by applying a strong barrier potential between FIs. 
Both FIs are given by thermal states at temperature $T_\alpha(t=0)$ with magnetization $M_\alpha(t=0)$.
At time $t=0$, one opens the channel and both FIs starts to exchange magnetization and heat.
Then, one observes the relaxation of the differences of magnetization and temperature, $\Delta M(t)=M_\rmL(t)-M_\rmR(t)$ and $\Delta T(t)=T_\rmL(t)-T_\rmR(t)$.
Suppose initial values of $M(0)$ and $T(0)$ are sufficiently small and the thermalization in each FIs is much faster than the relaxation, which is justified for a weak tunneling coupling $J_\rmT\ll J$.
In this case, one can analyze the relaxation dynamics with the quasi-stationary model where both FIs are regarded as thermalized states with $M_\alpha$ and $T_\alpha$ at $t>0$.
This model enables us to experimentally extract the conductance $L_{ij}$ of our interest.

The time-evolution equations in quasi-stationary model is derived from two equations.
The first one is given by the linear response in Eq.~\eqref{eq:linear currents}.
For convenience, we rewrite this response in terms of $\dv{t}(-\Delta M,\Delta S)=-2(I_\mathrm{S},I_\mathrm{H}/T)$, where $\Delta S=S_\rmL-S_\rmR$ and $S_\alpha$ is entropy in FI $\alpha=\rmL,\rmR$.
Then, we find
\begin{align}\label{eq:current_bias_relaxation}
\dv{t}
\begin{pmatrix}
-\Delta M\\
\Delta S
\end{pmatrix}
&=-2
    \begin{pmatrix}
    L_{11}&L_{12}\\
    L_{12}&L_{22}/T
    \end{pmatrix}
    \begin{pmatrix}
    \Delta h\\
    \Delta T
    \end{pmatrix},
\end{align}
where the Onsager reciprocal relations $L_{21}=L_{12}T$ was used.
The second one is thermodynamic relations, which are obtained by expanding $\Delta M$ and $\Delta S$ with respect to the biases:
\begin{align}\label{eq:thermodynamic_relation}
    \begin{pmatrix}
    -\Delta M\\
    \Delta S
    \end{pmatrix}
    &=
    \begin{pmatrix}
    \qty(\pdv{M}{h})_T&-\qty(\pdv{M}{T})_h\\
    - \qty(\pdv{S}{h})_T&\qty(\pdv{S}{T})_h\\
    \end{pmatrix}
    \begin{pmatrix}
    \Delta h\\
    \Delta T
    \end{pmatrix}.
\end{align}
By combining this with Eq.~\eqref{eq:current_bias_relaxation}, we obtain the following time evolution equation for $\Delta M$ and $\Delta T$:
\begin{align}\label{eq:relaxation}
   \tau_0\dv{t}
    \begin{pmatrix}
    -\Delta M/\kappa\\
    \Delta T    
    \end{pmatrix}
    =-
    \begin{pmatrix}
    1&-\alpha\\
    -\frac{\alpha}{l}&\frac{\mathcal{L}+\alpha^2}{l}
    \end{pmatrix}
    \begin{pmatrix}
    -\Delta M/\kappa\\
    \Delta T  
    \end{pmatrix},
\end{align}
where $\kappa=\qty(\pdv{M}{h})_T,$
$\tau_0=\kappa/(2L_{11})$,
$\alpha=L_{12}/L_{11}-\alpha_r$,
$\alpha_r=-\qty(\pdv{S}{M})_T,$ $l=\frac{1}{\kappa}\qty(\pdv{S}{T})_{M}$, and $\mathcal{L}$ is given by Eq.~\eqref{eq:Lorenz number}.
As shown in Appendix~\ref{appendix:quasistationary model}, the solutions of $\Delta M$ and $\Delta T$ are found in terms of $L_{ij}$ and equilibrium thermodynamic quantities appearing in the matrix in Eq.~\eqref{eq:thermodynamic_relation}.
Furthermore, the thermodynamic quantities can be experimentally determined independently of the relaxation.
As a result, by observing the relaxation dynamics and fitting them to the solutions of Eq.~\eqref{eq:relaxation} using the thermodynamic quantities measured in equilibrium experiments, the experimental values of the conductances $L_{ij}$ can be extracted.
It is worthy of mentioning that $\alpha$ in Eq.~\eqref{eq:relaxation} is normally nonzero and dynamics of spin and heat mutually affect one another.

\subsection{\label{sec:anomalous_relaxation}Anomalous relaxation dynamics in MLJ}
Let us now demonstrate the dynamics through MLJ using SWA, and discuss anomalous relaxation resulting from the junction geometry.
Here, we exhibit remarkable relaxation dynamics in the highly spin-polarized regime where magnons govern relaxation: Time evolutions of spins and temperatures through MLJ are completely decoupled from each other and share the same decay time $\tau_0$ insensitive to $h$ and $T$.
By applying SWA and solving Eq.~\eqref{eq:relaxation} (see Appendix~\ref{appendix:quasistationary model} for details), we obtain
\begin{align}\label{eq:Delta_M,T(t)}
\begin{pmatrix}
\Delta M(t)\\
\Delta T(t)
\end{pmatrix}
=
\begin{pmatrix}
e^{-t/\tau_0}&0\\
0&e^{-t/\tau_0}
\end{pmatrix}
\begin{pmatrix}
\Delta M(0)\\
\Delta T(0)
\end{pmatrix},
\end{align}
In addition, the relaxation time is independent of the thermodynamic variables $h$ and $T$:
\begin{align}\label{eq:tau_0^LJ}
\tau_0&=\frac{2}{\pi}\frac{(\N/\N_\mathrm{T})(J/J_\mathrm{T})^2}{J}=\textrm{constant}.
\end{align}
Therefore, time evolution of $\Delta M(t)$ [$\Delta T(t)$] depends only on its initial value $\Delta M(0)$ [$\Delta T(0)$] from the quantum regime ($h/T\ll1$) to the Boltzmann regime ($h/T\gg1$).
We find that such phenomena occur because the frequency power laws of the magnon density of states $\rho_\alpha(\omega)\propto\sqrt{\omega}$ characterizing bulk FIs is identical with that of magnon transmittance $\T(\omega)\propto\sqrt{\omega}$ responsible for transport (expect for prefactors).
This correspondence of the power laws makes the matrices in Eqs.~\eqref{eq:current_bias_relaxation} and ~\eqref{eq:thermodynamic_relation} proportional to each other.

We stress that the above dynamics in MLJ with the decoupling of spin and heat and constant $\tau_0$ are quite different from those in MQPC.
In MQPC, the decoupling occurs only in the quantum regime $h/T\ll1$ because of the existence of the magnonic critical point at $h=0$.
In this case, the magnonic criticality also makes the magnetization decay time divergently large $\tau_0\sim1/|\sqrt{h}\ln h|$, leading to the slowing down of spin relaxation.
This originates from a divergently large differential spin susceptibility corresponding to a capacity, and thus it is regarded as a significant capacitance effect.
On the other hand, in MLJ, the critical behavior of the spin conductance $L_{11}\sim 1/\sqrt{h}$ [Eq.~\eqref{eq:L_11_critical}] cancels that of the differential spin susceptibility $\kappa\sim1/\sqrt{h}$ corresponding to a capacity, leading to $\tau_0=\kappa/(2L_{11})=\textrm{constant}$.

It would also be worth mentioning 
a possible breakdown of the decoupled spin and heat relaxation in MLJ.
First, $\Delta M(t)$ and $\Delta T(t)$ would couple with each other when there are differences in external Zeeman fields or magnetic anisotropy between FIs.
The decoupled solution in Eq.~\eqref{eq:Delta_M,T(t)} relies on the coincidence of parameters between FIs (except for the slight differences of $h_\alpha$ and $T_\alpha$ to generate biases).
If two FIs feel different strengths of external Zeeman fields or magnetic anisotropy, the mismatch of the magnon band structure between FIs would break the relation $\mathcal{T}(\omega)\propto\rho_\alpha(\omega)$, so that Eq.~\eqref{eq:Delta_M,T(t)} is no longer a solution.
Second, magnon-magnon interactions relevant to dense magnon gases at $h_\alpha\ll T_\alpha\sim J$ would couple the spin and heat dynamics.
In this parameter regime, the interaction could modify the relation $\mathcal{T}(\omega)\propto\rho(\omega)$ between the transmittance and the density of states at the frequency scale $\omega\sim T$, so that the decoupled dynamics is not guaranteed.
The importance of the magnon-magnon interaction at $h_\alpha\ll T_\alpha\sim J$ is also discussed in Sec.~\ref{sec:interaction}.

\section{\label{sec:summary}Summary}
\begin{table*}[tb]
\centering
\caption{\label{tab:results}
Impact on junction geometry on magnon transport. 
The different power laws in magnon transmittance $\T(\omega)$ affected by shape of interface result in significant differences in tunneling transport.}
\begin{tabular}{c|cccccc}
\hline\hline
Ref.&Junction &Transmittance 
    &Spin conductance &Magnonic WF law &Nonlinear spin current & Spin relaxation time\\
    && $\T(\omega)$&$L_{11}$&
    $\mathcal{L}$
    &$I_\mathrm{S}$&
    $\tau_0$
    \\
\hline
Present&MLJ&
    $\sim \sqrt{\omega}$
    &$\sim 1/\sqrt{h}$
    &$\simeq 3/2$
    &$\sim\sqrt{\Delta h}$&
    $\sim$ constant
    \\
Ref.\cite{Sekino2024-vp}&MQPC&
    $\sim \omega$
    &$\sim |\ln h|$
    &$\simeq 2$
    &$\sim\Delta h|\ln \Delta h|$
    &$\sim 1/|\sqrt{h}\ln h|$
    \\
\hline\hline
\end{tabular}
\end{table*}
In this paper, we analyzed magnon-mediated spin and heat transport in an MLJ setup and demonstrated the significant effects of the interface geometry on the dynamics.
The obtained results, along with their comparison to those in MQPC, are summarized in Table~\ref{tab:results}, highlighting the unique behaviors observed in the MLJ setup.
By employing the nonequilibrium Green’s function approach combined with SWA, we revealed that magnonic criticality under weak Zeeman fields causes anomalous conductance enhancements and a breakdown of Ohm's law in spin currents, with behaviors influenced by linear interface geometries.
The magnonic WF law is found to sensitive to the junction geometry in the classical Boltzmann regime and is broken in the quantum regime, which is in a sharp contrast to Fermi liquids.
We also clarified that magnetization and temperature differences between FIs in MLJ independently equilibrate, exhibiting robust decay properties that are insensitive to the mean temperature or effective Zeeman field.
These anomalous thermomagnetic dynamics results from the sensitivity of magnons to the junction geometry of MLJ.
Our findings shed light on the unique thermomagnetic behaviors of magnons, enabled by their Bose-Einstein statistics, and offer a pathway for advancing thermomagnetic transport studies using the tunability of cold atomic systems.

\begin{acknowledgments}
The authors thank Ryotaro Sano for his fruitful comments on the magnonic Wiedemann-Franz laws.
We acknowledge JSPS KAKENHI for Grants (No. JP21K03436, No. JP21H01800, No. JP23H01839, and No. JP24H00322) from MEXT, Japan. 
Y. S. is supported by JST ERATO Grant Number JPMJER2302, Japan and by the RIKEN TRIP initiative (RIKEN Quantum).
S. U. is supported  by JST PRESTO (JPMJPR2351) and Matsuo Foundation.
M. M. is supported by the National Natural Science Foundation of China (NSFC) under Grant No. 12374126 and
by the Priority Program of the Chinese Academy of Sciences under Grant No. XDB28000000.
\end{acknowledgments}

\appendix

\section{\label{appendix:SWA} Spin-wave approximation}
This appendix is devoted to a review on the spin-wave approximation (SWA) describing highly spin polarized FIs at temperature below the Curie temperature.
We initially consider the general magnitude of spin $\S=1/2,1,\cdots$ for convenience~\cite{holstein1940field}, and set $\S=1/2$ at the end of the calculations.
As in Fig.~\ref{fig:system}, the spins are aligned along the $+z$ direction.
In this case, we can perform the Holstein-Primakoff expansion for spin operators, which is regarded as expansion with respect to $1/\S$~\cite{holstein1940field}:
\begin{align}
    \label{eq_app:HP_S^-}
    \hat{s}_{\r_i\alpha}^-
    &=\hat{b}_{\r_i\alpha}^\+(2\S-\hat{b}_{\r_i\alpha}^\+\hat{b}_{\r_i\alpha})^{1/2}
    \simeq\sqrt{2\S}\hat{b}_{\r_i\alpha}^\+
    ,\\
    \label{eq_app:HP_S^+}
    \hat{s}_{\r_i\alpha}^+
    &=(2\S-\hat{b}_{\r_i\alpha}^\+\hat{b}_{\r_i\alpha})^{1/2}\hat{b}_{\r_i\alpha}
    \simeq\sqrt{2\S}\hat{b}_{\r_i\alpha}
    ,\\
    \label{eq_app:HP_S^z}
    \hat{s}_{\r_i\alpha}^z
    &=\S-\hat{b}_{\r_i\alpha}^\+\hat{b}_{\r_i\alpha},
\end{align}
where the bosonic fields $\hat{b}_{\r_i\alpha}$ and $\hat{b}_{\r_i\alpha}^\+$ describes magnon degrees of freedom.
The higher-order terms $\order{\S^{-1/2}}$ discribing interactions between magnons are negligible when the density of magnons is sufficiently small due to a large magnetization along $+z$ direction.
Substituting Eqs.~\eqref{eq_app:HP_S^-}--\eqref{eq_app:HP_S^z} into Eq.~\eqref{eq:K_alpha}, the (grandcanonical) Heisenberg Hamiltonian of FIs reduces to 
\begin{align}\label{eq_app:K_SWA}
\hat{K}_\alpha
&=\sum_{\k_\alpha} E_{\k_\alpha}\hat{b}_{\k_\alpha}^\+\hat{b}_{\k_\alpha}+E_\alpha^{(\mathrm{GS})}+\order{\S^{-1}},
\end{align}
where $\hat{b}_{\r_i\alpha}=\frac{1}{\sqrt{\N}}\sum_\k e^{i\k\cdot\r_i}\hat{b}_{\k_\alpha}$, $\N$ is the number of lattice sites in each FI, 
$E_{\k_\alpha}=h_\alpha+2J\S[3-\sum_{\mu=x,y,z} \cos(k^\mu_\alpha )]$
is the energy of a magnon with momentum $\k$, and $E_\alpha^{(\mathrm{GS})}=-3J\,\S^2\N-h_\alpha\S\N$ is the ground state energy.
At sufficiently low temperature $T_\alpha\ll J$, we can use the quadratic form of the magnon energy $E_{\k_\alpha}\simeq h_\alpha+J\S\k_\alpha^2$.

At the end of this appendix, we evaluate the thermodynamic quantities of the FIs relevant to relaxation dynamics discussed in Sec.~\ref{sec:relaxation} and Appendix~\ref{appendix:quasistationary model}.
Substituting Eq.~\eqref{eq_app:K_SWA} into $\Omega_\alpha=-T_\alpha\ln \mathrm{Tr}[e^{-\hat{K_\alpha}/T_\alpha}]$ at temperature $T_\alpha>0$, we obtain the grand potential in SWA as
\begin{align}\label{eq_app:Omega}
    \Omega_\alpha
    &=E_\alpha^{(\mathrm{GS})}-T\sum_\k\ln \frac{1}{1-e^{-E_\k/T}}.
\end{align}
Evaluating the magnetization $M_\alpha=-(\partial\Omega_\alpha/\partial h_\alpha)_{T_\alpha}$ and entropy $S_\alpha=-(\partial\Omega_\alpha/\partial T_\alpha)_{h_\alpha}$ yields
\begin{align}\label{eq_app:M_SWA}
    M_\alpha&=\S\N-N_\alpha,\\
    \label{eq_app:S_SWA}
    S_\alpha&=\sum_\k\ln \frac{1}{1-e^{-E_\k/T}}+\frac{U_\alpha + h_\alpha N_\alpha}{T},
\end{align}
where $N_\alpha=\sum_{\k_\alpha} n_{\mathrm{B},\alpha}(E_{\k_\alpha})$ is the magnon number, $U_\alpha=\sum_{\k_\alpha} \omega_{\k_\alpha} n_{\mathrm{B},\alpha}(E_{\k_\alpha})$ is the internal energy for the magnon gas, $\omega_{\k_\alpha}=E_{\k_\alpha}-h_\alpha=J\S\k_\alpha^2$ is the kinetic energy of a single magnon with momentum $\k_\alpha$ and  $n_{\mathrm{B},\alpha} (\omega)=1/(e^{\omega/T_\alpha}-1)$ is the Bose-Einstein distribution function of magnons.
At low temperature, we can use $\sum_\k\cdots=\frac{\N}{(2\pi)^3}\int_{\mathbb{R}^3} \dd\k\cdots$ and then obtain
\begin{align}\label{eq_app:N_SWA}
    N_\alpha
    &=B\,T_\alpha^{3/2}F_{3/2}(h_\alpha/T_\alpha),\\
    \label{eq_app:U_SWA}
    U_\alpha
    &=\frac{3}{2}B\,T_\alpha^{5/2}F_{5/2}(h_\alpha/T_\alpha),
\end{align}
where $B=\N/(4\pi J\S)^{3/2}$ and $F_d(x)$ is the Bose-Einstein integral defined by Eq.~\eqref{eq:F_d(x)}.
The above results of the particle number and internal energy are the same as those for the ideal Bose gas above the transition temperature of Bose-Einstein condensation.

\section{\label{appendix:currents} Details to derive the expressions of tunneling currents}
Here, we provide the details to derive Eqs.~\eqref{eqs:I_SWA}.
Spin, energy, and heat currents flowing from the left to right FIs are described by the operators
\begin{align}\label{eq_app:hat_I_S}
    \hat{I}_\mathrm{S}&\equiv\d_t \hat{M}_{\rmL}=-i[\hat{M}_{\rmL},\hat{\H}]=-i[\hat{M}_{\rmL},\hat{H}_\rmT],\\
    \label{eq_app:hat_I_E}
    \hat{I}_\mathrm{E}&\equiv-\d_t \hat{H}_{\rmL}=i[\hat{H}_{\rmL},\hat{\H}]=-i[\hat{H}_\rmT,\hat{H}_{\rmL}],\\
    \label{eq_app:hat_I_H}
    \hat{I}_\mathrm{H}&\equiv-\d_t \hat{K}_{\rmL}=\hat{I}_\mathrm{E}+h_\rmL\hat{I}_\mathrm{S},
\end{align}
respectively.
By using Eq.~\eqref{eq:H_T} and the commutation relations between $\hat{s}_{\r}^\pm=\hat{s}_{\r}^x\pm \hat{s}_{\r}^y$ and  $\hat{s}_\r^z$,
Eq.~\eqref{eq_app:hat_I_S} reduces to
\begin{align}\label{eq_app:hat_I_S_2}
    \hat{I}_\mathrm{S}&=\sum_{\<\r_\rmL,\r_\rmR\>'}\frac{-J_\rmT}{2}\qty[i\hat{s}_{\r_\rmL}^-\hat{s}_{\r_\rmR}^+-i\hat{s}_{\r_\rmR}^-\hat{s}_{\r_\rmL}^+].
\end{align}
On the other hand, the expression of $\hat{I}_\mathrm{E}$ becomes more complex.
By defining $\d_{\rmL/\rmT}\hat{O}\equiv-i[\hat{O},\hat{H}_{\rmL/\rmT}]$ and using $\d_\rmL \hat{\bm{s}}_{\r_\rmL}=\d_t \hat{\bm{s}}_{\r_\rmL}-\d_\rmT\hat{\bm{s}}_{\r_\rmL}$ as well as Eq.~\eqref{eq:H_T}, $\hat{I}_\mathrm{E}=\d_\rmL \hat{H}_\rmT$ in Eq.~\eqref{eq_app:hat_I_E} is rewritten as
\begin{align}\label{eq_app:hat_I_E_2}
    \hat{I}_\mathrm{E}
    &=\hat{I}_\mathrm{E}^{(+-)}+\hat{I}_\mathrm{E}^{(zz)}+\hat{I}_\mathrm{E}^{(\mathrm{S})},
\end{align}
where
\begin{subequations}
\begin{align}
    \hat{I}_\mathrm{E}^{(+-)}&=
    -\frac{J_\rmT}{2}\sum_{\<\r_\rmL,\r_\rmR\>'}\qty[
    (\d_t\hat{s}_{\r_\rmL}^-)\hat{s}_{\r_\rmR}^+
    +\hat{s}_{\r_\rmR}^-(\d_t\hat{s}_{\r_\rmL}^+)
    ],\\
    \hat{I}_\mathrm{E}^{(zz)}&=
    -\frac{J_\rmT}{2}\sum_{\<\r_\rmL,\r_\rmR\>'}\qty[
    (\d_t\hat{s}_{\r_\rmL}^z)\hat{s}_{\r_\rmR}^z
    +\hat{s}_{\r_\rmR}^z(\d_t\hat{s}_{\r_\rmL}^z)],\\
    \label{eq_app:hat_I_E^S}
    \hat{I}_\mathrm{E}^{(\mathrm{S})}&=
    \frac{J_\rmT}{2}\sum_{\<\r_\rmL,\r_\rmR\>'}\qty[
    (\d_\rmT\hat{s}_{\r_\rmL}^-)\hat{s}_{\r_\rmR}^+
    +\hat{s}_{\r_\rmR}^-(\d_\rmT\hat{s}_{\r_\rmL}^+)]
    \nonumber\\
    &\quad+\frac{J_\rmT}{2}\sum_{\<\r_\rmL,\r_\rmR\>'}\qty[
    (\d_\rmT\hat{s}_{\r_\rmL}^z)\hat{s}_{\r_\rmR}^z
    +\hat{s}_{\r_\rmR}^z(\d_\rmT\hat{s}_{\r_\rmL}^z)].
\end{align}
\end{subequations}
The operators $\d_\rmT \hat{s}_{\r_\rmL}^\pm$ and $\d_\rmT \hat{s}_{\r_\rmL}^z$ are rewritten as
\begin{subequations}
    \begin{align*}
        \d_\rmT \hat{s}_{\r_\rmL}^\pm&=\pm i J_\rmT(\hat{s}_{\r_\rmL}^z\hat{s}_{\r_\rmR}^\pm-\hat{s}_{\r_\rmL}^\pm\hat{s}_{\r_\rmR}^z),\\
        \d_\rmT \hat{s}_{\r_\rmL}^z&=\frac{iJ_\rmT}{2}(\hat{s}_{\r_\rmL}^+\hat{s}_{\r_\rmR}^--\hat{s}_{\r_\rmL}^-\hat{s}_{\r_\rmR}^+).
    \end{align*}
\end{subequations}
These relations combined with the commutation relation of $\hat{s}_\r$ and Eqs.~\eqref{eq_app:hat_I_S_2} and \eqref{eq_app:hat_I_E^S} show that $\hat{I}_\mathrm{E}^{(S)}$ is proportional to the spin current operator:
\begin{align}\label{eq_app:hat_I_E^S_2}
    \hat{I}_\mathrm{E}^{(\mathrm{S})}
    &=-\frac{J_\rmT}{2}\hat{I}_\mathrm{S}.
\end{align}
We note that $\hat{I}_\mathrm{E}^{(zz)}$ and $\hat{I}_\mathrm{E}^{(\mathrm{S})}$ vanish when the  $\hat{s}_{\r_\rmL}^z\hat{s}_{\r_\rmR}^z$ term in $\hat{H}_\rmT$ is absent.

We now compute the expectation values of the currents.
For convenience, we define the shorthand notation
\begin{align}
    \<\hat{A}_1(t_1)\cdots \hat{A}_n(t_n)\>_{\hat{K}'}^{\hat{H}'}&\equiv\Tr[\hat{\rho}_{\hat{K}'}\hat{A}_{1;\hat{H}'}(t_1)\cdots \hat{A}_{n;\hat{H}'}(t_n)],\\
    \hat{A}_{j;\hat{H}'}(t_j)&=e^{i\hat{H}' t_j}\hat{A}_{j}e^{-i\hat{H}' t_j},
\end{align}
where the time evolution of operators are governed by a hermitian operator $\hat{H}'$ and the average is taken with respect to the density operator $\hat{\rho}_{\hat{K}'}$.
By taking the average of $\hat{I}_\mathrm{S}(t)$, $\hat{I}_\mathrm{E}(t)$, and, $\hat{I}_\mathrm{H}(t)$ with respect to $\hat{\rho}_{\hat{K}}$, spin, energy, and heat currents are given by
\begin{subequations}\label{eqs_app:I}
    \begin{align}
    I_\mathrm{S}(t)&=\<\hat{I}_\mathrm{S}(t)\>_{\hat{K}}^{\hat{\H}},\\
    I_\mathrm{E}(t)
    &=\<\hat{I}_\mathrm{E}(t)\>_{\hat{K}}^{\hat{\H}},\\
    I_\mathrm{H}(t)&=I_\mathrm{E}(t)+h_\rmL I_\mathrm{S}(t),
    \end{align}
\end{subequations}
respectively.
Because of Eqs.~\eqref{eq_app:hat_I_E_2}--\eqref{eq_app:hat_I_E^S_2} and \eqref{eqs_app:I}, the energy current has three terms as
\begin{align}\label{eq_app:I_E_2}
    I_\mathrm{E}(t)=I_\mathrm{E}^{(+-)}(t)+I_\mathrm{E}^{(zz)}(t)-\frac{J_\rmT}{2}I_\mathrm{S}(t).
\end{align}
Introducing the two correlation functions
\begin{subequations}\label{eqs_app:I^<}
\begin{align}
    \I_{+-}^<(t_1,t_2)&\equiv 
    J_\rmT\sum_{\<\r_\rmL,\r_\rmR\>'}\expval{ \hat{s}_{\r_\rmL}^-(t_2)\hat{s}_{\r_\rmR}^+(t_1)}_{\hat{K}}^{\hat{\H}},\\
    \I_{zz}^<(t_1,t_2)&\equiv 
    J_\rmT\sum_{\<\r_\rmL,\r_\rmR\>'}\expval{ \hat{s}_{\r_\rmL}^z(t_2)\hat{s}_{\r_\rmR}^z(t_1)}_{\hat{K}}^{\hat{\H}},
\end{align}
\end{subequations}
and using Eqs.~\eqref{eq_app:hat_I_S_2}--\eqref{eq_app:hat_I_E^S_2} and \eqref{eqs_app:I}, we obtain
\begin{subequations}\label{eqs_app:I_I^>}
\begin{align}
    \label{eq_app:<I_S>(t)>}
    I_\mathrm{S}(t)
    &=\Im[\I_{+-}^<(t,t)],\\
    \label{eq_app:<I_E^+->(t)>}
    I_\mathrm{E}^{(+-)}(t)
    &=-\Re[\d_{t_2}\I_{+-}^<(t_1,t_2)]_{t_1,t_2\to t},\\
    \label{eq_app:<I_E^zz>(t)>}
    I_\mathrm{E}^{(zz)}(t)
    &=-\Re[\d_{t_2}\I_{zz}^<(t_1,t_2)]_{t_1,t_2\to t}.
\end{align}
\end{subequations}

\begin{figure}[tb]
\centering
\includegraphics[width=0.6\columnwidth,clip]{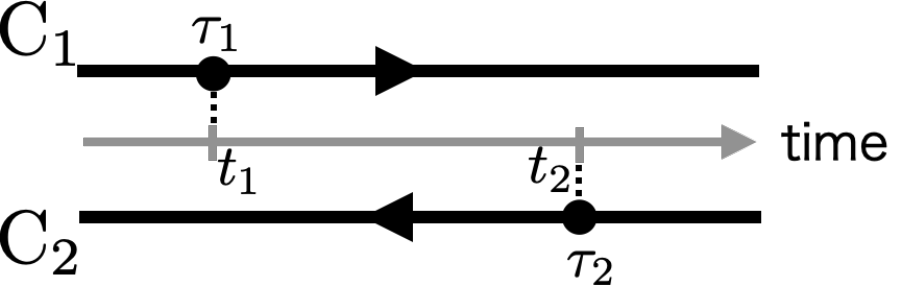}
\caption{\label{fig:Contour1}
The Keldysh contour $\mathrm{C}=\mathrm{C}_1 + \mathrm{C}_2$.}
\end{figure}

We evaluate the currents within the second-order perturbation in $J_\rmT$.
To this end, we utilize the Schwinger-Keldysh formalism, regarding $\I_{+-}^<(t_1,t_2)$ and $\I_{zz}^<(t_1,t_2)$ as the lesser components of the time-ordered contour correlation functions given by
\begin{align}\label{eq_app:I_munu^<}
    \I_{\mu\nu}^<(t_1,t_2)&=J_\rmT\sum_{\<\r_\rmL,\r_\rmR\>'}\expval{T_\rmC \hat{s}_{\r_\rmR}^\mu(\tau_1)\hat{s}_{\r_\rmL}^\nu(\tau_2)}_{\hat{K}}^{\hat{\H}},
\end{align}
where $(\mu,\nu)=(+,-), (z,z)$, $\tau_2=(t_1,\rmC_1)$ and $\tau_1=(t_2,\rmC_2)$ are the contour variables in the Keldysh contour $\rmC=\rmC_1+\rmC_2$ as in Fig.~\ref{fig:Contour1}, and $T_\rmC$ is the time-ordering operator over $\rmC$.
Because of $I_\mathrm{S}=\order{(J_\rmT)^2}$ as shown below, the last term $\sim\order{(J_\rmT)^3}$ in Eq.~\eqref{eq_app:I_E_2} is negligible.

\subsection{Perturbative expansions}
By applying the formal expression of the perturbative expansions in small $J_\rmT$, $\I_{\mu\nu}^<(t_1,t_2)$ Eq.~\eqref{eq_app:I_munu^<} is rewritten as
\begin{align}\label{eq_app:I_munu^<_2}
    &\I_{\mu\nu}^<(t_1,t_2)\nonumber\\
    &=J_\rmT\sum_{\<\r_\rmL,\r_\rmR\>'}\expval{T_\rmC \hat{s}_{\r_\rmR}^\mu(\tau_1)\hat{s}_{\r_\rmL}^\nu(\tau_2)\exp\qty[-i\int_\rmC \dd{\xi}\hat{H}_\rmT(\xi)]}_{\hat{K}}^{\hat{H}}\nonumber\\
    &=\frac{i^3(J_\rmT)^2}{2^{n_{\mu\nu}}}\sum_{\<\r_\rmL,\r_\rmR\>'}\sum_{\<\r_\rmL',\r_\rmR'\>'}\int_\rmC \dd{\xi}
    \nonumber\\
    &\quad\times \G_{\mu\nu;\r_\rmR\r_\rmR'}(\tau_1,\xi)
    \G_{\mu\nu;\r_\rmL'\r_\rmL}(\xi,\tau_2) + \order{(J_\rmT)^3},
\end{align}
where time evolution of the operators in $\<\cdots\>_{\hat{K}}^{\hat{H}}$ are governed by $\hat{H}=\hat{H}_\rmL+\hat{H}_\rmR$, $n_{\mu\nu}$ takes $n_{+-}=1$ and $n_{zz}=0$, and 
$
    i\G_{\mu\nu;\r_\alpha \r_\alpha'}(\tau,\tau')\equiv\expval{T_\rmC \hat{s}_{\r_\alpha}^\mu(\tau)        {{\hat{s}}}_{\r_\alpha'}^\nu(\tau')}_{\hat{K}_\alpha}^{\hat{H}_\alpha}
$
is the time-ordered spin correlation function.
By using 
$\hat{s}_{\r_\alpha;\hat{H}_\alpha}^\pm(t)=\hat{s}_{\r_\alpha;\hat{K}_\alpha}^\pm(t)e^{\pm ih_\alpha t}$ and $\hat{s}_{\r_\alpha;\hat{H}_\alpha}^z(t)=\hat{s}_{\r_\alpha;\hat{K}_\alpha}^z(t)$, 
the spin correlation function reads
\begin{align}\label{eqs_app:mathrm_G_to_G}
    \G_{\mu\nu;\r_\alpha \r_\alpha'}(\tau,\tau')
    &=G_{\mu\nu;\r_\alpha \r_\alpha'}(\tau,\tau')e^{ih_\alpha (t-t')n_{\mu\nu}},
\end{align}
where 
\begin{align}\label{eq_app:G}
    iG_{\mu\nu;\r_\alpha \r_\alpha'}(\tau,\tau')\equiv
    \expval{T_\rmC \hat{s}_{\r_\alpha}^\mu(\tau)        {{\hat{s}}}_{\r_\alpha'}^\nu(\tau')}_{\hat{K}_\alpha}^{\hat{K}_\alpha},
\end{align}
and $t$ and $t'$ are time variable corresponding to $\tau$ and $\tau'$, respectively.
By using the Langreth rule for the lesser component with $\tau_1=(t_1,\rmC_1)$ and $\tau_2=(t_2,\rmC_2)$ for Eq~Eq.~\eqref{eq_app:I_munu^<_2} as well as Eq.~\eqref{eqs_app:mathrm_G_to_G}, we obtain
\begin{align}\label{eq_app:I_munu^<_3}
    &\I_{\mu\nu}^<(t_1,t_2)\nonumber\\
    &=\frac{i^3(J_\rmT)^2}{2^{n_{\mu\nu}}}\sum_{\<\r_\rmL,\r_\rmR\>'}\sum_{\<\r_\rmL',\r_\rmR'\>'}\int_{-\infty}^{\infty}\dd{s}
    e^{[ih_\rmR(t_1-s)+ih_\rmL(s-t_2)]n_{\mu\nu}}\nonumber\\
    &\qquad\times
    \left[G_{\mu\nu;\r_\rmR\r_\rmR'}^\ret(t_1-s)G_{\mu\nu;\r_\rmL'\r_\rmL}^<(s-t_2)\right.\nonumber\\
    &\qquad\qquad\left.+G_{\mu\nu;\r_\rmR\r_\rmR'}^<(t_1-s)G_{\mu\nu;\r_\rmL'\r_\rmL}^\adv(s-t_2)\right],
\end{align}
where $G_{\mu\nu;\r_\rmR\r_\rmR'}^\ret(t)$, $G_{\mu\nu;\r_\rmR\r_\rmR'}^\adv(t)$, and $G_{\mu\nu;\r_\rmL'\r_\rmL}^<(t)$ are the retarded, advanced, and lesser components of $G_{\mu\nu;\r_\rmR\r_\rmR'}(\tau,\tau')$, respectively.
To derive the above result, we used the fact that $G_{\mu\nu;\r_\alpha\r_\alpha'}^\beta(t,t')=G_{\mu\nu;\r_\alpha\r_\alpha'}^\beta(t-t')$ with $\beta=\ret, \adv, <$ for a thermal state depends on time only via the difference $t-t'$.

We note that Eq.~\eqref{eq_app:I_munu^<_3} provides the leading behavior of the tunneling current with respect to $J_\rmT$ and applicable beyond SWA.

\subsection{Bulk spin correlation functions in SWA}
To evaluate the tunneling currents governed by magnons, we now employ SWA to compute $G_{\mu\nu;\r_\alpha\r_\alpha'}^\beta(t)$.
These correlation functions are given by
\begin{subequations}\label{eqs:G_r}
\begin{align}
iG_{\mu\nu;\r_\alpha \r_\alpha'}^\ret(t)&=\theta(t)\<[\hat{s}_{\r_\alpha}^\mu(t),{{\hat{s}}}_{\r_\alpha'}^\nu(0)]\>_{\hat{K}_\alpha}^{\hat{K}_\alpha},\\
iG_{\mu\nu;\r_\alpha \r_\alpha'}^\adv(t)&
=-\theta(-t)\<[\hat{s}_{\r_\alpha}^\mu(t),{{\hat{s}}}_{\r_\alpha'}^\nu(0)]\>_{\hat{K}_\alpha}^{\hat{K}_\alpha},\\
iG_{\mu\nu;\r_\alpha \r_\alpha'}^<(t)&=\<{{\hat{s}}}_{\r_\alpha'}^\nu(0)\hat{s}_{\r_\alpha}^\mu(t)\>_{\hat{K}_\alpha}^{\hat{K}_\alpha}.
\end{align}
\end{subequations}

We first evaluate $G_{+-;\r_\alpha\r_\alpha'}^\beta(t)$ corresponding to single-magnon correlation functions.
Since the bulk Hamiltonian in Eq.~\eqref{eq_app:K_SWA} respects momentum conservation, we can perform the Fourier transformation
\begin{align}\label{eq_app:G_+-_Fourier}
    G_{+-;\r_\alpha \r_\alpha'}^\beta(t)
    &=\frac{1}{\N}\sum_{\k_\alpha}\int\frac{\dd{\tilde{\omega}_\alpha}}{2\pi}e^{i\k_\alpha\cdot(\r_\alpha-\r_\alpha')-i\tilde{\omega}_\alpha t}\nonumber\\
    &\qquad\times G_{+-}^\beta(\k_\alpha,\tilde{\omega}_\alpha).
\end{align}
We note that the lattice constants of the cubic lattices in FIs are set unity and thus the momentum $\k_\alpha$ is dimensionless in this paper.
By using Eq.~\eqref{eq_app:HP_S^-} and \eqref{eq_app:HP_S^+}, the correlation functions in frequency-momentum space are given by
\begin{subequations}\label{eqs_app:G_+-_SWA}
\begin{align}
\label{eq:G^R(k,omega)_SWA}
G_{+-}^\ret({\k_\alpha},\tilde{\omega}_\alpha)&=[G_{+-}^\adv({\k_\alpha},\tilde{\omega}_\alpha)]^*=\frac{2\S}{\tilde{\omega}_\alpha^+- E_{\k_\alpha}},\\
G_{+-}^<(\k_\alpha,\tilde{\omega}_\alpha)&=n_{\mathrm{B},\alpha}(\tilde{\omega}_\alpha)2i\Im\,G_{+-}^\ret(\k_\alpha,\tilde{\omega}_\alpha),
\end{align}
\end{subequations}
where $\tilde{\omega}_\alpha^+=\tilde{\omega}_\alpha+i0^+$.

On the other hand, the functions $G_{zz;\r_\alpha\r_\alpha'}^\beta(t)$ corresponding to density-density correlations of magnons in SWA are found to be
\begin{subequations}\label{eqs_app:G_zz_SWA}
\begin{align}
G_{zz;\r_\alpha \r_\alpha'}^\ret(t)
&=\order{\S^0}
,\\
G_{zz;\r_\alpha \r_\alpha'}^\adv(t)
&=\order{\S^0}
,\\
G_{zz;\r_\alpha \r_\alpha'}^<(t)
&=-i\S^2-2i\S\rho_\alpha+\order{\S^0},
\end{align}
\end{subequations}
where $\rho_\alpha=\<\hat{b}_{\r_\alpha}^\dagger\hat{b}_{\r_\alpha}\>_{\hat{K}_\alpha}^{\hat{K}_\alpha}=\sum_{\k_\alpha}n_{\mathrm{B},\alpha}(E_{\k_\alpha})/\N$ is the number density of magnons.

\subsection{Application of SWA to currents}
We next evaluate $\I_{+-}^<(t_1,t_2)$ in Eq.~\eqref{eq_app:I_munu^<_3} within SWA to derive the expression of spin and heat currents in Eqs.~\eqref{eqs:I_SWA}.
We first evaluate $\I_{+-}^<(t_1,t_2)$ by substituting Eq.~\eqref{eq_app:G_+-_Fourier} into Eq.~\eqref{eq_app:I_munu^<_3}.
Recalling that the bonds on the interface denoted by $\<\r_\rmL,\r_\rmR\>'$ align along the $z$ axis as in Fig.~\ref{fig:system}, the summation $\sum_{\<\r_\rmL,\r_\rmR\>'}$ is replaced with $\sum_{z}$, where $\r_{\rmL,\rmR}=(0,0,z)+\r_{\rmL,\rmR}^{(0)}$ and $\r_{\rmL,\rmR}^{(0)}$ are edges of a reference bond.
As a result, $\I_{+-}^<(t_1,t_2)$ in 
Eq.~\eqref{eq_app:I_munu^<_3} reads
\begin{align}\label{eq_app:I_+-^<}
    &\I_{+-}^<(t_1,t_2)\nonumber\\
    &=\frac{i^3(J_\rmT)^2}{2}\int_{-\infty}^{\infty}\frac{\dd{\omega}}{2\pi}    e^{-i\omega(t_1-t_2)}\frac{(\N_z)^2}{\N^2}\sum_{\k_\rmL\k_\rmR}\delta_{k_\rmL^z,k_\rmR^z}\nonumber\\
    &\qquad\times\left[G_{+-}^\ret(\k_\rmR,\omega_\rmR)G_{+-}^<(\k_\rmL,\omega_\rmL)\right.\nonumber\\
    &\qquad\left.+G_{+-}^<(\k_\rmR,\omega_\rmR)G_{+-}^\adv(\k_\rmL,\omega_\rmL)\right]+\order{(J_\rmT)^3},
\end{align}
where $\omega_\alpha=\omega+h_\alpha$ and $\N_z$ is the number of lattice sites along the $z$ direction.
Using Eqs.~\eqref{eqs_app:G_+-_SWA} as well as $\A(\k_\alpha,\omega_\alpha)=-\Im\,G_{+-}^\ret(\k_\alpha,\omega_\alpha)$, we obtain
\begin{align}
    \I_{+-}^<(t_1,t_2)
    &=\int_{-\infty}^{\infty}\dd{\omega}e^{-i\omega(t_1-t_2)}[i\T(\omega)\Delta n_\mathrm{B}(\omega)\nonumber\\
    &\qquad+\text{(real-valued terms)}]+\order{(J_\rmT)^3},
\end{align}
where $\Delta n_\mathrm{B}(\omega)=n_{\mathrm{B},\rmL}(\omega_\rmL)-n_{\mathrm{B},\rmR}(\omega_\rmR)$ and the magnon transmittance $\T(\omega)$ is given in Eq.~\eqref{eq:T_+-}.
Therefore, Eqs.~\eqref{eq_app:<I_S>(t)>} and \eqref{eq_app:<I_E^+->(t)>} are found to be
\begin{align}\label{eq_app:<I_S(t)>_SWA}
    I_\mathrm{S}(t)
    &=\int_{-\infty}^{\infty}\dd{\omega}\T(\omega)\Delta n_\mathrm{B}(\omega),\\
    \label{eq_app:<I_E^+-(t)>_SWA}
    I_\mathrm{E}^{(+-)}(t)
    &=\int_{-\infty}^{\infty}\dd{\omega}\omega\T(\omega)\Delta n_\mathrm{B}(\omega).
\end{align}
These quantities are $\order{\S^2}$ because of $G_{+-}^\ret(\k_\alpha,\omega_\alpha)=\order{\S}$ as shown in Eq.~\eqref{eq:G^R(k,omega)_SWA}.

Finally, we estimate $\I_{zz}^<(t_1,t_2)$ and show that 
$I_\mathrm{E}^{(zz)}(t)$ in Eq.~\eqref{eq_app:<I_E^+->(t)>} contributing to the energy and heat currents is $\order{\S^0}$ and thus negligible within our accuracy of SWA because it is much smaller than $I_\mathrm{E}^{(+-)}(t)=\order{\S^2}$.
This can be proven as follows.
From Eqs.~\eqref{eqs_app:G_zz_SWA}, we see that time dependent parts of $G_{zz;\r_\alpha\r_\alpha'}^\ret(t)$, $G_{zz;\r_\alpha\r_\alpha'}^\adv(t)$, and $G_{zz;\r_\alpha\r_\alpha'}^<(t)$ are $\order{\S^0}$.
Combining these with Eqs.~\eqref{eq_app:<I_E^zz>(t)>} and ~\eqref{eq_app:I_munu^<_3}, we find $I_\mathrm{E}^{(zz)}(t)=\order{\S^0}$.

By substituting Eqs.~\eqref{eq_app:<I_S(t)>_SWA} and\eqref{eq_app:<I_E^+-(t)>_SWA} into Eqs.~\eqref{eqs_app:I}, 
the currents up to order $\order{(J_\rmT)^2,\S^2}$ are found to be
\begin{subequations}\label{eqs_app:I_SWA}
    \begin{align}
    I_\mathrm{S}&=\int_{-\infty}^{\infty}\dd{\omega}\T(\omega)\Delta n_\mathrm{B}(\omega),\\
    I_\mathrm{E}
    &=\int_{-\infty}^{\infty}\dd{\omega}\omega\T(\omega)\Delta n_\mathrm{B}(\omega),\\
    I_\mathrm{H}
    &=\int_{-\infty}^{\infty}\dd{\omega}(\omega+h_\rmL)\T(\omega)\Delta n_\mathrm{B}(\omega).
    \end{align}
\end{subequations}

\section{\label{appendix:fermionic Lorenz number}Lorenz number for free fermions}
\begin{figure}
    \centering
    \includegraphics[width=0.95\columnwidth,clip]{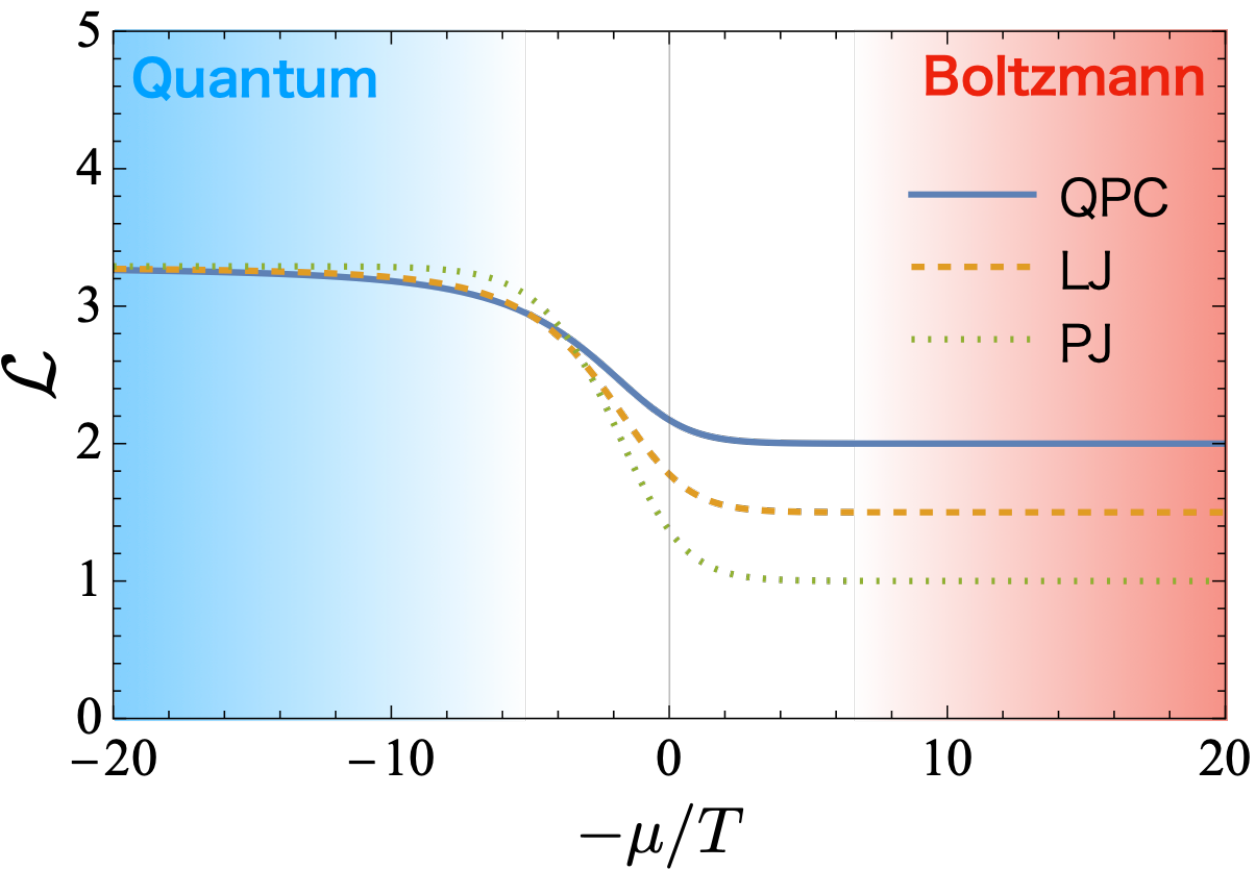}
    \caption{\label{fig:L for fermions}
    Lorenz number $\mathcal{L}$ for junction systems consisting of two noninteracting Fermi gases.
    In the Boltzmann limit $-\mu/T\to\infty$, $\mathcal{L}$ in different junctions approach different constants, while all $\mathcal{L}$ in the quantum degenerate limit $-\mu/T\to-\infty$ go to the same value $\pi^2/3=3.29$, manifesting the universal WF law for Fermi liquids.
    }
\end{figure}

Here, we discuss the Lorenz number $\mathcal{L}$ for free Fermi gases in the cases of a QPC, linear junction (LJ), and planar junction (PJ).
We explicitly show the following two points: (i) $\mathcal{L}\to(d+1)$ in the Boltzmann limit approaches a constant sensitive to the junction geometry via the power $d$ in $\T(\omega)\sim\omega^d$ (ii) $\mathcal{L}\to\pi^2/3$ in the quantum limit approaches to a constant insensitive to the junction geometry.
The results are summarized in Fig.~\ref{fig:L for fermions}.

We consider tunneling transport in the junction systems consisting of two Fermi gases with quadratic dispersion~\cite{mahan2000many}.
For convenience, we set the charge of fermions $e$ and the Boltzmann constant $k_B$ unity ($e=k_B=1$).
When the two reservoirs are weakly connected with each other, the particle (heat) current $I_\mathrm{P}$ ($I_\mathrm{H}$) flowing between two Fermi gases up to the second order of tunneling coupling are obtained by replacing $n_{\mathrm{B},\alpha}(\omega+h_\alpha)$ with $n_{\mathrm{F},\alpha}(\omega-\mu_\alpha)=1/[e^{(\omega-\mu_\alpha)/T_\alpha}+1]$ in Eq.~\eqref{eqs:I_SWA} [Eq.~\eqref{eq:I_H_SWA}], where $\mu_\alpha$ and $T_\alpha$ are a chemical potential and temperature in the reservoir $\alpha=\rmL,\rmR$, respectively.
The transmittance of fermions obeys power law $\T(\omega)=\tilde{A}\omega^d/\Gamma(d+1)$ with 
\begin{align}
    d=
    \begin{cases}
    1&(\mathrm{QPC})\\
    1/2&(\mathrm{LJ})\\
    0&(\mathrm{PJ})
    \end{cases}.
\end{align}
(We do not mention the details of the coefficient $\tilde{A}>0$ irrelevant to the following discussion.)
By expanding $I_\mathrm{P}$ and $I_\mathrm{H}$ with respect to small biases $\Delta \mu=\mu_\rmL-\mu_\rmR$ and $\Delta T=T_\rmL-T_\rmR$, we obtain the conductances as
\begin{align}
    \frac{L_{11}}{\tilde{A}T^{d}}&=\tilde{F}_{d}(x),\\
    \frac{L_{12}}{\tilde{A}T^{d}}&=\frac{L_{21}}{\tilde{A}T^{d+1}}=(d+1)\,\tilde{F}_{d+1}(x)
    +x\,\tilde{F}_{d}(x),\\
    \frac{L_{22}}{\tilde{A}T^{d+1}}&=(d+2)(d+1)\,\tilde{F}_{d+2}(x)\nonumber\\
    &\quad+2(d+1)x\,\tilde{F}_{d+1}(x)+x^2\,\tilde{F}_{d}(x),
\end{align}
where $x=-\mu/T$, $\mu=(\mu_\rmL+\mu_\rmR)/2$, $T=(T_\rmL+T_\rmR)/2$, and 
\begin{align}
    \tilde{F}_{d'}(x)
    &=\int_0^\infty\!\!\dd{y}\,\frac{y^{d'-1}}{\Gamma(d')}\frac{1}{e^{y+x}+1}=-\mathrm{Li}_{d'}(-e^{-x})
\end{align}
is the Fermi-Dirac integral.
The thermal conductance is given by
\begin{align}
K=L_{22}-\frac{L_{12}L_{21}}{L_{11}}=L_{22}-\frac{T(L_{12})^2}{L_{11}}.
\end{align}

Figure~\ref{fig:L for fermions} shows the Lorenz numbers $\mathcal{L}=K/(L_{11}T)$ for QPC, LJ, and PJ as functions of $x=-\mu/T$.
As for magnons in the main text, $x\to+\infty$ corresponds to the Boltzmann limit.
In this case, the Lorenz numbers approach to constant values sensitive to the junction geometry as discussed in Sec.~\ref{sec:Lorenz number}:
\begin{align}\label{eq_app:Lorenz_boltzmann}
    \mathcal{L}\to(d+1).
\end{align}
In the LJ case with $d=1/2$, we can recover Eq.~\eqref{eq:Lorenz_classical} in the main text.

The other limit $x\to-\infty$ corresponds to the quantum degenerate regime of fermions.
This is because a chemical potential approaches the Fermi energy $\mu\to\epsilon_\mathrm{F}$ in the low temperature limit with fixed fermion density.
In this case, the Lorenz numbers approach another constant value independent from the junction geometry:
\begin{align}\label{eq_app:Lorenz_degenerate_fermions}
    \mathcal{L}\to\pi^2/3=3.29.
\end{align}
This insensitivity to the junction geometry results from the strong constraints on degenerate fermions, for which the Fermi surface allows the replacement, $\T(\omega)\simeq T(\epsilon_\mathrm{F})$ and $\omega\T(\omega)\simeq \epsilon_\mathrm{F} \T(\epsilon_\mathrm{F})$, in the evaluation of currents.
As a result, the tunneling currents are governed only by Fermi distribution functions of reservoirs independently from the interface geometry.
For the same reason, $\mathcal{L}$ in Eq.~\eqref{eq_app:Lorenz_degenerate_fermions} takes the same value as its bulk counterpart given by the ratio of a thermal conductivity to an electrical conductivity, which is originally found by 
Wiedemann and Franz~\cite{Franz1853-do} and theoretically derived by Sommerfeld with consideration of the quantum degeneracy effect~\cite{Sommerfeld1928-re}.
We note that the WF law is broken for strongly interacting fermions in the unitary limit~\cite{husmann2018breakdown}, where the above replacement of transmittance is not justified~\cite{sekino2020mesoscopic}.

\section{\label{appendix:quasistationary model}Quasi-stationary model}
In this appendix, we review the quasi-stationary model describing relaxation dynamics near equilibrium, which is originally proposed in the tunneling transport for ultracold Fermi gases in Refs.~\cite{brantut2013thermoelectric,grenier2012probing} and extended to spin systems~\cite{Sekino2024-vp}.
We follow the formalism in these previous works.
We first introduce the model without the assumptions of junction geometry and of the high spin polarization of FIs, which is applicable beyond SWA.

We first relate the dynamics of spin and entropy currents to biases.
Using Eq.~\eqref{eq:linear currents}, the time derivative of magnetization and entropy differences between FIs, $\Delta M=M_\rmL-M_\rmR$ and $\Delta S=S_\rmL-S_\rmR$, are expressed as
\begin{align}\label{eq_app:current-bias3}
\dv{t}
\begin{pmatrix}
-\Delta M\\
\Delta S
\end{pmatrix}
&=-2
\begin{pmatrix}
I_{S}\\
I_\mathrm{H}/T
\end{pmatrix}
\nonumber\\
&=-G
\begin{pmatrix}
1&\alpha_{ch}\\
\alpha_{ch}&\mathcal{L}+\alpha_{ch}^2
\end{pmatrix}
\begin{pmatrix}
\Delta h\\
\Delta T
\end{pmatrix},
\end{align}
where we define
\begin{align}\label{eq_app:transport}
G=2L_{11},\quad\alpha_{ch}=\frac{L_{12}}{L_{11}},\quad 
\mathcal{L}=\frac{L_{22}}{TL_{11}}-\qty(\frac{L_{12}}{L_{11}})^2.
\end{align}
We note that the value of these transport coefficients generally depend on the geometry of the junction~\cite{Sekino2024-vp}.

When the system is in a quasi-stationary state, characterized by the rapid thermalization within each FI compared to the slower relaxation of $\Delta M$ and $\Delta T$ through MLJ, $\Delta M$ and $\Delta S$ are respond to the small biases.
By expanding $M_{\rmL/\rmR}(h\mp\Delta h/2,T\pm\Delta T/2)$ and $S_{\rmL/\rmR}(h\mp\Delta h/2,T\pm\Delta T/2)$ with respect to $\Delta h$ and $\Delta T$, we can obtain the following thermodynamic relation:
\begin{align}\label{eq_app:thermodynamics}
    \begin{pmatrix}
    -\Delta M\\
    \Delta S
    \end{pmatrix}
    &=
    \begin{pmatrix}
    \qty(\pdv{M}{h})_T&-\qty(\pdv{M}{T})_h\\
    - \qty(\pdv{S}{h})_T&\qty(\pdv{S}{T})_h\\
    \end{pmatrix}
    \begin{pmatrix}
    \Delta h\\
    \Delta T
    \end{pmatrix}\\
    &=\kappa
    \begin{pmatrix}
    1&\alpha_{r}\\
    \alpha_{r}&l+\alpha_{r}^2
    \end{pmatrix}
    \begin{pmatrix}
    \Delta h\\
    \Delta T
    \end{pmatrix},
\end{align}
where
\begin{subequations}\label{eq_app:thermodynamic_coefficients}
\begin{align}
    \kappa&=\qty(\pdv{M}{h})_T,\\
    \alpha_r&=-\qty(\pdv{S}{M})_T,\\
    l&=\frac{C_{M}}{\kappa T}=\frac{1}{\kappa}\qty(\pdv{S}{T})_{M}.
\end{align}
    
\end{subequations}
In Eqs.~\eqref{eq_app:thermodynamics} and \eqref{eq_app:thermodynamic_coefficients}, thermodynamic quantities are evaluated with the mean values: $M=(M_\rmL+M_\rmR)/2$, $S=(S_\rmL+S_\rmR)/2$, $h=(h_\rmL+h_\rmR)/2$, and $T=(T_\rmL+T_\rmR)/2$.
In terms of spins, $\kappa$ is the differential spin susceptibility and $C_{M}$ is the heat capacity with fixed magnetization.
In the context of the magnon gas, $\kappa$ and $\alpha_r$ are regarded as the isothermal compressibility and the dilatation coefficient, respectively.

The combination of Eq.\eqref{eq_app:current-bias3} with Eq.\eqref{eq_app:thermodynamics} results in the following differential equation describing the dynamics:
\begin{align}\label{eq_app:relaxation}
   \tau_0\dv{t}
    \begin{pmatrix}
    -\Delta M/\kappa\\
    \Delta T    
    \end{pmatrix}
    =-
    \begin{pmatrix}
    1&-\alpha\\
    -\frac{\alpha}{l}&\frac{\mathcal{L}+\alpha^2}{l}
    \end{pmatrix}
    \begin{pmatrix}
    -\Delta M/\kappa\\
    \Delta T  
    \end{pmatrix},
\end{align}
where
\begin{align}\label{eq_app:tau_0}
\tau_0=\frac{\kappa}{2L_{11}},\qquad
\alpha=\alpha_r-\alpha_{ch}.
\end{align}
We can find the solution of Eq.~\eqref{eq_app:relaxation} of the form
\begin{align}\label{eq_app:Delta_N,T(t)}
\begin{pmatrix}
-\Delta M(t)/\N\\
\Delta T(t)/T
\end{pmatrix}
=
\begin{pmatrix}
\tilde{\Lambda}_{11}(t)&\tilde{\Lambda}_{12}(t)\\
\tilde{\Lambda}_{21}(t)&\tilde{\Lambda}_{22}(t)
\end{pmatrix}
\begin{pmatrix}
-\Delta M(0)/\N\\
\Delta T(0)/T
\end{pmatrix},
\end{align}
where
\begin{align}\label{eq_app:Lambda_11}
\tilde{\Lambda}_{11}(t)
&=\frac{1}{2}\left\{\qty(e^{-t/\tau_-}+e^{-t/\tau_+})\right.\nonumber\\
&\qquad+\left.\frac{\frac{\mathcal{L}+\alpha^2}{l}-1}{\lambda_+-\lambda_-}\qty(e^{-t/\tau_-}-e^{-t/\tau_+})\right\},\\
\label{eq_app:Lambda_12}
\tilde{\Lambda}_{12}(t)&=\qty(\frac{T\kappa}{\N})^2l\tilde{\Lambda}_{21}(t)\nonumber\\
&=\frac{T}{\N}\frac{\alpha \kappa}{\lambda_+-\lambda_-}\qty(e^{-t/\tau_-}-e^{-t/\tau_+}),\\
\label{eq_app:Lambda_22}
\tilde{\Lambda}_{22}(t)&=\frac{1}{2}\left\{\qty(e^{-t/\tau_-}+e^{-t/\tau_+})\right.\nonumber\\
&\qquad\left.-\frac{\frac{\mathcal{L}+\alpha^2}{l}-1}{\lambda_+-\lambda_-}\qty(e^{-t/\tau_-}-e^{-t/\tau_+})\right\}.
\end{align}
The relaxation times $\tau_\pm$ and $\lambda_\pm$ are determined as 
\begin{align}\label{eq_app:lambda_pm}
\tau_\pm &= \tau_0/\lambda_\pm,\\
\lambda_\pm&=\frac{1}{2}\qty(1+\frac{\mathcal{L}+\alpha^2}{l})\pm\sqrt{\frac{\alpha^2}{l}+\frac{1}{4}\qty(1-\frac{\mathcal{L}+\alpha^2}{l})^2}.
\end{align}
Note that the signs before $\frac{\frac{\mathcal{L}+\alpha^2}{l}-1}{\lambda_+-\lambda_-}$ in Eqs.~\eqref{eq_app:Lambda_11} and \eqref{eq_app:Lambda_22} are consistent with those in Ref.~\cite{grenier2012probing} but opposite to those in Ref.~\cite{brantut2013thermoelectric}.
It is noteworthy that this solution for relaxation dynamics remains valid for any $h$, $T$, and junction geometry as long as the biases $\Delta h, \Delta T$ are small and the quasi-stationary condition is satisfied, even beyond the regime where the SWA is valid.

Let us now discuss the relaxation dynamics between highly spin-polarized FIs by applying SWA.
By using Eqs.~\eqref{eq_app:M_SWA}--\eqref{eq_app:U_SWA} for $\S=1/2$, we can evaluate the thermodynamic coefficients in Eqs.~\eqref{eq_app:thermodynamic_coefficients}.
Surprisingly, in this case, Eqs.~\eqref{eq_app:thermodynamic_coefficients} characterizing thermodynamics of bulk are identical with $L_{ij}$ for tunneling transport [in Eqs.~\eqref{eqs:L_ij}] in the MLJ, leading to
\begin{align}\label{eq:kappa_0}
    \frac{\kappa}{B\sqrt{T}}
    &=\frac{L_{11}}{A\sqrt{T}}
    =F_{1/2}(x),\\
    \label{eq:alpha_r_0}
    \alpha_r&=\alpha_{ch}
    =\frac{3}{2}\frac{F_{3/2}(x)}{F_{1/2}(x)}+x,\\
    l&=\mathcal{L}
    =\frac{15}{4}\frac{F_{5/2}\left(x\right)}{F_{1/2}\left(x\right)}-\frac{9}{4}\qty(\frac{F_{3/2}(x)}{F_{1/2}(x)})^2.
\end{align}
This comes from the fact that the magnon density of states $\rho(\omega)\propto\sqrt{\omega}$ in FIs follows the same power law as the magnon transmittance $\T(\omega)$ in MLJ [see Eq.~\eqref{eq:T_+-_power}].
As a result, the dynamics of $\Delta M$ and $\Delta T$, which are governed by the differential equation~\eqref{eq_app:relaxation}, are decoupled from each other:
\begin{align}
   \tau_0\dv{t}
    \begin{pmatrix}
    -\Delta M\\
    \Delta T    
    \end{pmatrix}
    =-
    \begin{pmatrix}
    1&0\\
    0&1
    \end{pmatrix}
    \begin{pmatrix}
    -\Delta M\\
    \Delta T  
    \end{pmatrix}.
\end{align}
where the relaxation time
\begin{align}\label{eq_app:tau_0^LJ}
\tau_0&=\frac{2}{\pi}\frac{(\N/\N_\mathrm{T})(J/J_\mathrm{T})^2}{J}=\textrm{constant}
\end{align}
is independent of $h$ and $T$.
The dynamics of spin and heat independently decay in the same relaxation time
\begin{align}
\begin{pmatrix}
\Delta M(t)\\
\Delta T(t)
\end{pmatrix}
=
\begin{pmatrix}
e^{-t/\tau_0}&0\\
0&e^{-t/\tau_0}
\end{pmatrix}
\begin{pmatrix}
\Delta M(0)\\
\Delta T(0)
\end{pmatrix},
\end{align}
Because of $J/J_\mathrm{T},\, J/T,\, \N/\N_\mathrm{T}\gg1$ in our case, $\tau_0^\mathrm{(LJ)}$ in Eq.~\eqref{eq_app:tau_0^LJ} is much larger than the typical thermalization time scale $1/J$.

\bibliography{masterbib}

\begin{thebibliography}{70}%
\makeatletter
\providecommand \@ifxundefined [1]{%
 \@ifx{#1\undefined}
}%
\providecommand \@ifnum [1]{%
 \ifnum #1\expandafter \@firstoftwo
 \else \expandafter \@secondoftwo
 \fi
}%
\providecommand \@ifx [1]{%
 \ifx #1\expandafter \@firstoftwo
 \else \expandafter \@secondoftwo
 \fi
}%
\providecommand \natexlab [1]{#1}%
\providecommand \enquote  [1]{``#1''}%
\providecommand \bibnamefont  [1]{#1}%
\providecommand \bibfnamefont [1]{#1}%
\providecommand \citenamefont [1]{#1}%
\providecommand \href@noop [0]{\@secondoftwo}%
\providecommand \href [0]{\begingroup \@sanitize@url \@href}%
\providecommand \@href[1]{\@@startlink{#1}\@@href}%
\providecommand \@@href[1]{\endgroup#1\@@endlink}%
\providecommand \@sanitize@url [0]{\catcode `\\12\catcode `\$12\catcode `\&12\catcode `\#12\catcode `\^12\catcode `\_12\catcode `\%12\relax}%
\providecommand \@@startlink[1]{}%
\providecommand \@@endlink[0]{}%
\providecommand \url  [0]{\begingroup\@sanitize@url \@url }%
\providecommand \@url [1]{\endgroup\@href {#1}{\urlprefix }}%
\providecommand \urlprefix  [0]{URL }%
\providecommand \Eprint [0]{\href }%
\providecommand \doibase [0]{https://doi.org/}%
\providecommand \selectlanguage [0]{\@gobble}%
\providecommand \bibinfo  [0]{\@secondoftwo}%
\providecommand \bibfield  [0]{\@secondoftwo}%
\providecommand \translation [1]{[#1]}%
\providecommand \BibitemOpen [0]{}%
\providecommand \bibitemStop [0]{}%
\providecommand \bibitemNoStop [0]{.\EOS\space}%
\providecommand \EOS [0]{\spacefactor3000\relax}%
\providecommand \BibitemShut  [1]{\csname bibitem#1\endcsname}%
\let\auto@bib@innerbib\@empty
\bibitem [{\citenamefont {Sidorenkov}\ \emph {et~al.}(2013)\citenamefont {Sidorenkov}, \citenamefont {Tey}, \citenamefont {Grimm}, \citenamefont {Hou}, \citenamefont {Pitaevskii},\ and\ \citenamefont {Stringari}}]{sidorenkov2013second}%
  \BibitemOpen
  \bibfield  {author} {\bibinfo {author} {\bibfnamefont {L.~A.}\ \bibnamefont {Sidorenkov}}, \bibinfo {author} {\bibfnamefont {M.~K.}\ \bibnamefont {Tey}}, \bibinfo {author} {\bibfnamefont {R.}~\bibnamefont {Grimm}}, \bibinfo {author} {\bibfnamefont {Y.-H.}\ \bibnamefont {Hou}}, \bibinfo {author} {\bibfnamefont {L.}~\bibnamefont {Pitaevskii}},\ and\ \bibinfo {author} {\bibfnamefont {S.}~\bibnamefont {Stringari}},\ }\bibfield  {title} {\bibinfo {title} {Second sound and the superfluid fraction in a fermi gas with resonant interactions},\ }\href@noop {} {\bibfield  {journal} {\bibinfo  {journal} {Nature}\ }\textbf {\bibinfo {volume} {498}},\ \bibinfo {pages} {78} (\bibinfo {year} {2013})}\BibitemShut {NoStop}%
\bibitem [{\citenamefont {Brantut}\ \emph {et~al.}(2013)\citenamefont {Brantut}, \citenamefont {Grenier}, \citenamefont {Meineke}, \citenamefont {Stadler}, \citenamefont {Krinner}, \citenamefont {Kollath}, \citenamefont {Esslinger},\ and\ \citenamefont {Georges}}]{brantut2013thermoelectric}%
  \BibitemOpen
  \bibfield  {author} {\bibinfo {author} {\bibfnamefont {J.-P.}\ \bibnamefont {Brantut}}, \bibinfo {author} {\bibfnamefont {C.}~\bibnamefont {Grenier}}, \bibinfo {author} {\bibfnamefont {J.}~\bibnamefont {Meineke}}, \bibinfo {author} {\bibfnamefont {D.}~\bibnamefont {Stadler}}, \bibinfo {author} {\bibfnamefont {S.}~\bibnamefont {Krinner}}, \bibinfo {author} {\bibfnamefont {C.}~\bibnamefont {Kollath}}, \bibinfo {author} {\bibfnamefont {T.}~\bibnamefont {Esslinger}},\ and\ \bibinfo {author} {\bibfnamefont {A.}~\bibnamefont {Georges}},\ }\bibfield  {title} {\bibinfo {title} {{A Thermoelectric Heat Engine with Ultracold Atoms}},\ }\href {https://doi.org/10.1126/science.1242308} {\bibfield  {journal} {\bibinfo  {journal} {Science}\ }\textbf {\bibinfo {volume} {342}},\ \bibinfo {pages} {713} (\bibinfo {year} {2013})}\BibitemShut {NoStop}%
\bibitem [{\citenamefont {Krinner}\ \emph {et~al.}(2017)\citenamefont {Krinner}, \citenamefont {Esslinger},\ and\ \citenamefont {Brantut}}]{krinner2017}%
  \BibitemOpen
  \bibfield  {author} {\bibinfo {author} {\bibfnamefont {S.}~\bibnamefont {Krinner}}, \bibinfo {author} {\bibfnamefont {T.}~\bibnamefont {Esslinger}},\ and\ \bibinfo {author} {\bibfnamefont {J.-P.}\ \bibnamefont {Brantut}},\ }\bibfield  {title} {\bibinfo {title} {Two-terminal transport measurements with cold atoms},\ }\href@noop {} {\bibfield  {journal} {\bibinfo  {journal} {Journal of Physics: Condensed Matter}\ }\textbf {\bibinfo {volume} {29}},\ \bibinfo {pages} {343003} (\bibinfo {year} {2017})}\BibitemShut {NoStop}%
\bibitem [{\citenamefont {Husmann}\ \emph {et~al.}(2018)\citenamefont {Husmann}, \citenamefont {Lebrat}, \citenamefont {H\"{a}usler}, \citenamefont {Brantut}, \citenamefont {Corman},\ and\ \citenamefont {Esslinger}}]{husmann2018breakdown}%
  \BibitemOpen
  \bibfield  {author} {\bibinfo {author} {\bibfnamefont {D.}~\bibnamefont {Husmann}}, \bibinfo {author} {\bibfnamefont {M.}~\bibnamefont {Lebrat}}, \bibinfo {author} {\bibfnamefont {S.}~\bibnamefont {H\"{a}usler}}, \bibinfo {author} {\bibfnamefont {J.-P.}\ \bibnamefont {Brantut}}, \bibinfo {author} {\bibfnamefont {L.}~\bibnamefont {Corman}},\ and\ \bibinfo {author} {\bibfnamefont {T.}~\bibnamefont {Esslinger}},\ }\bibfield  {title} {\bibinfo {title} {Breakdown of the wiedemann--franz law in a unitary fermi gas},\ }\href {https://doi.org/10.1073/pnas.1803336115} {\bibfield  {journal} {\bibinfo  {journal} {Proceedings of the National Academy of Sciences}\ }\textbf {\bibinfo {volume} {115}},\ \bibinfo {pages} {8563} (\bibinfo {year} {2018})}\BibitemShut {NoStop}%
\bibitem [{\citenamefont {Guardado-Sanchez}\ \emph {et~al.}(2020)\citenamefont {Guardado-Sanchez}, \citenamefont {Morningstar}, \citenamefont {Spar}, \citenamefont {Brown}, \citenamefont {Huse},\ and\ \citenamefont {Bakr}}]{PhysRevX.10.011042}%
  \BibitemOpen
  \bibfield  {author} {\bibinfo {author} {\bibfnamefont {E.}~\bibnamefont {Guardado-Sanchez}}, \bibinfo {author} {\bibfnamefont {A.}~\bibnamefont {Morningstar}}, \bibinfo {author} {\bibfnamefont {B.~M.}\ \bibnamefont {Spar}}, \bibinfo {author} {\bibfnamefont {P.~T.}\ \bibnamefont {Brown}}, \bibinfo {author} {\bibfnamefont {D.~A.}\ \bibnamefont {Huse}},\ and\ \bibinfo {author} {\bibfnamefont {W.~S.}\ \bibnamefont {Bakr}},\ }\bibfield  {title} {\bibinfo {title} {Subdiffusion and heat transport in a tilted two-dimensional fermi-hubbard system},\ }\href {https://doi.org/10.1103/PhysRevX.10.011042} {\bibfield  {journal} {\bibinfo  {journal} {Phys. Rev. X}\ }\textbf {\bibinfo {volume} {10}},\ \bibinfo {pages} {011042} (\bibinfo {year} {2020})}\BibitemShut {NoStop}%
\bibitem [{\citenamefont {Christodoulou}\ \emph {et~al.}(2021)\citenamefont {Christodoulou}, \citenamefont {Ga{\l}ka}, \citenamefont {Dogra}, \citenamefont {Lopes}, \citenamefont {Schmitt},\ and\ \citenamefont {Hadzibabic}}]{christodoulou2021observation}%
  \BibitemOpen
  \bibfield  {author} {\bibinfo {author} {\bibfnamefont {P.}~\bibnamefont {Christodoulou}}, \bibinfo {author} {\bibfnamefont {M.}~\bibnamefont {Ga{\l}ka}}, \bibinfo {author} {\bibfnamefont {N.}~\bibnamefont {Dogra}}, \bibinfo {author} {\bibfnamefont {R.}~\bibnamefont {Lopes}}, \bibinfo {author} {\bibfnamefont {J.}~\bibnamefont {Schmitt}},\ and\ \bibinfo {author} {\bibfnamefont {Z.}~\bibnamefont {Hadzibabic}},\ }\bibfield  {title} {\bibinfo {title} {Observation of first and second sound in a bkt superfluid},\ }\href@noop {} {\bibfield  {journal} {\bibinfo  {journal} {Nature}\ }\textbf {\bibinfo {volume} {594}},\ \bibinfo {pages} {191} (\bibinfo {year} {2021})}\BibitemShut {NoStop}%
\bibitem [{\citenamefont {Hoffmann}\ \emph {et~al.}(2021)\citenamefont {Hoffmann}, \citenamefont {Singh}, \citenamefont {Paintner}, \citenamefont {J{\"a}ger}, \citenamefont {Limmer}, \citenamefont {Mathey},\ and\ \citenamefont {Hecker~Denschlag}}]{hoffmann2021second}%
  \BibitemOpen
  \bibfield  {author} {\bibinfo {author} {\bibfnamefont {D.~K.}\ \bibnamefont {Hoffmann}}, \bibinfo {author} {\bibfnamefont {V.~P.}\ \bibnamefont {Singh}}, \bibinfo {author} {\bibfnamefont {T.}~\bibnamefont {Paintner}}, \bibinfo {author} {\bibfnamefont {M.}~\bibnamefont {J{\"a}ger}}, \bibinfo {author} {\bibfnamefont {W.}~\bibnamefont {Limmer}}, \bibinfo {author} {\bibfnamefont {L.}~\bibnamefont {Mathey}},\ and\ \bibinfo {author} {\bibfnamefont {J.}~\bibnamefont {Hecker~Denschlag}},\ }\bibfield  {title} {\bibinfo {title} {{Second sound in the crossover from the Bose-Einstein condensate to the Bardeen-Cooper-Schrieffer superfluid}},\ }\href@noop {} {\bibfield  {journal} {\bibinfo  {journal} {Nature Communications}\ }\textbf {\bibinfo {volume} {12}},\ \bibinfo {pages} {7074} (\bibinfo {year} {2021})}\BibitemShut {NoStop}%
\bibitem [{\citenamefont {Yan}\ \emph {et~al.}(2022)\citenamefont {Yan}, \citenamefont {Patel}, \citenamefont {Mukherjee}, \citenamefont {Vale}, \citenamefont {Fletcher},\ and\ \citenamefont {Zwierlein}}]{yan2022thermography}%
  \BibitemOpen
  \bibfield  {author} {\bibinfo {author} {\bibfnamefont {Z.}~\bibnamefont {Yan}}, \bibinfo {author} {\bibfnamefont {P.~B.}\ \bibnamefont {Patel}}, \bibinfo {author} {\bibfnamefont {B.}~\bibnamefont {Mukherjee}}, \bibinfo {author} {\bibfnamefont {C.~J.}\ \bibnamefont {Vale}}, \bibinfo {author} {\bibfnamefont {R.~J.}\ \bibnamefont {Fletcher}},\ and\ \bibinfo {author} {\bibfnamefont {M.}~\bibnamefont {Zwierlein}},\ }\bibfield  {title} {\bibinfo {title} {{Thermography of the superfluid transition in a strongly interacting Fermi gas}},\ }\href {https://doi.org/10.48550/arXiv.2212.13752} {\bibfield  {journal} {\bibinfo  {journal} {arXiv preprint arXiv:2212.13752}\ } (\bibinfo {year} {2022})}\BibitemShut {NoStop}%
\bibitem [{\citenamefont {Krinner}\ \emph {et~al.}(2016)\citenamefont {Krinner}, \citenamefont {Lebrat}, \citenamefont {Husmann}, \citenamefont {Grenier}, \citenamefont {Brantut},\ and\ \citenamefont {Esslinger}}]{Krinner2016-gd}%
  \BibitemOpen
  \bibfield  {author} {\bibinfo {author} {\bibfnamefont {S.}~\bibnamefont {Krinner}}, \bibinfo {author} {\bibfnamefont {M.}~\bibnamefont {Lebrat}}, \bibinfo {author} {\bibfnamefont {D.}~\bibnamefont {Husmann}}, \bibinfo {author} {\bibfnamefont {C.}~\bibnamefont {Grenier}}, \bibinfo {author} {\bibfnamefont {J.-P.}\ \bibnamefont {Brantut}},\ and\ \bibinfo {author} {\bibfnamefont {T.}~\bibnamefont {Esslinger}},\ }\bibfield  {title} {\bibinfo {title} {Mapping out spin and particle conductances in a quantum point contact},\ }\href@noop {} {\bibfield  {journal} {\bibinfo  {journal} {Proc. Natl. Acad. Sci. U. S. A.}\ }\textbf {\bibinfo {volume} {113}},\ \bibinfo {pages} {8144} (\bibinfo {year} {2016})}\BibitemShut {NoStop}%
\bibitem [{\citenamefont {Enss}\ and\ \citenamefont {Thywissen}(2019)}]{enss2019universal}%
  \BibitemOpen
  \bibfield  {author} {\bibinfo {author} {\bibfnamefont {T.}~\bibnamefont {Enss}}\ and\ \bibinfo {author} {\bibfnamefont {J.~H.}\ \bibnamefont {Thywissen}},\ }\bibfield  {title} {\bibinfo {title} {Universal spin transport and quantum bounds for unitary fermions},\ }\href@noop {} {\bibfield  {journal} {\bibinfo  {journal} {Annual Review of Condensed Matter Physics}\ }\textbf {\bibinfo {volume} {10}},\ \bibinfo {pages} {85} (\bibinfo {year} {2019})}\BibitemShut {NoStop}%
\bibitem [{\citenamefont {Sch{\"a}fer}\ \emph {et~al.}(2020)\citenamefont {Sch{\"a}fer}, \citenamefont {Fukuhara}, \citenamefont {Sugawa}, \citenamefont {Takasu},\ and\ \citenamefont {Takahashi}}]{Schafer2020tools}%
  \BibitemOpen
  \bibfield  {author} {\bibinfo {author} {\bibfnamefont {F.}~\bibnamefont {Sch{\"a}fer}}, \bibinfo {author} {\bibfnamefont {T.}~\bibnamefont {Fukuhara}}, \bibinfo {author} {\bibfnamefont {S.}~\bibnamefont {Sugawa}}, \bibinfo {author} {\bibfnamefont {Y.}~\bibnamefont {Takasu}},\ and\ \bibinfo {author} {\bibfnamefont {Y.}~\bibnamefont {Takahashi}},\ }\bibfield  {title} {\bibinfo {title} {{Tools for quantum simulation with ultracold atoms in optical lattices}},\ }\href {https://doi.org/10.1038/s42254-020-0195-3} {\bibfield  {journal} {\bibinfo  {journal} {Nat. Rev. Phys.}\ }\textbf {\bibinfo {volume} {2}},\ \bibinfo {pages} {411} (\bibinfo {year} {2020})}\BibitemShut {NoStop}%
\bibitem [{\citenamefont {Nichols}\ \emph {et~al.}(2019)\citenamefont {Nichols}, \citenamefont {Cheuk}, \citenamefont {Okan}, \citenamefont {Hartke}, \citenamefont {Mendez}, \citenamefont {Senthil}, \citenamefont {Khatami}, \citenamefont {Zhang},\ and\ \citenamefont {Zwierlein}}]{Nichols2019}%
  \BibitemOpen
  \bibfield  {author} {\bibinfo {author} {\bibfnamefont {M.~A.}\ \bibnamefont {Nichols}}, \bibinfo {author} {\bibfnamefont {L.~W.}\ \bibnamefont {Cheuk}}, \bibinfo {author} {\bibfnamefont {M.}~\bibnamefont {Okan}}, \bibinfo {author} {\bibfnamefont {T.~R.}\ \bibnamefont {Hartke}}, \bibinfo {author} {\bibfnamefont {E.}~\bibnamefont {Mendez}}, \bibinfo {author} {\bibfnamefont {T.}~\bibnamefont {Senthil}}, \bibinfo {author} {\bibfnamefont {E.}~\bibnamefont {Khatami}}, \bibinfo {author} {\bibfnamefont {H.}~\bibnamefont {Zhang}},\ and\ \bibinfo {author} {\bibfnamefont {M.~W.}\ \bibnamefont {Zwierlein}},\ }\bibfield  {title} {\bibinfo {title} {{Spin transport in a Mott insulator of ultracold fermions}},\ }\href {https://doi.org/10.1126/science.aat4387} {\bibfield  {journal} {\bibinfo  {journal} {Science}\ }\textbf {\bibinfo {volume} {363}},\ \bibinfo {pages} {383} (\bibinfo {year} {2019})}\BibitemShut {NoStop}%
\bibitem [{\citenamefont {Fukuhara}\ \emph {et~al.}(2013{\natexlab{a}})\citenamefont {Fukuhara}, \citenamefont {Kantian}, \citenamefont {Endres}, \citenamefont {Cheneau}, \citenamefont {Schau{\ss}}, \citenamefont {Hild}, \citenamefont {Bellem}, \citenamefont {Schollw{\"o}ck}, \citenamefont {Giamarchi}, \citenamefont {Gross}, \citenamefont {Bloch},\ and\ \citenamefont {Kuhr}}]{Fukuhara2013-jv}%
  \BibitemOpen
  \bibfield  {author} {\bibinfo {author} {\bibfnamefont {T.}~\bibnamefont {Fukuhara}}, \bibinfo {author} {\bibfnamefont {A.}~\bibnamefont {Kantian}}, \bibinfo {author} {\bibfnamefont {M.}~\bibnamefont {Endres}}, \bibinfo {author} {\bibfnamefont {M.}~\bibnamefont {Cheneau}}, \bibinfo {author} {\bibfnamefont {P.}~\bibnamefont {Schau{\ss}}}, \bibinfo {author} {\bibfnamefont {S.}~\bibnamefont {Hild}}, \bibinfo {author} {\bibfnamefont {D.}~\bibnamefont {Bellem}}, \bibinfo {author} {\bibfnamefont {U.}~\bibnamefont {Schollw{\"o}ck}}, \bibinfo {author} {\bibfnamefont {T.}~\bibnamefont {Giamarchi}}, \bibinfo {author} {\bibfnamefont {C.}~\bibnamefont {Gross}}, \bibinfo {author} {\bibfnamefont {I.}~\bibnamefont {Bloch}},\ and\ \bibinfo {author} {\bibfnamefont {S.}~\bibnamefont {Kuhr}},\ }\bibfield  {title} {\bibinfo {title} {Quantum dynamics of a mobile spin impurity},\ }\href@noop {} {\bibfield  {journal} {\bibinfo  {journal} {Nat. Phys.}\ }\textbf {\bibinfo {volume} {9}},\ \bibinfo {pages} {235} (\bibinfo {year}
  {2013}{\natexlab{a}})}\BibitemShut {NoStop}%
\bibitem [{\citenamefont {Fukuhara}\ \emph {et~al.}(2013{\natexlab{b}})\citenamefont {Fukuhara}, \citenamefont {Schau{\ss}}, \citenamefont {Endres}, \citenamefont {Hild}, \citenamefont {Cheneau}, \citenamefont {Bloch},\ and\ \citenamefont {Gross}}]{Fukuhara2013-tj}%
  \BibitemOpen
  \bibfield  {author} {\bibinfo {author} {\bibfnamefont {T.}~\bibnamefont {Fukuhara}}, \bibinfo {author} {\bibfnamefont {P.}~\bibnamefont {Schau{\ss}}}, \bibinfo {author} {\bibfnamefont {M.}~\bibnamefont {Endres}}, \bibinfo {author} {\bibfnamefont {S.}~\bibnamefont {Hild}}, \bibinfo {author} {\bibfnamefont {M.}~\bibnamefont {Cheneau}}, \bibinfo {author} {\bibfnamefont {I.}~\bibnamefont {Bloch}},\ and\ \bibinfo {author} {\bibfnamefont {C.}~\bibnamefont {Gross}},\ }\bibfield  {title} {\bibinfo {title} {Microscopic observation of magnon bound states and their dynamics},\ }\href@noop {} {\bibfield  {journal} {\bibinfo  {journal} {Nature}\ }\textbf {\bibinfo {volume} {502}},\ \bibinfo {pages} {76} (\bibinfo {year} {2013}{\natexlab{b}})}\BibitemShut {NoStop}%
\bibitem [{\citenamefont {Hild}\ \emph {et~al.}(2014)\citenamefont {Hild}, \citenamefont {Fukuhara}, \citenamefont {Schau{\ss}}, \citenamefont {Zeiher}, \citenamefont {Knap}, \citenamefont {Demler}, \citenamefont {Bloch},\ and\ \citenamefont {Gross}}]{Hild2014-hy}%
  \BibitemOpen
  \bibfield  {author} {\bibinfo {author} {\bibfnamefont {S.}~\bibnamefont {Hild}}, \bibinfo {author} {\bibfnamefont {T.}~\bibnamefont {Fukuhara}}, \bibinfo {author} {\bibfnamefont {P.}~\bibnamefont {Schau{\ss}}}, \bibinfo {author} {\bibfnamefont {J.}~\bibnamefont {Zeiher}}, \bibinfo {author} {\bibfnamefont {M.}~\bibnamefont {Knap}}, \bibinfo {author} {\bibfnamefont {E.}~\bibnamefont {Demler}}, \bibinfo {author} {\bibfnamefont {I.}~\bibnamefont {Bloch}},\ and\ \bibinfo {author} {\bibfnamefont {C.}~\bibnamefont {Gross}},\ }\bibfield  {title} {\bibinfo {title} {Far-from-equilibrium spin transport in heisenberg quantum magnets},\ }\href@noop {} {\bibfield  {journal} {\bibinfo  {journal} {Phys. Rev. Lett.}\ }\textbf {\bibinfo {volume} {113}},\ \bibinfo {pages} {147205} (\bibinfo {year} {2014})}\BibitemShut {NoStop}%
\bibitem [{\citenamefont {Jepsen}\ \emph {et~al.}(2020)\citenamefont {Jepsen}, \citenamefont {Amato-Grill}, \citenamefont {Dimitrova}, \citenamefont {Ho}, \citenamefont {Demler},\ and\ \citenamefont {Ketterle}}]{Jepsen2020-fg}%
  \BibitemOpen
  \bibfield  {author} {\bibinfo {author} {\bibfnamefont {P.~N.}\ \bibnamefont {Jepsen}}, \bibinfo {author} {\bibfnamefont {J.}~\bibnamefont {Amato-Grill}}, \bibinfo {author} {\bibfnamefont {I.}~\bibnamefont {Dimitrova}}, \bibinfo {author} {\bibfnamefont {W.~W.}\ \bibnamefont {Ho}}, \bibinfo {author} {\bibfnamefont {E.}~\bibnamefont {Demler}},\ and\ \bibinfo {author} {\bibfnamefont {W.}~\bibnamefont {Ketterle}},\ }\bibfield  {title} {\bibinfo {title} {Spin transport in a tunable heisenberg model realized with ultracold atoms},\ }\href@noop {} {\bibfield  {journal} {\bibinfo  {journal} {Nature}\ }\textbf {\bibinfo {volume} {588}},\ \bibinfo {pages} {403} (\bibinfo {year} {2020})}\BibitemShut {NoStop}%
\bibitem [{\citenamefont {Jepsen}\ \emph {et~al.}(2021)\citenamefont {Jepsen}, \citenamefont {Ho}, \citenamefont {Amato-Grill}, \citenamefont {Dimitrova}, \citenamefont {Demler},\ and\ \citenamefont {Ketterle}}]{Jepsen2021-ew}%
  \BibitemOpen
  \bibfield  {author} {\bibinfo {author} {\bibfnamefont {P.~N.}\ \bibnamefont {Jepsen}}, \bibinfo {author} {\bibfnamefont {W.~W.}\ \bibnamefont {Ho}}, \bibinfo {author} {\bibfnamefont {J.}~\bibnamefont {Amato-Grill}}, \bibinfo {author} {\bibfnamefont {I.}~\bibnamefont {Dimitrova}}, \bibinfo {author} {\bibfnamefont {E.}~\bibnamefont {Demler}},\ and\ \bibinfo {author} {\bibfnamefont {W.}~\bibnamefont {Ketterle}},\ }\bibfield  {title} {\bibinfo {title} {Transverse spin dynamics in the anisotropic heisenberg model realized with ultracold atoms},\ }\href@noop {} {\bibfield  {journal} {\bibinfo  {journal} {Phys. Rev. X}\ }\textbf {\bibinfo {volume} {11}},\ \bibinfo {pages} {041054} (\bibinfo {year} {2021})}\BibitemShut {NoStop}%
\bibitem [{\citenamefont {Jepsen}\ \emph {et~al.}(2022)\citenamefont {Jepsen}, \citenamefont {Lee}, \citenamefont {Lin}, \citenamefont {Dimitrova}, \citenamefont {Margalit}, \citenamefont {Ho},\ and\ \citenamefont {Ketterle}}]{Jepsen2022-sf}%
  \BibitemOpen
  \bibfield  {author} {\bibinfo {author} {\bibfnamefont {P.~N.}\ \bibnamefont {Jepsen}}, \bibinfo {author} {\bibfnamefont {Y.~K.~e.}\ \bibnamefont {Lee}}, \bibinfo {author} {\bibfnamefont {H.}~\bibnamefont {Lin}}, \bibinfo {author} {\bibfnamefont {I.}~\bibnamefont {Dimitrova}}, \bibinfo {author} {\bibfnamefont {Y.}~\bibnamefont {Margalit}}, \bibinfo {author} {\bibfnamefont {W.~W.}\ \bibnamefont {Ho}},\ and\ \bibinfo {author} {\bibfnamefont {W.}~\bibnamefont {Ketterle}},\ }\bibfield  {title} {\bibinfo {title} {Long-lived phantom helix states in heisenberg quantum magnets},\ }\href@noop {} {\bibfield  {journal} {\bibinfo  {journal} {Nat. Phys.}\ }\textbf {\bibinfo {volume} {18}},\ \bibinfo {pages} {899} (\bibinfo {year} {2022})}\BibitemShut {NoStop}%
\bibitem [{\citenamefont {Wei}\ \emph {et~al.}(2022)\citenamefont {Wei}, \citenamefont {Rubio-Abadal}, \citenamefont {Ye}, \citenamefont {Machado}, \citenamefont {Kemp}, \citenamefont {Srakaew}, \citenamefont {Hollerith}, \citenamefont {Rui}, \citenamefont {Gopalakrishnan}, \citenamefont {Yao}, \citenamefont {Bloch},\ and\ \citenamefont {Zeiher}}]{Wei2022-qx}%
  \BibitemOpen
  \bibfield  {author} {\bibinfo {author} {\bibfnamefont {D.}~\bibnamefont {Wei}}, \bibinfo {author} {\bibfnamefont {A.}~\bibnamefont {Rubio-Abadal}}, \bibinfo {author} {\bibfnamefont {B.}~\bibnamefont {Ye}}, \bibinfo {author} {\bibfnamefont {F.}~\bibnamefont {Machado}}, \bibinfo {author} {\bibfnamefont {J.}~\bibnamefont {Kemp}}, \bibinfo {author} {\bibfnamefont {K.}~\bibnamefont {Srakaew}}, \bibinfo {author} {\bibfnamefont {S.}~\bibnamefont {Hollerith}}, \bibinfo {author} {\bibfnamefont {J.}~\bibnamefont {Rui}}, \bibinfo {author} {\bibfnamefont {S.}~\bibnamefont {Gopalakrishnan}}, \bibinfo {author} {\bibfnamefont {N.~Y.}\ \bibnamefont {Yao}}, \bibinfo {author} {\bibfnamefont {I.}~\bibnamefont {Bloch}},\ and\ \bibinfo {author} {\bibfnamefont {J.}~\bibnamefont {Zeiher}},\ }\bibfield  {title} {\bibinfo {title} {Quantum gas microscopy of {Kardar-Parisi-Zhang} superdiffusion},\ }\href@noop {} {\bibfield  {journal} {\bibinfo  {journal} {Science}\ }\textbf {\bibinfo {volume} {376}},\ \bibinfo {pages} {716} (\bibinfo
  {year} {2022})}\BibitemShut {NoStop}%
\bibitem [{\citenamefont {Sommer}\ \emph {et~al.}(2011{\natexlab{a}})\citenamefont {Sommer}, \citenamefont {Ku}, \citenamefont {Roati},\ and\ \citenamefont {Zwierlein}}]{sommer2011universal}%
  \BibitemOpen
  \bibfield  {author} {\bibinfo {author} {\bibfnamefont {A.}~\bibnamefont {Sommer}}, \bibinfo {author} {\bibfnamefont {M.}~\bibnamefont {Ku}}, \bibinfo {author} {\bibfnamefont {G.}~\bibnamefont {Roati}},\ and\ \bibinfo {author} {\bibfnamefont {M.~W.}\ \bibnamefont {Zwierlein}},\ }\bibfield  {title} {\bibinfo {title} {{Universal spin transport in a strongly interacting Fermi gas}},\ }\href@noop {} {\bibfield  {journal} {\bibinfo  {journal} {Nature}\ }\textbf {\bibinfo {volume} {472}},\ \bibinfo {pages} {201} (\bibinfo {year} {2011}{\natexlab{a}})}\BibitemShut {NoStop}%
\bibitem [{\citenamefont {Sommer}\ \emph {et~al.}(2011{\natexlab{b}})\citenamefont {Sommer}, \citenamefont {Ku},\ and\ \citenamefont {Zwierlein}}]{sommer2011spin}%
  \BibitemOpen
  \bibfield  {author} {\bibinfo {author} {\bibfnamefont {A.}~\bibnamefont {Sommer}}, \bibinfo {author} {\bibfnamefont {M.}~\bibnamefont {Ku}},\ and\ \bibinfo {author} {\bibfnamefont {M.~W.}\ \bibnamefont {Zwierlein}},\ }\bibfield  {title} {\bibinfo {title} {{Spin transport in polaronic and superfluid Fermi gases}},\ }\href@noop {} {\bibfield  {journal} {\bibinfo  {journal} {New Journal of Physics}\ }\textbf {\bibinfo {volume} {13}},\ \bibinfo {pages} {055009} (\bibinfo {year} {2011}{\natexlab{b}})}\BibitemShut {NoStop}%
\bibitem [{\citenamefont {Uchida}\ \emph {et~al.}(2008)\citenamefont {Uchida}, \citenamefont {Takahashi}, \citenamefont {Harii}, \citenamefont {Ieda}, \citenamefont {Koshibae}, \citenamefont {Ando}, \citenamefont {Maekawa},\ and\ \citenamefont {Saitoh}}]{uchida2008observation}%
  \BibitemOpen
  \bibfield  {author} {\bibinfo {author} {\bibfnamefont {K.-I.}\ \bibnamefont {Uchida}}, \bibinfo {author} {\bibfnamefont {S.}~\bibnamefont {Takahashi}}, \bibinfo {author} {\bibfnamefont {K.}~\bibnamefont {Harii}}, \bibinfo {author} {\bibfnamefont {J.}~\bibnamefont {Ieda}}, \bibinfo {author} {\bibfnamefont {W.}~\bibnamefont {Koshibae}}, \bibinfo {author} {\bibfnamefont {K.}~\bibnamefont {Ando}}, \bibinfo {author} {\bibfnamefont {S.}~\bibnamefont {Maekawa}},\ and\ \bibinfo {author} {\bibfnamefont {E.}~\bibnamefont {Saitoh}},\ }\bibfield  {title} {\bibinfo {title} {{Observation of the spin Seebeck effect}},\ }\href@noop {} {\bibfield  {journal} {\bibinfo  {journal} {Nature}\ }\textbf {\bibinfo {volume} {455}},\ \bibinfo {pages} {778} (\bibinfo {year} {2008})}\BibitemShut {NoStop}%
\bibitem [{\citenamefont {Jaworski}\ \emph {et~al.}(2010)\citenamefont {Jaworski}, \citenamefont {Yang}, \citenamefont {Mack}, \citenamefont {Awschalom}, \citenamefont {Heremans},\ and\ \citenamefont {Myers}}]{jaworski2010observation}%
  \BibitemOpen
  \bibfield  {author} {\bibinfo {author} {\bibfnamefont {C.}~\bibnamefont {Jaworski}}, \bibinfo {author} {\bibfnamefont {J.}~\bibnamefont {Yang}}, \bibinfo {author} {\bibfnamefont {S.}~\bibnamefont {Mack}}, \bibinfo {author} {\bibfnamefont {D.}~\bibnamefont {Awschalom}}, \bibinfo {author} {\bibfnamefont {J.}~\bibnamefont {Heremans}},\ and\ \bibinfo {author} {\bibfnamefont {R.}~\bibnamefont {Myers}},\ }\bibfield  {title} {\bibinfo {title} {{Observation of the spin-Seebeck effect in a ferromagnetic semiconductor}},\ }\href@noop {} {\bibfield  {journal} {\bibinfo  {journal} {Nature materials}\ }\textbf {\bibinfo {volume} {9}},\ \bibinfo {pages} {898} (\bibinfo {year} {2010})}\BibitemShut {NoStop}%
\bibitem [{\citenamefont {Uchida}\ \emph {et~al.}(2010)\citenamefont {Uchida}, \citenamefont {Xiao}, \citenamefont {Adachi}, \citenamefont {Ohe}, \citenamefont {Takahashi}, \citenamefont {Ieda}, \citenamefont {Ota}, \citenamefont {Kajiwara}, \citenamefont {Umezawa}, \citenamefont {Kawai} \emph {et~al.}}]{uchida2010spin}%
  \BibitemOpen
  \bibfield  {author} {\bibinfo {author} {\bibfnamefont {K.-i.}\ \bibnamefont {Uchida}}, \bibinfo {author} {\bibfnamefont {J.}~\bibnamefont {Xiao}}, \bibinfo {author} {\bibfnamefont {H.}~\bibnamefont {Adachi}}, \bibinfo {author} {\bibfnamefont {J.-i.}\ \bibnamefont {Ohe}}, \bibinfo {author} {\bibfnamefont {S.}~\bibnamefont {Takahashi}}, \bibinfo {author} {\bibfnamefont {J.}~\bibnamefont {Ieda}}, \bibinfo {author} {\bibfnamefont {T.}~\bibnamefont {Ota}}, \bibinfo {author} {\bibfnamefont {Y.}~\bibnamefont {Kajiwara}}, \bibinfo {author} {\bibfnamefont {H.}~\bibnamefont {Umezawa}}, \bibinfo {author} {\bibfnamefont {H.}~\bibnamefont {Kawai}}, \emph {et~al.},\ }\bibfield  {title} {\bibinfo {title} {Spin seebeck insulator},\ }\href@noop {} {\bibfield  {journal} {\bibinfo  {journal} {Nature materials}\ }\textbf {\bibinfo {volume} {9}},\ \bibinfo {pages} {894} (\bibinfo {year} {2010})}\BibitemShut {NoStop}%
\bibitem [{\citenamefont {Xiao}\ \emph {et~al.}(2010)\citenamefont {Xiao}, \citenamefont {Bauer}, \citenamefont {Uchida}, \citenamefont {Saitoh}, \citenamefont {Maekawa} \emph {et~al.}}]{xiao2010theory}%
  \BibitemOpen
  \bibfield  {author} {\bibinfo {author} {\bibfnamefont {J.}~\bibnamefont {Xiao}}, \bibinfo {author} {\bibfnamefont {G.~E.}\ \bibnamefont {Bauer}}, \bibinfo {author} {\bibfnamefont {K.-c.}\ \bibnamefont {Uchida}}, \bibinfo {author} {\bibfnamefont {E.}~\bibnamefont {Saitoh}}, \bibinfo {author} {\bibfnamefont {S.}~\bibnamefont {Maekawa}}, \emph {et~al.},\ }\bibfield  {title} {\bibinfo {title} {Theory of magnon-driven spin seebeck effect},\ }\href@noop {} {\bibfield  {journal} {\bibinfo  {journal} {Physical Review B}\ }\textbf {\bibinfo {volume} {81}},\ \bibinfo {pages} {214418} (\bibinfo {year} {2010})}\BibitemShut {NoStop}%
\bibitem [{\citenamefont {Adachi}\ \emph {et~al.}(2011)\citenamefont {Adachi}, \citenamefont {Ohe}, \citenamefont {Takahashi},\ and\ \citenamefont {Maekawa}}]{adachi2011linear}%
  \BibitemOpen
  \bibfield  {author} {\bibinfo {author} {\bibfnamefont {H.}~\bibnamefont {Adachi}}, \bibinfo {author} {\bibfnamefont {J.-i.}\ \bibnamefont {Ohe}}, \bibinfo {author} {\bibfnamefont {S.}~\bibnamefont {Takahashi}},\ and\ \bibinfo {author} {\bibfnamefont {S.}~\bibnamefont {Maekawa}},\ }\bibfield  {title} {\bibinfo {title} {Linear-response theory of spin seebeck effect in ferromagnetic insulators},\ }\href@noop {} {\bibfield  {journal} {\bibinfo  {journal} {Physical Review B}\ }\textbf {\bibinfo {volume} {83}},\ \bibinfo {pages} {094410} (\bibinfo {year} {2011})}\BibitemShut {NoStop}%
\bibitem [{\citenamefont {Adachi}\ \emph {et~al.}(2013)\citenamefont {Adachi}, \citenamefont {Uchida}, \citenamefont {Saitoh},\ and\ \citenamefont {Maekawa}}]{adachi2013theory}%
  \BibitemOpen
  \bibfield  {author} {\bibinfo {author} {\bibfnamefont {H.}~\bibnamefont {Adachi}}, \bibinfo {author} {\bibfnamefont {K.-i.}\ \bibnamefont {Uchida}}, \bibinfo {author} {\bibfnamefont {E.}~\bibnamefont {Saitoh}},\ and\ \bibinfo {author} {\bibfnamefont {S.}~\bibnamefont {Maekawa}},\ }\bibfield  {title} {\bibinfo {title} {Theory of the spin seebeck effect},\ }\href@noop {} {\bibfield  {journal} {\bibinfo  {journal} {Reports on Progress in Physics}\ }\textbf {\bibinfo {volume} {76}},\ \bibinfo {pages} {036501} (\bibinfo {year} {2013})}\BibitemShut {NoStop}%
\bibitem [{\citenamefont {Ohnuma}\ \emph {et~al.}(2017)\citenamefont {Ohnuma}, \citenamefont {Matsuo},\ and\ \citenamefont {Maekawa}}]{ohnuma2017theory}%
  \BibitemOpen
  \bibfield  {author} {\bibinfo {author} {\bibfnamefont {Y.}~\bibnamefont {Ohnuma}}, \bibinfo {author} {\bibfnamefont {M.}~\bibnamefont {Matsuo}},\ and\ \bibinfo {author} {\bibfnamefont {S.}~\bibnamefont {Maekawa}},\ }\bibfield  {title} {\bibinfo {title} {Theory of the spin peltier effect},\ }\href@noop {} {\bibfield  {journal} {\bibinfo  {journal} {Physical Review B}\ }\textbf {\bibinfo {volume} {96}},\ \bibinfo {pages} {134412} (\bibinfo {year} {2017})}\BibitemShut {NoStop}%
\bibitem [{\citenamefont {Matsuo}\ \emph {et~al.}(2018)\citenamefont {Matsuo}, \citenamefont {Ohnuma}, \citenamefont {Kato},\ and\ \citenamefont {Maekawa}}]{matsuo2018spin}%
  \BibitemOpen
  \bibfield  {author} {\bibinfo {author} {\bibfnamefont {M.}~\bibnamefont {Matsuo}}, \bibinfo {author} {\bibfnamefont {Y.}~\bibnamefont {Ohnuma}}, \bibinfo {author} {\bibfnamefont {T.}~\bibnamefont {Kato}},\ and\ \bibinfo {author} {\bibfnamefont {S.}~\bibnamefont {Maekawa}},\ }\bibfield  {title} {\bibinfo {title} {Spin current noise of the spin seebeck effect and spin pumping},\ }\href@noop {} {\bibfield  {journal} {\bibinfo  {journal} {Physical review letters}\ }\textbf {\bibinfo {volume} {120}},\ \bibinfo {pages} {037201} (\bibinfo {year} {2018})}\BibitemShut {NoStop}%
\bibitem [{\citenamefont {Kato}\ \emph {et~al.}(2019)\citenamefont {Kato}, \citenamefont {Ohnuma}, \citenamefont {Matsuo}, \citenamefont {Rech}, \citenamefont {Jonckheere},\ and\ \citenamefont {Martin}}]{kato2019microscopic}%
  \BibitemOpen
  \bibfield  {author} {\bibinfo {author} {\bibfnamefont {T.}~\bibnamefont {Kato}}, \bibinfo {author} {\bibfnamefont {Y.}~\bibnamefont {Ohnuma}}, \bibinfo {author} {\bibfnamefont {M.}~\bibnamefont {Matsuo}}, \bibinfo {author} {\bibfnamefont {J.}~\bibnamefont {Rech}}, \bibinfo {author} {\bibfnamefont {T.}~\bibnamefont {Jonckheere}},\ and\ \bibinfo {author} {\bibfnamefont {T.}~\bibnamefont {Martin}},\ }\bibfield  {title} {\bibinfo {title} {Microscopic theory of spin transport at the interface between a superconductor and a ferromagnetic insulator},\ }\href {https://doi.org/10.1103/PhysRevB.99.144411} {\bibfield  {journal} {\bibinfo  {journal} {Phys. Rev. B}\ }\textbf {\bibinfo {volume} {99}},\ \bibinfo {pages} {144411} (\bibinfo {year} {2019})}\BibitemShut {NoStop}%
\bibitem [{\citenamefont {Flipse}\ \emph {et~al.}(2014)\citenamefont {Flipse}, \citenamefont {Dejene}, \citenamefont {Wagenaar}, \citenamefont {Bauer}, \citenamefont {Ben~Youssef},\ and\ \citenamefont {van Wees}}]{Flipse2014-ne}%
  \BibitemOpen
  \bibfield  {author} {\bibinfo {author} {\bibfnamefont {J.}~\bibnamefont {Flipse}}, \bibinfo {author} {\bibfnamefont {F.~K.}\ \bibnamefont {Dejene}}, \bibinfo {author} {\bibfnamefont {D.}~\bibnamefont {Wagenaar}}, \bibinfo {author} {\bibfnamefont {G.~E.~W.}\ \bibnamefont {Bauer}}, \bibinfo {author} {\bibfnamefont {J.}~\bibnamefont {Ben~Youssef}},\ and\ \bibinfo {author} {\bibfnamefont {B.~J.}\ \bibnamefont {van Wees}},\ }\bibfield  {title} {\bibinfo {title} {Observation of the spin peltier effect for magnetic insulators},\ }\href@noop {} {\bibfield  {journal} {\bibinfo  {journal} {Phys. Rev. Lett.}\ }\textbf {\bibinfo {volume} {113}},\ \bibinfo {pages} {027601} (\bibinfo {year} {2014})}\BibitemShut {NoStop}%
\bibitem [{\citenamefont {Daimon}\ \emph {et~al.}(2016)\citenamefont {Daimon}, \citenamefont {Iguchi}, \citenamefont {Hioki}, \citenamefont {Saitoh},\ and\ \citenamefont {Uchida}}]{Daimon2016-vy}%
  \BibitemOpen
  \bibfield  {author} {\bibinfo {author} {\bibfnamefont {S.}~\bibnamefont {Daimon}}, \bibinfo {author} {\bibfnamefont {R.}~\bibnamefont {Iguchi}}, \bibinfo {author} {\bibfnamefont {T.}~\bibnamefont {Hioki}}, \bibinfo {author} {\bibfnamefont {E.}~\bibnamefont {Saitoh}},\ and\ \bibinfo {author} {\bibfnamefont {K.-I.}\ \bibnamefont {Uchida}},\ }\bibfield  {title} {\bibinfo {title} {Thermal imaging of spin peltier effect},\ }\href@noop {} {\bibfield  {journal} {\bibinfo  {journal} {Nat Commun}\ }\textbf {\bibinfo {volume} {7}},\ \bibinfo {pages} {1} (\bibinfo {year} {2016})}\BibitemShut {NoStop}%
\bibitem [{\citenamefont {Bauer}\ \emph {et~al.}(2012)\citenamefont {Bauer}, \citenamefont {Saitoh},\ and\ \citenamefont {Van~Wees}}]{bauer2012spin}%
  \BibitemOpen
  \bibfield  {author} {\bibinfo {author} {\bibfnamefont {G.~E.}\ \bibnamefont {Bauer}}, \bibinfo {author} {\bibfnamefont {E.}~\bibnamefont {Saitoh}},\ and\ \bibinfo {author} {\bibfnamefont {B.~J.}\ \bibnamefont {Van~Wees}},\ }\bibfield  {title} {\bibinfo {title} {Spin caloritronics},\ }\href@noop {} {\bibfield  {journal} {\bibinfo  {journal} {Nature materials}\ }\textbf {\bibinfo {volume} {11}},\ \bibinfo {pages} {391} (\bibinfo {year} {2012})}\BibitemShut {NoStop}%
\bibitem [{\citenamefont {Sekino}\ \emph {et~al.}(2024)\citenamefont {Sekino}, \citenamefont {Ominato}, \citenamefont {Tajima}, \citenamefont {Uchino},\ and\ \citenamefont {Matsuo}}]{Sekino2024-vp}%
  \BibitemOpen
  \bibfield  {author} {\bibinfo {author} {\bibfnamefont {Y.}~\bibnamefont {Sekino}}, \bibinfo {author} {\bibfnamefont {Y.}~\bibnamefont {Ominato}}, \bibinfo {author} {\bibfnamefont {H.}~\bibnamefont {Tajima}}, \bibinfo {author} {\bibfnamefont {S.}~\bibnamefont {Uchino}},\ and\ \bibinfo {author} {\bibfnamefont {M.}~\bibnamefont {Matsuo}},\ }\bibfield  {title} {\bibinfo {title} {Thermomagnetic anomalies by magnonic criticality in ultracold atomic transport},\ }\href@noop {} {\bibfield  {journal} {\bibinfo  {journal} {Phys. Rev. Lett.}\ }\textbf {\bibinfo {volume} {133}},\ \bibinfo {pages} {163402} (\bibinfo {year} {2024})}\BibitemShut {NoStop}%
\bibitem [{\citenamefont {Amico}\ \emph {et~al.}(2021)\citenamefont {Amico}, \citenamefont {Boshier}, \citenamefont {Birkl}, \citenamefont {Minguzzi}, \citenamefont {Miniatura}, \citenamefont {Kwek}, \citenamefont {Aghamalyan}, \citenamefont {Ahufinger}, \citenamefont {Anderson}, \citenamefont {Andrei}, \citenamefont {Arnold}, \citenamefont {Baker}, \citenamefont {Bell}, \citenamefont {Bland}, \citenamefont {Brantut}, \citenamefont {Cassettari}, \citenamefont {Chetcuti}, \citenamefont {Chevy}, \citenamefont {Citro}, \citenamefont {De~Palo}, \citenamefont {Dumke}, \citenamefont {Edwards}, \citenamefont {Folman}, \citenamefont {Fortagh}, \citenamefont {Gardiner}, \citenamefont {Garraway}, \citenamefont {Gauthier}, \citenamefont {G{\"u}nther}, \citenamefont {Haug}, \citenamefont {Hufnagel}, \citenamefont {Keil}, \citenamefont {Ireland}, \citenamefont {Lebrat}, \citenamefont {Li}, \citenamefont {Longchambon}, \citenamefont {Mompart}, \citenamefont {Morsch}, \citenamefont {Naldesi}, \citenamefont {Neely},
  \citenamefont {Olshanii}, \citenamefont {Orignac}, \citenamefont {Pandey}, \citenamefont {P{\'e}rez-Obiol}, \citenamefont {Perrin}, \citenamefont {Piroli}, \citenamefont {Polo}, \citenamefont {Pritchard}, \citenamefont {Proukakis}, \citenamefont {Rylands}, \citenamefont {Rubinsztein-Dunlop}, \citenamefont {Scazza}, \citenamefont {Stringari}, \citenamefont {Tosto}, \citenamefont {Trombettoni}, \citenamefont {Victorin}, \citenamefont {Klitzing}, \citenamefont {Wilkowski}, \citenamefont {Xhani},\ and\ \citenamefont {Yakimenko}}]{doi:10.1116/5.0026178}%
  \BibitemOpen
  \bibfield  {author} {\bibinfo {author} {\bibfnamefont {L.}~\bibnamefont {Amico}}, \bibinfo {author} {\bibfnamefont {M.}~\bibnamefont {Boshier}}, \bibinfo {author} {\bibfnamefont {G.}~\bibnamefont {Birkl}}, \bibinfo {author} {\bibfnamefont {A.}~\bibnamefont {Minguzzi}}, \bibinfo {author} {\bibfnamefont {C.}~\bibnamefont {Miniatura}}, \bibinfo {author} {\bibfnamefont {L.-C.}\ \bibnamefont {Kwek}}, \bibinfo {author} {\bibfnamefont {D.}~\bibnamefont {Aghamalyan}}, \bibinfo {author} {\bibfnamefont {V.}~\bibnamefont {Ahufinger}}, \bibinfo {author} {\bibfnamefont {D.}~\bibnamefont {Anderson}}, \bibinfo {author} {\bibfnamefont {N.}~\bibnamefont {Andrei}}, \bibinfo {author} {\bibfnamefont {A.~S.}\ \bibnamefont {Arnold}}, \bibinfo {author} {\bibfnamefont {M.}~\bibnamefont {Baker}}, \bibinfo {author} {\bibfnamefont {T.~A.}\ \bibnamefont {Bell}}, \bibinfo {author} {\bibfnamefont {T.}~\bibnamefont {Bland}}, \bibinfo {author} {\bibfnamefont {J.~P.}\ \bibnamefont {Brantut}}, \bibinfo {author} {\bibfnamefont
  {D.}~\bibnamefont {Cassettari}}, \bibinfo {author} {\bibfnamefont {W.~J.}\ \bibnamefont {Chetcuti}}, \bibinfo {author} {\bibfnamefont {F.}~\bibnamefont {Chevy}}, \bibinfo {author} {\bibfnamefont {R.}~\bibnamefont {Citro}}, \bibinfo {author} {\bibfnamefont {S.}~\bibnamefont {De~Palo}}, \bibinfo {author} {\bibfnamefont {R.}~\bibnamefont {Dumke}}, \bibinfo {author} {\bibfnamefont {M.}~\bibnamefont {Edwards}}, \bibinfo {author} {\bibfnamefont {R.}~\bibnamefont {Folman}}, \bibinfo {author} {\bibfnamefont {J.}~\bibnamefont {Fortagh}}, \bibinfo {author} {\bibfnamefont {S.~A.}\ \bibnamefont {Gardiner}}, \bibinfo {author} {\bibfnamefont {B.~M.}\ \bibnamefont {Garraway}}, \bibinfo {author} {\bibfnamefont {G.}~\bibnamefont {Gauthier}}, \bibinfo {author} {\bibfnamefont {A.}~\bibnamefont {G{\"u}nther}}, \bibinfo {author} {\bibfnamefont {T.}~\bibnamefont {Haug}}, \bibinfo {author} {\bibfnamefont {C.}~\bibnamefont {Hufnagel}}, \bibinfo {author} {\bibfnamefont {M.}~\bibnamefont {Keil}}, \bibinfo {author} {\bibfnamefont
  {P.}~\bibnamefont {Ireland}}, \bibinfo {author} {\bibfnamefont {M.}~\bibnamefont {Lebrat}}, \bibinfo {author} {\bibfnamefont {W.}~\bibnamefont {Li}}, \bibinfo {author} {\bibfnamefont {L.}~\bibnamefont {Longchambon}}, \bibinfo {author} {\bibfnamefont {J.}~\bibnamefont {Mompart}}, \bibinfo {author} {\bibfnamefont {O.}~\bibnamefont {Morsch}}, \bibinfo {author} {\bibfnamefont {P.}~\bibnamefont {Naldesi}}, \bibinfo {author} {\bibfnamefont {T.~W.}\ \bibnamefont {Neely}}, \bibinfo {author} {\bibfnamefont {M.}~\bibnamefont {Olshanii}}, \bibinfo {author} {\bibfnamefont {E.}~\bibnamefont {Orignac}}, \bibinfo {author} {\bibfnamefont {S.}~\bibnamefont {Pandey}}, \bibinfo {author} {\bibfnamefont {A.}~\bibnamefont {P{\'e}rez-Obiol}}, \bibinfo {author} {\bibfnamefont {H.}~\bibnamefont {Perrin}}, \bibinfo {author} {\bibfnamefont {L.}~\bibnamefont {Piroli}}, \bibinfo {author} {\bibfnamefont {J.}~\bibnamefont {Polo}}, \bibinfo {author} {\bibfnamefont {A.~L.}\ \bibnamefont {Pritchard}}, \bibinfo {author} {\bibfnamefont
  {N.~P.}\ \bibnamefont {Proukakis}}, \bibinfo {author} {\bibfnamefont {C.}~\bibnamefont {Rylands}}, \bibinfo {author} {\bibfnamefont {H.}~\bibnamefont {Rubinsztein-Dunlop}}, \bibinfo {author} {\bibfnamefont {F.}~\bibnamefont {Scazza}}, \bibinfo {author} {\bibfnamefont {S.}~\bibnamefont {Stringari}}, \bibinfo {author} {\bibfnamefont {F.}~\bibnamefont {Tosto}}, \bibinfo {author} {\bibfnamefont {A.}~\bibnamefont {Trombettoni}}, \bibinfo {author} {\bibfnamefont {N.}~\bibnamefont {Victorin}}, \bibinfo {author} {\bibfnamefont {W.~v.}\ \bibnamefont {Klitzing}}, \bibinfo {author} {\bibfnamefont {D.}~\bibnamefont {Wilkowski}}, \bibinfo {author} {\bibfnamefont {K.}~\bibnamefont {Xhani}},\ and\ \bibinfo {author} {\bibfnamefont {A.}~\bibnamefont {Yakimenko}},\ }\bibfield  {title} {\bibinfo {title} {Roadmap on atomtronics: State of the art and perspective},\ }\href {https://doi.org/10.1116/5.0026178} {\bibfield  {journal} {\bibinfo  {journal} {AVS Quantum Science}\ }\textbf {\bibinfo {volume} {3}},\ \bibinfo {pages}
  {039201} (\bibinfo {year} {2021})},\ \Eprint {https://arxiv.org/abs/https://doi.org/10.1116/5.0026178} {https://doi.org/10.1116/5.0026178} \BibitemShut {NoStop}%
\bibitem [{\citenamefont {Amico}\ \emph {et~al.}(2022)\citenamefont {Amico}, \citenamefont {Anderson}, \citenamefont {Boshier}, \citenamefont {Brantut}, \citenamefont {Kwek}, \citenamefont {Minguzzi},\ and\ \citenamefont {von Klitzing}}]{amico2022atomtronic}%
  \BibitemOpen
  \bibfield  {author} {\bibinfo {author} {\bibfnamefont {L.}~\bibnamefont {Amico}}, \bibinfo {author} {\bibfnamefont {D.}~\bibnamefont {Anderson}}, \bibinfo {author} {\bibfnamefont {M.}~\bibnamefont {Boshier}}, \bibinfo {author} {\bibfnamefont {J.-P.}\ \bibnamefont {Brantut}}, \bibinfo {author} {\bibfnamefont {L.-C.}\ \bibnamefont {Kwek}}, \bibinfo {author} {\bibfnamefont {A.}~\bibnamefont {Minguzzi}},\ and\ \bibinfo {author} {\bibfnamefont {W.}~\bibnamefont {von Klitzing}},\ }\bibfield  {title} {\bibinfo {title} {Colloquium: Atomtronic circuits: From many-body physics to quantum technologies},\ }\href {https://doi.org/10.1103/RevModPhys.94.041001} {\bibfield  {journal} {\bibinfo  {journal} {Rev. Mod. Phys.}\ }\textbf {\bibinfo {volume} {94}},\ \bibinfo {pages} {041001} (\bibinfo {year} {2022})}\BibitemShut {NoStop}%
\bibitem [{\citenamefont {van Houten}\ and\ \citenamefont {Beenakker}(1996)}]{van-Houten1996-sx}%
  \BibitemOpen
  \bibfield  {author} {\bibinfo {author} {\bibfnamefont {H.}~\bibnamefont {van Houten}}\ and\ \bibinfo {author} {\bibfnamefont {C.}~\bibnamefont {Beenakker}},\ }\bibfield  {title} {\bibinfo {title} {Quantum point contacts},\ }\href@noop {} {\bibfield  {journal} {\bibinfo  {journal} {Phys. Today}\ }\textbf {\bibinfo {volume} {49}},\ \bibinfo {pages} {22} (\bibinfo {year} {1996})}\BibitemShut {NoStop}%
\bibitem [{\citenamefont {Nazarov}\ and\ \citenamefont {Yaroslav}(2009)}]{Nazarov2009-ev}%
  \BibitemOpen
  \bibfield  {author} {\bibinfo {author} {\bibfnamefont {Y.~V.}\ \bibnamefont {Nazarov}}\ and\ \bibinfo {author} {\bibfnamefont {B.~M.}\ \bibnamefont {Yaroslav}},\ }\href@noop {} {\emph {\bibinfo {title} {Quantum transport: introduction to nanoscience}}}\ (\bibinfo  {publisher} {Cambridge Univ. Press},\ \bibinfo {address} {Cambridge},\ \bibinfo {year} {2009})\BibitemShut {NoStop}%
\bibitem [{\citenamefont {Nakata}\ \emph {et~al.}(2015{\natexlab{a}})\citenamefont {Nakata}, \citenamefont {Simon},\ and\ \citenamefont {Loss}}]{Nakata2015-ns}%
  \BibitemOpen
  \bibfield  {author} {\bibinfo {author} {\bibfnamefont {K.}~\bibnamefont {Nakata}}, \bibinfo {author} {\bibfnamefont {P.}~\bibnamefont {Simon}},\ and\ \bibinfo {author} {\bibfnamefont {D.}~\bibnamefont {Loss}},\ }\bibfield  {title} {\bibinfo {title} {{Wiedemann-Franz} law for magnon transport},\ }\href@noop {} {\bibfield  {journal} {\bibinfo  {journal} {Phys. Rev. B Condens. Matter}\ }\textbf {\bibinfo {volume} {92}},\ \bibinfo {pages} {134425} (\bibinfo {year} {2015}{\natexlab{a}})}\BibitemShut {NoStop}%
\bibitem [{\citenamefont {Nakata}\ \emph {et~al.}(2017{\natexlab{a}})\citenamefont {Nakata}, \citenamefont {Klinovaja},\ and\ \citenamefont {Loss}}]{Nakata2017-hu}%
  \BibitemOpen
  \bibfield  {author} {\bibinfo {author} {\bibfnamefont {K.}~\bibnamefont {Nakata}}, \bibinfo {author} {\bibfnamefont {J.}~\bibnamefont {Klinovaja}},\ and\ \bibinfo {author} {\bibfnamefont {D.}~\bibnamefont {Loss}},\ }\bibfield  {title} {\bibinfo {title} {Magnonic quantum hall effect and wiedemann-franz law},\ }\href@noop {} {\bibfield  {journal} {\bibinfo  {journal} {Phys. Rev. B}\ }\textbf {\bibinfo {volume} {95}},\ \bibinfo {pages} {125429} (\bibinfo {year} {2017}{\natexlab{a}})}\BibitemShut {NoStop}%
\bibitem [{\citenamefont {Nakata}\ \emph {et~al.}(2017{\natexlab{b}})\citenamefont {Nakata}, \citenamefont {Kim}, \citenamefont {Klinovaja},\ and\ \citenamefont {Loss}}]{Nakata2017-ot}%
  \BibitemOpen
  \bibfield  {author} {\bibinfo {author} {\bibfnamefont {K.}~\bibnamefont {Nakata}}, \bibinfo {author} {\bibfnamefont {S.~K.}\ \bibnamefont {Kim}}, \bibinfo {author} {\bibfnamefont {J.}~\bibnamefont {Klinovaja}},\ and\ \bibinfo {author} {\bibfnamefont {D.}~\bibnamefont {Loss}},\ }\bibfield  {title} {\bibinfo {title} {Magnonic topological insulators in antiferromagnets},\ }\href@noop {} {\bibfield  {journal} {\bibinfo  {journal} {Phys. Rev. B}\ }\textbf {\bibinfo {volume} {96}},\ \bibinfo {pages} {224414} (\bibinfo {year} {2017}{\natexlab{b}})}\BibitemShut {NoStop}%
\bibitem [{\citenamefont {Nakata}\ \emph {et~al.}(2017{\natexlab{c}})\citenamefont {Nakata}, \citenamefont {Simon},\ and\ \citenamefont {Loss}}]{Nakata2017-iy}%
  \BibitemOpen
  \bibfield  {author} {\bibinfo {author} {\bibfnamefont {K.}~\bibnamefont {Nakata}}, \bibinfo {author} {\bibfnamefont {P.}~\bibnamefont {Simon}},\ and\ \bibinfo {author} {\bibfnamefont {D.}~\bibnamefont {Loss}},\ }\bibfield  {title} {\bibinfo {title} {Spin currents and magnon dynamics in insulating magnets},\ }\href@noop {} {\bibfield  {journal} {\bibinfo  {journal} {J. Phys. D Appl. Phys.}\ }\textbf {\bibinfo {volume} {50}},\ \bibinfo {pages} {114004} (\bibinfo {year} {2017}{\natexlab{c}})}\BibitemShut {NoStop}%
\bibitem [{\citenamefont {Sano}\ and\ \citenamefont {Matsuo}(2023)}]{Sano2023-ff}%
  \BibitemOpen
  \bibfield  {author} {\bibinfo {author} {\bibfnamefont {R.}~\bibnamefont {Sano}}\ and\ \bibinfo {author} {\bibfnamefont {M.}~\bibnamefont {Matsuo}},\ }\bibfield  {title} {\bibinfo {title} {Breaking down the magnonic wiedemann-franz law in the hydrodynamic regime},\ }\href@noop {} {\bibfield  {journal} {\bibinfo  {journal} {Phys. Rev. Lett.}\ }\textbf {\bibinfo {volume} {130}},\ \bibinfo {pages} {166201} (\bibinfo {year} {2023})}\BibitemShut {NoStop}%
\bibitem [{\citenamefont {Mook}\ \emph {et~al.}(2018)\citenamefont {Mook}, \citenamefont {Göbel}, \citenamefont {Henk},\ and\ \citenamefont {Mertig}}]{Mook2018-oy}%
  \BibitemOpen
  \bibfield  {author} {\bibinfo {author} {\bibfnamefont {A.}~\bibnamefont {Mook}}, \bibinfo {author} {\bibfnamefont {B.}~\bibnamefont {Göbel}}, \bibinfo {author} {\bibfnamefont {J.}~\bibnamefont {Henk}},\ and\ \bibinfo {author} {\bibfnamefont {I.}~\bibnamefont {Mertig}},\ }\bibfield  {title} {\bibinfo {title} {Taking an electron-magnon duality shortcut from electron to magnon transport},\ }\href@noop {} {\bibfield  {journal} {\bibinfo  {journal} {Phys. Rev. B.}\ }\textbf {\bibinfo {volume} {97}},\ \bibinfo {pages} {140401} (\bibinfo {year} {2018})}\BibitemShut {NoStop}%
\bibitem [{\citenamefont {Nakata}\ \emph {et~al.}(2018)\citenamefont {Nakata}, \citenamefont {Ohnuma},\ and\ \citenamefont {Matsuo}}]{nakata2018magnonic}%
  \BibitemOpen
  \bibfield  {author} {\bibinfo {author} {\bibfnamefont {K.}~\bibnamefont {Nakata}}, \bibinfo {author} {\bibfnamefont {Y.}~\bibnamefont {Ohnuma}},\ and\ \bibinfo {author} {\bibfnamefont {M.}~\bibnamefont {Matsuo}},\ }\bibfield  {title} {\bibinfo {title} {{Magnonic noise and Wiedemann-Franz law}},\ }\href {https://doi.org/10.1103/PhysRevB.98.094430} {\bibfield  {journal} {\bibinfo  {journal} {Phys. Rev. B}\ }\textbf {\bibinfo {volume} {98}},\ \bibinfo {pages} {094430} (\bibinfo {year} {2018})}\BibitemShut {NoStop}%
\bibitem [{\citenamefont {Nakata}\ \emph {et~al.}(2022)\citenamefont {Nakata}, \citenamefont {Ohnuma},\ and\ \citenamefont {Kim}}]{Nakata2022-ep}%
  \BibitemOpen
  \bibfield  {author} {\bibinfo {author} {\bibfnamefont {K.}~\bibnamefont {Nakata}}, \bibinfo {author} {\bibfnamefont {Y.}~\bibnamefont {Ohnuma}},\ and\ \bibinfo {author} {\bibfnamefont {S.~K.}\ \bibnamefont {Kim}},\ }\bibfield  {title} {\bibinfo {title} {Violation of the magnonic wiedemann-franz law in the strong nonlinear regime},\ }\href@noop {} {\bibfield  {journal} {\bibinfo  {journal} {Phys. Rev. B.}\ }\textbf {\bibinfo {volume} {105}},\ \bibinfo {pages} {184409} (\bibinfo {year} {2022})}\BibitemShut {NoStop}%
\bibitem [{\citenamefont {Franz}\ and\ \citenamefont {Wiedemann}(1853)}]{Franz1853-do}%
  \BibitemOpen
  \bibfield  {author} {\bibinfo {author} {\bibfnamefont {R.}~\bibnamefont {Franz}}\ and\ \bibinfo {author} {\bibfnamefont {G.}~\bibnamefont {Wiedemann}},\ }\bibfield  {title} {\bibinfo {title} {Ueber die {Wärme‐Leitungsfähigkeit} der metalle},\ }\href@noop {} {\bibfield  {journal} {\bibinfo  {journal} {Ann. Phys.}\ }\textbf {\bibinfo {volume} {165}},\ \bibinfo {pages} {497} (\bibinfo {year} {1853})}\BibitemShut {NoStop}%
\bibitem [{\citenamefont {Kuklov}\ and\ \citenamefont {Svistunov}(2003)}]{PhysRevLett.90.100401}%
  \BibitemOpen
  \bibfield  {author} {\bibinfo {author} {\bibfnamefont {A.~B.}\ \bibnamefont {Kuklov}}\ and\ \bibinfo {author} {\bibfnamefont {B.~V.}\ \bibnamefont {Svistunov}},\ }\bibfield  {title} {\bibinfo {title} {{Counterflow Superfluidity of Two-Species Ultracold Atoms in a Commensurate Optical Lattice}},\ }\href {https://doi.org/10.1103/PhysRevLett.90.100401} {\bibfield  {journal} {\bibinfo  {journal} {Phys. Rev. Lett.}\ }\textbf {\bibinfo {volume} {90}},\ \bibinfo {pages} {100401} (\bibinfo {year} {2003})}\BibitemShut {NoStop}%
\bibitem [{\citenamefont {Duan}\ \emph {et~al.}(2003)\citenamefont {Duan}, \citenamefont {Demler},\ and\ \citenamefont {Lukin}}]{Duan2003Controlling}%
  \BibitemOpen
  \bibfield  {author} {\bibinfo {author} {\bibfnamefont {L.-M.}\ \bibnamefont {Duan}}, \bibinfo {author} {\bibfnamefont {E.}~\bibnamefont {Demler}},\ and\ \bibinfo {author} {\bibfnamefont {M.~D.}\ \bibnamefont {Lukin}},\ }\bibfield  {title} {\bibinfo {title} {{Controlling Spin Exchange Interactions of Ultracold Atoms in Optical Lattices}},\ }\href {https://doi.org/10.1103/PhysRevLett.91.090402} {\bibfield  {journal} {\bibinfo  {journal} {Phys. Rev. Lett.}\ }\textbf {\bibinfo {volume} {91}},\ \bibinfo {pages} {090402} (\bibinfo {year} {2003})}\BibitemShut {NoStop}%
\bibitem [{\citenamefont {Garc{\'\i}a-Ripoll}\ and\ \citenamefont {Cirac}(2003)}]{Garcia-Ripoll:2003aa}%
  \BibitemOpen
  \bibfield  {author} {\bibinfo {author} {\bibfnamefont {J.~J.}\ \bibnamefont {Garc{\'\i}a-Ripoll}}\ and\ \bibinfo {author} {\bibfnamefont {J.~I.}\ \bibnamefont {Cirac}},\ }\bibfield  {title} {\bibinfo {title} {Spin dynamics for bosons in an optical lattice},\ }\href {https://doi.org/10.1088/1367-2630/5/1/376} {\bibfield  {journal} {\bibinfo  {journal} {New Journal of Physics}\ }\textbf {\bibinfo {volume} {5}},\ \bibinfo {pages} {76} (\bibinfo {year} {2003})}\BibitemShut {NoStop}%
\bibitem [{\citenamefont {Altman}\ \emph {et~al.}(2003)\citenamefont {Altman}, \citenamefont {Hofstetter}, \citenamefont {Demler},\ and\ \citenamefont {Lukin}}]{Altman:2003aa}%
  \BibitemOpen
  \bibfield  {author} {\bibinfo {author} {\bibfnamefont {E.}~\bibnamefont {Altman}}, \bibinfo {author} {\bibfnamefont {W.}~\bibnamefont {Hofstetter}}, \bibinfo {author} {\bibfnamefont {E.}~\bibnamefont {Demler}},\ and\ \bibinfo {author} {\bibfnamefont {M.~D.}\ \bibnamefont {Lukin}},\ }\bibfield  {title} {\bibinfo {title} {Phase diagram of two-component bosons on an optical lattice},\ }\href {https://doi.org/10.1088/1367-2630/5/1/113} {\bibfield  {journal} {\bibinfo  {journal} {New Journal of Physics}\ }\textbf {\bibinfo {volume} {5}},\ \bibinfo {pages} {113} (\bibinfo {year} {2003})}\BibitemShut {NoStop}%
\bibitem [{\citenamefont {de~Hond}\ \emph {et~al.}(2022)\citenamefont {de~Hond}, \citenamefont {Xiang}, \citenamefont {Chung}, \citenamefont {Cruz-Col\'on}, \citenamefont {Chen}, \citenamefont {Burton}, \citenamefont {Kennedy},\ and\ \citenamefont {Ketterle}}]{De_Hond2022-xi}%
  \BibitemOpen
  \bibfield  {author} {\bibinfo {author} {\bibfnamefont {J.}~\bibnamefont {de~Hond}}, \bibinfo {author} {\bibfnamefont {J.}~\bibnamefont {Xiang}}, \bibinfo {author} {\bibfnamefont {W.~C.}\ \bibnamefont {Chung}}, \bibinfo {author} {\bibfnamefont {E.}~\bibnamefont {Cruz-Col\'on}}, \bibinfo {author} {\bibfnamefont {W.}~\bibnamefont {Chen}}, \bibinfo {author} {\bibfnamefont {W.~C.}\ \bibnamefont {Burton}}, \bibinfo {author} {\bibfnamefont {C.~J.}\ \bibnamefont {Kennedy}},\ and\ \bibinfo {author} {\bibfnamefont {W.}~\bibnamefont {Ketterle}},\ }\bibfield  {title} {\bibinfo {title} {Preparation of the spin-mott state: A spinful mott insulator of repulsively bound pairs},\ }\href {https://doi.org/10.1103/PhysRevLett.128.093401} {\bibfield  {journal} {\bibinfo  {journal} {Phys. Rev. Lett.}\ }\textbf {\bibinfo {volume} {128}},\ \bibinfo {pages} {093401} (\bibinfo {year} {2022})}\BibitemShut {NoStop}%
\bibitem [{\citenamefont {Mahan}(2000)}]{mahan2000many}%
  \BibitemOpen
  \bibfield  {author} {\bibinfo {author} {\bibfnamefont {G.~D.}\ \bibnamefont {Mahan}},\ }\href@noop {} {\emph {\bibinfo {title} {Many Particle Physics, Third Edition}}}\ (\bibinfo  {publisher} {Plenum},\ \bibinfo {address} {New York},\ \bibinfo {year} {2000})\BibitemShut {NoStop}%
\bibitem [{\citenamefont {Zhang}\ \emph {et~al.}(2024)\citenamefont {Zhang}, \citenamefont {Gao}, \citenamefont {Xu}, \citenamefont {Wang}, \citenamefont {Dong}, \citenamefont {Guo}, \citenamefont {Deng}, \citenamefont {Zhang}, \citenamefont {Chen}, \citenamefont {Xu}, \citenamefont {Wang}, \citenamefont {Wu}, \citenamefont {Zhang}, \citenamefont {Jin}, \citenamefont {Zhu}, \citenamefont {Zhang}, \citenamefont {Zou}, \citenamefont {Tan}, \citenamefont {Cui}, \citenamefont {Zhu}, \citenamefont {Shen}, \citenamefont {Li}, \citenamefont {Zhong}, \citenamefont {Bao}, \citenamefont {Zhao}, \citenamefont {Hao}, \citenamefont {Li}, \citenamefont {Wang}, \citenamefont {Song}, \citenamefont {Guo}, \citenamefont {Wang},\ and\ \citenamefont {Poletti}}]{Zhang2024-kz}%
  \BibitemOpen
  \bibfield  {author} {\bibinfo {author} {\bibfnamefont {P.}~\bibnamefont {Zhang}}, \bibinfo {author} {\bibfnamefont {Y.}~\bibnamefont {Gao}}, \bibinfo {author} {\bibfnamefont {X.}~\bibnamefont {Xu}}, \bibinfo {author} {\bibfnamefont {N.}~\bibnamefont {Wang}}, \bibinfo {author} {\bibfnamefont {H.}~\bibnamefont {Dong}}, \bibinfo {author} {\bibfnamefont {C.}~\bibnamefont {Guo}}, \bibinfo {author} {\bibfnamefont {J.}~\bibnamefont {Deng}}, \bibinfo {author} {\bibfnamefont {X.}~\bibnamefont {Zhang}}, \bibinfo {author} {\bibfnamefont {J.}~\bibnamefont {Chen}}, \bibinfo {author} {\bibfnamefont {S.}~\bibnamefont {Xu}}, \bibinfo {author} {\bibfnamefont {K.}~\bibnamefont {Wang}}, \bibinfo {author} {\bibfnamefont {Y.}~\bibnamefont {Wu}}, \bibinfo {author} {\bibfnamefont {C.}~\bibnamefont {Zhang}}, \bibinfo {author} {\bibfnamefont {F.}~\bibnamefont {Jin}}, \bibinfo {author} {\bibfnamefont {X.}~\bibnamefont {Zhu}}, \bibinfo {author} {\bibfnamefont {A.}~\bibnamefont {Zhang}}, \bibinfo {author} {\bibfnamefont
  {Y.}~\bibnamefont {Zou}}, \bibinfo {author} {\bibfnamefont {Z.}~\bibnamefont {Tan}}, \bibinfo {author} {\bibfnamefont {Z.}~\bibnamefont {Cui}}, \bibinfo {author} {\bibfnamefont {Z.}~\bibnamefont {Zhu}}, \bibinfo {author} {\bibfnamefont {F.}~\bibnamefont {Shen}}, \bibinfo {author} {\bibfnamefont {T.}~\bibnamefont {Li}}, \bibinfo {author} {\bibfnamefont {J.}~\bibnamefont {Zhong}}, \bibinfo {author} {\bibfnamefont {Z.}~\bibnamefont {Bao}}, \bibinfo {author} {\bibfnamefont {L.}~\bibnamefont {Zhao}}, \bibinfo {author} {\bibfnamefont {J.}~\bibnamefont {Hao}}, \bibinfo {author} {\bibfnamefont {H.}~\bibnamefont {Li}}, \bibinfo {author} {\bibfnamefont {Z.}~\bibnamefont {Wang}}, \bibinfo {author} {\bibfnamefont {C.}~\bibnamefont {Song}}, \bibinfo {author} {\bibfnamefont {Q.}~\bibnamefont {Guo}}, \bibinfo {author} {\bibfnamefont {H.}~\bibnamefont {Wang}},\ and\ \bibinfo {author} {\bibfnamefont {D.}~\bibnamefont {Poletti}},\ }\bibfield  {title} {\bibinfo {title} {Emergence of steady quantum transport in a
  superconducting processor},\ }\href@noop {} {\bibfield  {journal} {\bibinfo  {journal} {Nat. Commun.}\ }\textbf {\bibinfo {volume} {15}},\ \bibinfo {pages} {1} (\bibinfo {year} {2024})}\BibitemShut {NoStop}%
\bibitem [{\citenamefont {Papoular}\ \emph {et~al.}(2016)\citenamefont {Papoular}, \citenamefont {Pitaevskii},\ and\ \citenamefont {Stringari}}]{Papoular2016quantized}%
  \BibitemOpen
  \bibfield  {author} {\bibinfo {author} {\bibfnamefont {D.~J.}\ \bibnamefont {Papoular}}, \bibinfo {author} {\bibfnamefont {L.~P.}\ \bibnamefont {Pitaevskii}},\ and\ \bibinfo {author} {\bibfnamefont {S.}~\bibnamefont {Stringari}},\ }\bibfield  {title} {\bibinfo {title} {Quantized conductance through the quantum evaporation of bosonic atoms},\ }\href {https://doi.org/10.1103/PhysRevA.94.023622} {\bibfield  {journal} {\bibinfo  {journal} {Phys. Rev. A}\ }\textbf {\bibinfo {volume} {94}},\ \bibinfo {pages} {023622} (\bibinfo {year} {2016})}\BibitemShut {NoStop}%
\bibitem [{\citenamefont {Drude}(1900)}]{Drude1900-bi}%
  \BibitemOpen
  \bibfield  {author} {\bibinfo {author} {\bibfnamefont {P.}~\bibnamefont {Drude}},\ }\bibfield  {title} {\bibinfo {title} {Zur elektronentheorie der metalle; {II}. teil. galvanomagnetische und thermomagnetische effecte},\ }\href@noop {} {\bibfield  {journal} {\bibinfo  {journal} {Ann. Phys.}\ }\textbf {\bibinfo {volume} {308}},\ \bibinfo {pages} {369} (\bibinfo {year} {1900})}\BibitemShut {NoStop}%
\bibitem [{\citenamefont {Drude}(1902)}]{Drude1902-qx}%
  \BibitemOpen
  \bibfield  {author} {\bibinfo {author} {\bibfnamefont {P.}~\bibnamefont {Drude}},\ }\bibfield  {title} {\bibinfo {title} {Zur elektronentheorie der metalle},\ }\href@noop {} {\bibfield  {journal} {\bibinfo  {journal} {Ann. Phys.}\ }\textbf {\bibinfo {volume} {312}},\ \bibinfo {pages} {687} (\bibinfo {year} {1902})}\BibitemShut {NoStop}%
\bibitem [{\citenamefont {Ashcroft}\ and\ \citenamefont {Mermin}(1976)}]{Ashcroft1976-hn}%
  \BibitemOpen
  \bibfield  {author} {\bibinfo {author} {\bibfnamefont {N.~W.}\ \bibnamefont {Ashcroft}}\ and\ \bibinfo {author} {\bibfnamefont {N.~D.}\ \bibnamefont {Mermin}},\ }\href@noop {} {\emph {\bibinfo {title} {Solid State Physics}}}\ (\bibinfo  {publisher} {Holt-Saunders},\ \bibinfo {year} {1976})\BibitemShut {NoStop}%
\bibitem [{\citenamefont {Sommerfeld}(1928)}]{Sommerfeld1928-re}%
  \BibitemOpen
  \bibfield  {author} {\bibinfo {author} {\bibfnamefont {A.}~\bibnamefont {Sommerfeld}},\ }\bibfield  {title} {\bibinfo {title} {Zur elektronentheorie der metalle auf grund der fermischen statistik: {I}. teil: Allgemeines, stromungs- und austrittsvorgaange},\ }\href@noop {} {\bibfield  {journal} {\bibinfo  {journal} {Eur. Phys. J. A}\ }\textbf {\bibinfo {volume} {47}},\ \bibinfo {pages} {1} (\bibinfo {year} {1928})}\BibitemShut {NoStop}%
\bibitem [{\citenamefont {Li}\ and\ \citenamefont {Orignac}(2002)}]{Li2002-hg}%
  \BibitemOpen
  \bibfield  {author} {\bibinfo {author} {\bibfnamefont {M.-R.}\ \bibnamefont {Li}}\ and\ \bibinfo {author} {\bibfnamefont {E.}~\bibnamefont {Orignac}},\ }\bibfield  {title} {\bibinfo {title} {Heat conduction and wiedemann-franz law in disordered luttinger liquids},\ }\href@noop {} {\bibfield  {journal} {\bibinfo  {journal} {EPL}\ }\textbf {\bibinfo {volume} {60}},\ \bibinfo {pages} {432} (\bibinfo {year} {2002})}\BibitemShut {NoStop}%
\bibitem [{\citenamefont {Garg}\ \emph {et~al.}(2009)\citenamefont {Garg}, \citenamefont {Rasch}, \citenamefont {Shimshoni},\ and\ \citenamefont {Rosch}}]{Garg2009-hl}%
  \BibitemOpen
  \bibfield  {author} {\bibinfo {author} {\bibfnamefont {A.}~\bibnamefont {Garg}}, \bibinfo {author} {\bibfnamefont {D.}~\bibnamefont {Rasch}}, \bibinfo {author} {\bibfnamefont {E.}~\bibnamefont {Shimshoni}},\ and\ \bibinfo {author} {\bibfnamefont {A.}~\bibnamefont {Rosch}},\ }\bibfield  {title} {\bibinfo {title} {Large violation of the wiedemann-franz law in luttinger liquids},\ }\href@noop {} {\bibfield  {journal} {\bibinfo  {journal} {Phys. Rev. Lett.}\ }\textbf {\bibinfo {volume} {103}},\ \bibinfo {pages} {096402} (\bibinfo {year} {2009})}\BibitemShut {NoStop}%
\bibitem [{\citenamefont {Wakeham}\ \emph {et~al.}(2011)\citenamefont {Wakeham}, \citenamefont {Bangura}, \citenamefont {Xu}, \citenamefont {Mercure}, \citenamefont {Greenblatt},\ and\ \citenamefont {Hussey}}]{Wakeham2011-ob}%
  \BibitemOpen
  \bibfield  {author} {\bibinfo {author} {\bibfnamefont {N.}~\bibnamefont {Wakeham}}, \bibinfo {author} {\bibfnamefont {A.~F.}\ \bibnamefont {Bangura}}, \bibinfo {author} {\bibfnamefont {X.}~\bibnamefont {Xu}}, \bibinfo {author} {\bibfnamefont {J.-F.}\ \bibnamefont {Mercure}}, \bibinfo {author} {\bibfnamefont {M.}~\bibnamefont {Greenblatt}},\ and\ \bibinfo {author} {\bibfnamefont {N.~E.}\ \bibnamefont {Hussey}},\ }\bibfield  {title} {\bibinfo {title} {Gross violation of the wiedemann-franz law in a quasi-one-dimensional conductor},\ }\href@noop {} {\bibfield  {journal} {\bibinfo  {journal} {Nat. Commun.}\ }\textbf {\bibinfo {volume} {2}},\ \bibinfo {pages} {396} (\bibinfo {year} {2011})}\BibitemShut {NoStop}%
\bibitem [{\citenamefont {Mahajan}\ \emph {et~al.}(2013)\citenamefont {Mahajan}, \citenamefont {Barkeshli},\ and\ \citenamefont {Hartnoll}}]{Mahajan2013-kw}%
  \BibitemOpen
  \bibfield  {author} {\bibinfo {author} {\bibfnamefont {R.}~\bibnamefont {Mahajan}}, \bibinfo {author} {\bibfnamefont {M.}~\bibnamefont {Barkeshli}},\ and\ \bibinfo {author} {\bibfnamefont {S.~A.}\ \bibnamefont {Hartnoll}},\ }\bibfield  {title} {\bibinfo {title} {Non-fermi liquids and the wiedemann-franz law},\ }\href@noop {} {\bibfield  {journal} {\bibinfo  {journal} {Phys. Rev. B}\ }\textbf {\bibinfo {volume} {88}},\ \bibinfo {pages} {125107} (\bibinfo {year} {2013})}\BibitemShut {NoStop}%
\bibitem [{\citenamefont {Filippone}\ \emph {et~al.}(2016)\citenamefont {Filippone}, \citenamefont {Hekking},\ and\ \citenamefont {Minguzzi}}]{Filippone2016-an}%
  \BibitemOpen
  \bibfield  {author} {\bibinfo {author} {\bibfnamefont {M.}~\bibnamefont {Filippone}}, \bibinfo {author} {\bibfnamefont {F.}~\bibnamefont {Hekking}},\ and\ \bibinfo {author} {\bibfnamefont {A.}~\bibnamefont {Minguzzi}},\ }\bibfield  {title} {\bibinfo {title} {Violation of the wiedemann-franz law for one-dimensional ultracold atomic gases},\ }\href@noop {} {\bibfield  {journal} {\bibinfo  {journal} {Phys. Rev. A}\ }\textbf {\bibinfo {volume} {93}},\ \bibinfo {pages} {011602} (\bibinfo {year} {2016})}\BibitemShut {NoStop}%
\bibitem [{\citenamefont {Uchino}\ and\ \citenamefont {Brantut}(2020)}]{PhysRevResearch.2.023284}%
  \BibitemOpen
  \bibfield  {author} {\bibinfo {author} {\bibfnamefont {S.}~\bibnamefont {Uchino}}\ and\ \bibinfo {author} {\bibfnamefont {J.-P.}\ \bibnamefont {Brantut}},\ }\bibfield  {title} {\bibinfo {title} {Bosonic superfluid transport in a quantum point contact},\ }\href {https://doi.org/10.1103/PhysRevResearch.2.023284} {\bibfield  {journal} {\bibinfo  {journal} {Phys. Rev. Res.}\ }\textbf {\bibinfo {volume} {2}},\ \bibinfo {pages} {023284} (\bibinfo {year} {2020})}\BibitemShut {NoStop}%
\bibitem [{\citenamefont {Uchino}(2020)}]{PhysRevResearch.2.023340}%
  \BibitemOpen
  \bibfield  {author} {\bibinfo {author} {\bibfnamefont {S.}~\bibnamefont {Uchino}},\ }\bibfield  {title} {\bibinfo {title} {Role of nambu-goldstone modes in the fermionic-superfluid point contact},\ }\href {https://doi.org/10.1103/PhysRevResearch.2.023340} {\bibfield  {journal} {\bibinfo  {journal} {Phys. Rev. Res.}\ }\textbf {\bibinfo {volume} {2}},\ \bibinfo {pages} {023340} (\bibinfo {year} {2020})}\BibitemShut {NoStop}%
\bibitem [{\citenamefont {Nakata}\ \emph {et~al.}(2015{\natexlab{b}})\citenamefont {Nakata}, \citenamefont {Simon},\ and\ \citenamefont {Loss}}]{Nakata2015-ts}%
  \BibitemOpen
  \bibfield  {author} {\bibinfo {author} {\bibfnamefont {K.}~\bibnamefont {Nakata}}, \bibinfo {author} {\bibfnamefont {P.}~\bibnamefont {Simon}},\ and\ \bibinfo {author} {\bibfnamefont {D.}~\bibnamefont {Loss}},\ }\bibfield  {title} {\bibinfo {title} {Magnon transport through microwave pumping},\ }\href@noop {} {\bibfield  {journal} {\bibinfo  {journal} {Phys. Rev. B}\ }\textbf {\bibinfo {volume} {92}},\ \bibinfo {pages} {014422} (\bibinfo {year} {2015}{\natexlab{b}})}\BibitemShut {NoStop}%
\bibitem [{\citenamefont {Grenier}\ \emph {et~al.}(2012)\citenamefont {Grenier}, \citenamefont {Kollath},\ and\ \citenamefont {Georges}}]{grenier2012probing}%
  \BibitemOpen
  \bibfield  {author} {\bibinfo {author} {\bibfnamefont {C.}~\bibnamefont {Grenier}}, \bibinfo {author} {\bibfnamefont {C.}~\bibnamefont {Kollath}},\ and\ \bibinfo {author} {\bibfnamefont {A.}~\bibnamefont {Georges}},\ }\bibfield  {title} {\bibinfo {title} {Probing thermoelectric transport with cold atoms},\ }\bibfield  {journal} {\bibinfo  {journal} {arXiv preprint arXiv:1209.3942}\ }\href {https://doi.org/10.48550/arXiv.1209.3942} {10.48550/arXiv.1209.3942} (\bibinfo {year} {2012})\BibitemShut {NoStop}%
\bibitem [{\citenamefont {Holstein}\ and\ \citenamefont {Primakoff}(1940)}]{holstein1940field}%
  \BibitemOpen
  \bibfield  {author} {\bibinfo {author} {\bibfnamefont {T.}~\bibnamefont {Holstein}}\ and\ \bibinfo {author} {\bibfnamefont {H.}~\bibnamefont {Primakoff}},\ }\bibfield  {title} {\bibinfo {title} {{Field Dependence of the Intrinsic Domain Magnetization of a Ferromagnet}},\ }\href {https://doi.org/10.1103/PhysRev.58.1098} {\bibfield  {journal} {\bibinfo  {journal} {Phys. Rev.}\ }\textbf {\bibinfo {volume} {58}},\ \bibinfo {pages} {1098} (\bibinfo {year} {1940})}\BibitemShut {NoStop}%
\bibitem [{\citenamefont {Sekino}\ \emph {et~al.}(2020)\citenamefont {Sekino}, \citenamefont {Tajima},\ and\ \citenamefont {Uchino}}]{sekino2020mesoscopic}%
  \BibitemOpen
  \bibfield  {author} {\bibinfo {author} {\bibfnamefont {Y.}~\bibnamefont {Sekino}}, \bibinfo {author} {\bibfnamefont {H.}~\bibnamefont {Tajima}},\ and\ \bibinfo {author} {\bibfnamefont {S.}~\bibnamefont {Uchino}},\ }\bibfield  {title} {\bibinfo {title} {Mesoscopic spin transport between strongly interacting fermi gases},\ }\href {https://doi.org/10.1103/PhysRevResearch.2.023152} {\bibfield  {journal} {\bibinfo  {journal} {Phys. Rev. Research}\ }\textbf {\bibinfo {volume} {2}},\ \bibinfo {pages} {023152} (\bibinfo {year} {2020})}\BibitemShut {NoStop}%
\end{thebibliography}%

\end{document}